\documentclass{emulateapj}
\usepackage{subfigure}
\usepackage{color}
\usepackage[colorlinks,linkcolor=blue,citecolor=blue,
urlcolor=magenta]{hyperref}
\usepackage{amsmath}
\usepackage{cases}
\usepackage{natbib}
\usepackage{ccaption}

\shorttitle{Origin of the Stellar Mass--Stellar Metallicity Relation}
\shortauthors{Xia \& Yu}
\slugcomment{Accepted to The Astrophysical Journal}

\def\be{\begin{equation}}
\def\ee{\end{equation}}

\def\kms{{\rm \,km\,s^{-1}}}

\def\Mpc{{\rm \,Mpc}}

\def\msun{{\,M_\odot}}
\def\hot{{_{\rm hot}}}

\def\disk{{_{\rm disk}}}
\def\reheat{{_{\rm reheat}}}

\def\SN{_{\rm SN}}
\def\halo{{_{\rm halo}}}
\def\vir{_{\rm vir}}
\def\E{{_{\rm E}}}

\usepackage{verbatim}
\usepackage{etoolbox}

\begin{document}

\title{The Origin of the Stellar Mass--Stellar Metallicity Relation in the Milky Way Satellites and beyond}

\author{Moran Xia, \& Qingjuan Yu$^{\dagger}$}

\affil{Kavli Institute for Astronomy and Astrophysics, and School of Physics, Peking University, Beijing 100871, China; $^{\dagger}$yuqj@pku.edu.cn}


\begin{abstract}

\noindent

Observations and semianalytical galaxy formation and evolution models (SAMs)
have suggested the existence of a stellar mass--stellar metallicity relation
(MZR), which is shown to be universal for different types of galaxies over a
large range of stellar masses ($M_*\sim 10^3$--$10^{11}\msun$) and dark matter
(DM) halo masses ($M_{\rm halo}\sim 10^9$--$10^{15}h^{-1}\msun$). In this work,
we construct a chemical evolution model to investigate the origin of the MZR,
including both the effects of gas inflows and outflows in galaxies. We solve
the MZR from the chemical evolution model, by assuming that the cold gas mass
($M_{\rm cold}$) and the stellar feedback efficiency ($\beta$) follow
some power-law scaling relationships with $M_*$ during the growth of a galaxy,
i.e., $M_{\rm cold}\propto M_*^{\alpha_{\rm gs}}$ and $\beta\propto
M_*^{\alpha_{\beta{\rm s}}}$.  We use the SAM to obtain these power-law scaling
relations, which appear to be roughly universal over a large range of stellar
masses for both satellites and central galaxies within a large range of halo
masses.  The range of the MZRs produced by our models is in a narrow space,
which provides support to the universality of the MZRs. The formation of the
MZR is a result caused jointly by that the cold gas fraction decreases with
increasing $M_*$ and by that the stellar feedback efficiency decreases with
increasing $M_*$ in the galaxy growth, and the exponent in the MZR is around
$-\alpha_{\beta{\rm s}}$ or $1-\alpha_{\rm gs}$.  The MZR represents an
``average'' evolutional track for the stellar metallicity of a galaxy.  The
comparison of our model with some previous models for the origin of MZRs is
also discussed.

\end{abstract}

\section{Introduction}
\label{sec:intro}

The stellar mass and the stellar metallicity are among the most basic physical
properties of a galaxy.  Both observations and semi-analytical galaxy formation
and evolution model (SAM) simulations reveal that the stellar metallicity
correlates with the stellar mass in both dwarf satellites and normal galaxies.
For example, \citet{Kirbyetal13} find that there exists a universal stellar
mass--stellar metallicity relation in the Milky Way/M31 dwarf satellites and
some other dwarf irregular galaxies in the Local Group, $Z_*\propto
M_*^{0.30\pm 0.02}$.  The SDSS spectra of over 40,000 galaxies show that local
galaxies with stellar masses in the range of $10^9 \la M_*/\msun \la 10^{12}$
also indicate that stellar masses have a correlation with stellar metallicities
or gas metallicities
(\citealt{gallazzi05,gallazzi06,tremonti04,Panteretal08,mannucci10,GDetal14,Leeetal06}).
Hereafter, for simplicity, we refer to the correlation between the stellar mass
and stellar metallicity as ``MZR'' (see also review in \citealt{MM19}).

The MZR in the dwarf satellites of the
Milky Way ($10^3\msun<M_*<10^8\msun$ or $10^9\msun$) was reproduced in the SAMs
(e.g., \citealt{font2011,Lietal10,Luetal14,Luetal17}; \citealt{HYL14},
hereafter HYL14). 
By using SAMs, \citet{Y13}
obtain the stellar mass--cold gas metallicity relation for central galaxies
with relatively high stellar mass range $M_*\sim 10^9$--$10^{11}\msun$,
and \citet{Somervilleetal15} present the stellar mass--cold gas metallicity
relation for central galaxies with stellar mass range $M_*\sim
10^7$--$10^{11}\msun$. 
\citet{XY19} use the SAM to produce the MZRs in galaxies with a
larger range of stellar masses $M_*\sim 10^3$--$10^{11}\msun$, including both
dwarf satellites and normal galaxies within a larger range of host halo masses
($M_{\rm halo}\sim 10^9-10^{15}h^{-1}\msun$). In this paper, we investigate the
origin of the universal MZRs, by constructing an analytical chemical evolution
model of galaxies and applying the reference results from the SAM.  The study
of the origin will help us to understand the physical processes involving
stellar formation and evolution, gas inflow and outflow, and chemical evolution
in galaxies.

One key to understanding the origin of the MZR is to model the galaxy metal
enrichment processes.  The modeling of galaxy metal enrichment has evolved
from a simple ``closed-box'' model to relatively complex ones including the
effects of gas inflow and outflow in the past fifty years. In the
``closed-box'' model \citep{schmidt63,TA71,SS72}, an isolated galaxy formation
process starts from pure metal-free gas,
and no gas escapes from or is accreted onto the galaxy.  Some later modified
models such as the ``leaky-box'' model, the ``accretion'' model \citep{lb75},
and the ``pre-enriched'' model \citep{pagel97}, include gas
outflow from a galaxy, gas inflow, and initial non-zero metallicity of
inflowing gas, respectively. The effects of both gas inflow and outflow are
included in the models of \citet{FD08}, \citet{Lillyetal13}, and \citet{PM14}.

Metal-enriched gas outflow is proposed to play an important role in shaping the
MZR. For example, the positive correlation in the MZR can be produced in the
following cases. (1) The outflows can be stronger for small galaxies due to
their shallower gravity potential wells, and thus small galaxies lose more
metals than larger ones \citep{DS86}. (2) The outflow wind strength is
ubiquitous for all the galaxies, but the wind metallicity is larger than the
metallicity in the inter-stellar medium (ISM) of smaller galaxies
\citep{kobayashi07}. Though gas inflows from host halos may dilute the
metallicity in the ISM, \citet{dalcanton07} points out that the inflow dilution
in the metallicity cannot contribute much to the low effective yield of small
galaxies, neither can the metal-unenriched outflow. \citet{FD08} propose an
equilibrium metallicity to explain the MZR, where the metallicity is obtained
by assuming that the gas inflow rate is balanced by the gas consumption rate
caused by star formation plus the gas outflow rate. \citet{Lillyetal13} relax
the assumption of gas processing equilibrium, but assume an enrichment
equilibrium to obtain the equilibrium metallicity of the gas, where metal
dilution balances metal enrichment in gas metallicity (see also a review
in \citealt{F16}).
 
In addition to the above general proposals to shape the MZR by using gas
outflow and inflow, there are also some models specifically involving star
formation rates and histories to explain the MZR. For example, the positive
correlation in the MZR is explained in \citet{Brooksetal07} by relatively low
star formation efficiencies in low-mass galaxies and in \citet{VAetal09} by
that the majority of the stars form earlier in large galaxies and the metal
enrichment in large galaxies proceeds earlier.  \citet{koppen07} suggests that
the positive correlation in the MZR can be produced, if large galaxies have a
top-heavy initial stellar mass function and there are more massive stars in
large galaxies to produce large effective yields. 

In this paper, we present a chemical evolution model to explain the origin of
the MZR. Previously, we have shown in HYL14 that the MZR in the satellites of
MW-like galaxies produced from the SAM matches the observation, given a
specific set of parameters (i.e., the fiducial model parameters in HYL14), such
as on the physical processes of feedback and the reionization of the universe.
Then we fix these parameters and explore the variations as a result of changing
halo masses ($M_{\rm halo}\sim 10^9$--$10^{15}h^{-1}\msun$), central galaxies
($M_*\sim 10^3$--$10^{11}\msun$), and dwarf satellites, and we find that the
MZR is quite independent of changing these variations as long as the fiducial
model parameters are fixed \citep{XY19}.  In this work, by using the results
obtained from the SAM to serve as the quantitative reference to some physical
galaxy properties involved in the chemical evolution model, we present an
explanation to the universal MZR.  This paper is organized as follows. In
Section~\ref{sec:method}, we give a short overview of the SAM used in this work
(for more details, see HYL14 and \citealt{XY19}). In Section~\ref{sec:model},
we present our analytic model in the galaxy metallicity evolution. In this
chemical evolution model, the evolution of the cold gas mass and the stellar
feedback efficiency with the growth of the stellar mass are involved and
assumed to follow some power-law scaling relationships. In
Section~\ref{sec:result}, we show these power-law scaling relations obtained
from the SAM, and apply them to the chemical evolution model constructed in
Section~\ref{sec:model}. Then we use the chemical evolution model results to
give an explanation to the universal MZRs.  A summary is given in
Section~\ref{sec:conclusions}. 

In this paper we set the Hubble constant as $H_0=100\,h\kms\Mpc$, and the
cosmological model used is $(\Omega_{\rm
m},\Omega_\Lambda,h,\sigma_8)=(0.25,0.75,0.70,0.90)$.

\section{The semi-analytical galaxy formation and evolution model (SAM)}
\label{sec:method}

In this section, we briefly describe the SAM used in this work, which will
provide some quantitative reference to some physical properties involved in the
chemical evolution model and help to explore the origin of the MZR of MW dwarf
satellites and beyond. 

The SAM used in this work is based on the GALFORM (\citealt{Coleetal00}; see
also \citealt{WhiteFrenk91,Kauffmann93,Somerville99,Somervilleetal08}), with
several modifications by HYL14 (e.g., on gas cooling, stellar feedback, and metal
enrichment). The model
includes planting DM halo merger trees and then incorporating semi-analytical
recipes of the baryonic physical processes into the halo merger trees, to trace
the hierarchical galaxy formation and evolution until redshift $z=0$
(\citealt{Coleetal00}; see also
\citealt{WhiteFrenk91,Kauffmann93,Somerville99,Somervilleetal08}. The DM halo
merger trees used in this work are generated from a modified version of the
extended Press-Schechter formula \citep{PS74,Bondetal91,LC93,Somervilleetal08},
developed by \citet{parkinson08} (see also
\citealt{WhiteFrenk91,Kauffmann93,SK99,Coleetal00}).  For each DM halo mass at
$z=0$, we use the Monte Carlo method to generate a number of merger trees
(e.g., from 10 to 1000) in order to do a statistical study. The progenitor
halos of each tree planted in this work can be dated back to high redshift
$z=20$. As shown in \citet{XY19}, to address the universality of the MZR, we
studied galaxies and DM halos with different masses by using the SAM and the
Monte Carlo method, so that the physical properties of galaxies with different
masses and their satellites in different environments can be explored.  

The SAM recipes are modified in HYL14 to make the model suitable for studying
low-mass dwarf galaxies/satellites. Some relevant recipes are briefly described
below. More details can be found in HYL14 (or Section 2 in \citealt{XY19}).

\begin{itemize}

\item Gas cooling:
The initial temperature of the hot gas in the newly formed halo is assumed to
be the same as the virial temperature of the DM halo.  The cooling rate of the
hot gas depends on its temperature, mass density, and metallicity, and we adopt
the cooling rate for atomic cooling from \citet{SD93}.  The detailed cooling
recipe is based on the GALFORM model by calculating how much hot halo gas can
cool down and ``free-fall'' to the halo center within a certain time, and
modified by HYL14 by including the mixing of the gas reheated by stellar
feedback with hot halo gas (mainly occurred in low-mass galaxies).

The molecular hydrogen cooling of pristine gas is also included in mini-halos
before the completeness of the reionization of the universe (see Equations
21-29 in \citealt{Benson10}; see also \citealt{galli1998}). As pointed out in
HYL14 that the MZR of the satellites in MW-like galaxies is not sensitive to
the molecular hydrogen cooling process.

\item Star formation and stellar feedback: 
The star formation rate is proportional to the mass of cold gas in the disk,
given by
\begin{eqnarray}
&& \psi =M_{\rm cold}/\tau _* \\
&& \tau _*=\epsilon _*^{-1}\tau\disk(V\disk/200\kms)^{\alpha _*},
\label{eq:sim_sf}
\end{eqnarray}
\noindent
where $\psi$ is the instantaneous star formation rate, $\tau_*$ is the star
formation time scale, $\tau\disk$ is the dynamic time scale of the galaxy
disk, $V\disk$ is the disk rotation velocity, and $\epsilon_*=0.005$ and
$\alpha_*=-1.5$ are two parameters (see eq.\ 4.14 in \citealt{Coleetal00}).

SNe and stellar winds will heat up the cold gas in the disk and possibly expel
it out of the galaxy as outflows.  During time interval $dt$, the reheated gas
mass is given by:
\begin{eqnarray}
&& dM\reheat=\beta \psi dt,
\label{eq:dMreheat}
\\
&& \beta =(V\disk/V\hot)^{-\alpha\hot},
\label{eq:sim_fb}
\end{eqnarray}
where $\beta$ is the feedback efficiency, and $V\hot$ and $\alpha\hot$ are the
two parameters defining the strength of the feedback.
The cold gas mass that can be expelled from the disk is constrained by the
energy released by SNe and coupling to the IGM (see also \citealt{Guoetal11}),
with the following inequality:
\begin{equation} dE\SN-\frac{1}{2}V\vir^2dM\reheat>0,
\label{eq:ener_condi} \end{equation} where
$dE\SN=\epsilon\halo\times\frac{1}{2}V\SN^2\psi dt$ is the total energy
released by SNe and coupling to the IGM during time $dt$, $V\vir$ is the virial
velocity of the halo, $\frac{1}{2}V\SN^2$ is the total energy released per unit
mass by SNe with $V\SN=630\kms$, and $\epsilon\halo=0.05$ is the fraction of
the energy that couples to the cold gas in the disk.

If inequality (\ref{eq:ener_condi}) is satisfied, then all the reheated gas can
be expelled out of the disk (part or all of the reheated gas can even escape
the DM halo). In this case, $\beta$ in the chemical evolution model described
in Section~\ref{sec:model} (Equations~\ref{eq:df2} and \ref{eq:df3}) can still
be obtained through Equation (\ref{eq:sim_fb}).  If inequality
(\ref{eq:ener_condi}) is not satisfied (normally for low-mass galaxies or the
progenitors of satellites, e.g., $M_*\la 10^8\msun$), the mass of the reheated
gas that can be expelled out of the disk during time interval $dt$ is given by
$\beta\E\psi dt$, where \begin{equation} \beta\E\equiv
dE\SN/(\frac{1}{2}V^2\vir\psi dt)=\epsilon\halo (V\vir/V\SN)^{-2},
\label{eq:betaE} \end{equation} and the expelled gas stays in the halo (see
more details in section 2 in HYL14).  In this case, $\beta$ in the chemical
evolution model described in Section~\ref{sec:model} is $\beta=\beta\E$.

Note that for satellites, $\beta$ in the chemical evolution model described in
Section~\ref{sec:model} is also obtained through Equation (\ref{eq:sim_fb}) for
the following reason. In the assembling history of a halo, a satellite of a
central galaxy may be the host galaxy of a small isolated halo before it fell
into the big halo at an early time.  After it fell into the big halo, the
original halo of the satellite can be largely tidally disrupted along its
motion in the big host halo, and thus we assume that all the reheated gas from
satellites (with mass expected by Equations~\ref{eq:dMreheat} and
\ref{eq:sim_fb}) is expelled into the big host halo. The tidal stripping and
disruption of the stellar and cold gas components of the satellites are not
considered in our model, as they are located in a smaller central region
compared with their original halo size.

The parameters in Equation (\ref{eq:sim_fb}) are set to be $V\hot=200\kms$ and
$\alpha\hot=3.2$, i.e. the fiducial model in HYL14, which can reproduce some
observational properties of the MW dwarfs (e.g., the satellite luminosity
function, the MZR) better than the other models (see also Figure 9 in HYL14 for
the changes to the MZRs caused by other sets of the two parameters).  Although
these parameters were set to be compatible with the observational properties of
the dwarf satellites in HYL14, the MZR for galaxies covering a large stellar
mass range ($10^{3}\msun\la M_*\la 10^{13}\msun$) can also be reproduced with
the same sets of the parameters in \citet{XY19}.

Through cosmological zoom-in numerical simulations for single central galaxies,
\citet{Maetal16} show that a large fraction of metals ejected by stellar
feedback can be retained inside a halo (as illustrated by Figure 11 therein)
even for low-mass galaxies down to $M_*\sim 10^6\msun$ and the retained metals
can further rain down back to galaxy disks), which provides an alternative way
to meet with both the observational stellar mass function and the MZR of dwarf
galaxies. This scenario has a similar effect as the application of Equation
(\ref{eq:ener_condi}) or (\ref{eq:betaE}) in the sense of retaining metals
inside halos. Satellites are not included in the work of \citet{Maetal16}.

\item Metallicity enrichment: The metals ejected by SNe are assumed to first
instantaneously and homogeneously mixed with the ISM in the galaxy, and after
the mixture, some metals can be ejected out of the galaxy along with the mixed
ISM that is ejected out by SN explosions. The ejected gas is assumed to have
the same metallicity as the cold gas. In this work, the Fe yield of SNe II is
adopted from tables 2–3 in \citet{SN_II_pattern}, and the Fe yield of SNe Ia is
from \citet{SN_Ia_pattern}.  As done in HYL14, we assume that SN II explosions
are instantaneous after the formation of SN II progenitors, and SN Ia
explosions have a nonnegligible time delay after the formation of SN Ia
progenitors. The SN Ia event rate adopted in HYL14 is based on the
observational results by \citet{MSG10}, in which a minimum time delay of
0.1~Gyr since the birth of a stellar population is reported. 

\item The reionization of the universe: the reionization in the early universe
reduces the baryon fraction of a DM halo to be below the cosmic average (e.g.,
\citealt{gnedin00}).
The reduced fraction can be modeled through a mass scale called the filtering
mass. In this work (also in HYL14), we adopt the recipes by
\citet{kravtsov04} (see also \citealt{gnedin00,Okamoto08}) to calculate the
filtering mass and model the effects of the reionization.

The filtering mass is a function of redshift, as well as a function of the
starting redshift $z_0$ (when the first HII bubble formed) and the completing
redshift $z_r$ of the reionization history.  In this work, we set $z_0=15$ and
$z_r=10$, as done in the fiducial model in HYL14. This relatively earlier
reionization of the universe produces a stronger effect in reducing the
baryonic fraction in low-mass DM halos, resulting in a relatively shallower
slope or a smaller exponent in the MZR at the low-mass end (which agrees with
the observation better, according to HYL14).  Throughout the paper, the word
``slope'' and the word ``exponent'' of a power-law relationship are
equivalently used.  \citet{Luetal17} also discuss the importance of preventive
feedback in meeting with both the observational stellar mass function and the
MZR of dwarf galaxies, which requires a strong prevention of baryons from
collapsing into low-mass halos in the first place and has a similar effect as
this early reionization scenario.

\end{itemize}

The parameters used in the recipes are chosen from the fiducial model in HYL14.

In \citet{XY19}, we use the above SAM to investigate the universality of the
MZRs for both satellites and central galaxies in a large number of different
host DM halos covering a large mass range of
$10^9$--$10^{15}h^{-1}\msun$. We find that the satellites in those simulated
host halos follow a similar relation, with the exponent $\alpha$ being in the
same range $\sim$0.2--0.4. The simulated central host galaxies with
$10^3\msun\la M_*\la 10^{11}\msun$ follow a similar relation.  That study
further shows that a double power law provides a better fit to the MZR than the
above single power law for both satellites and central galaxies, which gives
$\alpha\sim$0.2--0.4 at $10^3\msun\la M_*\la10^{8}\msun$ and a relatively
higher $\alpha\sim0.5$ at $10^8\msun\la M_*\la10^{11}\msun$. The stellar
metallicity in massive galaxies with $M_*\ga10^{11}\msun$ becomes roughly
constant, close to the metal yield.  The difference in the best-fit
normalizations of the MZRs is within a small factor (e.g., $\sim 2$ at
$M_*=10^6\msun$ and $\sim4$ at $M_*=10^3\msun$).  The relatively narrow space
of the fit slopes and normalizations revealed a universality in the MZRs, which
suggests the common physical processes in the stellar formation and chemical
evolution of the galaxies can be unified with a large range of galaxy masses
and halo masses.

\section{Analytic chemical evolution Models}
\label{sec:model}

To investigate the metallicity evolution in a galaxy, we start with the
following conservation laws of the total stellar mass, cold gas mass, and metal
mass within the galaxy (see eq.~4.6-4.11 in \citealt{Coleetal00}):
\begin{eqnarray}
&&\dot{M}_*=(1-R)\psi, \label{eq:df1}\\
&&\dot{M}_{\text{cold}}=\dot{M}_{\text{cool}}-(1-R+\beta)\psi, \label{eq:df2}\\
&&\dot{M}_*^Z=(1-R)Z_{\rm cold}\psi, \label{eq:dfmstarz} \\
&&\dot{M}_{\text{cold}}^Z=\dot{M}_{\text{cool}}Z_{\text{hot}}+[p-(1-R+\beta
)Z_{\text{cold}}]\psi, \label{eq:df3}
\end{eqnarray}
where $M_*$, $M_{\rm cold}$, $M_*^Z$, and $M_{\text{cold}}^Z$ are
the total stellar mass, the total cold gas mass, the total metal mass in stars,
and the total metal mass in cold gas, respectively, the overdot $\dot{}$ above
a variable represents its derivative with respect to time $d/dt$, i.e., the
changing rates of the corresponding variable, $\dot{M}_{\text{cool}}$
represents the gas mass cooling rate from the hot gas in the DM
halo to the cold gas in the galaxy disk, $\psi$ is the instantaneous star
formation rate, $R$ is the fraction of mass recycled by stars to the
ISM (through stellar winds and SNe), $Z_{\text{cold}}\equiv
M_{\text{cold}}^Z/M_{\text{cold}}$ is the metallicity of the cold
gas, $Z_{\text{hot}}$ is the metallicity of the hot gas in the DM
halo, $\dot{M}_{\text{cool}}Z_{\text{hot}}$ represents the cooling rate of the
gas in metals, $p$ denotes the yield (the fraction of mass converted into stars
that is returned to the ISM in the form of metals), and $\beta$ is the
efficiency of stellar feedback.

The stellar mass $M_*$ is a monotonically increasing function with time $t$,
and thus a physical variable (e.g., metallicity) as a function of time can also
be expressed as a function of $M_*$. One advantage of using $M_*$ as the
independent variable instead of $t$ is that some complexity in the difference
of star formation histories for different galaxies (which can be expressed
through the function $M_*(t)$) can be partly removed in understanding the
origin of the MZR. Another advantage of using $M_*$ as the independent variable
is that mass is a physical variable of a system relating to its gravitational
potential and some variables (e.g., $\beta$) is constrained by the
gravitational potential of the system.

The mean metallicity in stars
$Z_*\equiv M_*^Z/M_*$ as a function of stellar mass can be obtained by a
combination of Equations (\ref{eq:df1}) and (\ref{eq:dfmstarz}) and
an integration over the star formation history of the galaxy as follows,
\begin{eqnarray}
Z_*(M_*)& = & \frac{1}{M_*}\int_0^{M{_*^Z}} d M{_*^{Z}}'\\
& = & \frac{1}{M_*}\int _0^{M_*}Z_{\text{cold}}(M_*{'})dM_*{'}, \\
& =& \frac{M_{*,0}Z_{*,0}}{M_*}+\frac{1}{M_*}\int_{M_{*,0}}^{M_*} Z_{\text{cold}}\left(M_*^{\prime }\right) dM_*^{\prime }, \label{eq:zs}
\end{eqnarray}
where $M_{*,0}$ and $Z_{*,0}$ are the initial stellar mass and the initial
metallicity of the system, respectively.

A combination of Equations (\ref{eq:df1}) and (\ref{eq:df2}) gives
\begin{equation}
\dot{M}_{\text{cool}}=\dot{M}_{\text{cold}}+\frac{1-R+\beta }{1-R}\dot{M}_*, \label{eq:df4}
\end{equation}
and then applying Equation (\ref{eq:df4}) into Equation (\ref{eq:df3}) yields
\begin{eqnarray}
d(M_{\text{cold}}Z_{\text{cold}})-Z_{\text{hot}}\cdot dM_{\text{cold}}=\frac{p}{1-R} dM_* \nonumber\\
-\frac{1-R+\beta }{1-R}(Z_{\text{cold}}-Z_{\text{hot}}) dM_*. \label{eq:mother}
\end{eqnarray}
Below we show how $Z_{\rm cold}$ can be solved from the above equation,
given some assumption about $Z_{\text{hot}}$ or its explicit form (e.g.,
related to $Z_{\rm cold}$ or $M_*$).

\subsection{Some previous analytical models}
\label{sec:pm}

Equation (\ref{eq:mother}) is a basic equation to describe the chemical
evolution in a galaxy, which can be reduced to many previous
well-known metallicity enrichment models, e.g. the ``closed-box'' model
(\citealt{schmidt63,TA71,SS72}), the ``leaky-box'' model, the
``pre-enriched'' model (\citealt{pagel97}), and the ``accretion'' model
(\citealt{lb75}). In most of those models, a constant (or zero) hot halo gas
metallicity $z_{\rm hot}$ is assumed.
 
For the simple case of $Z_{\text{hot}}=0$, where the hot gas in the DM
halo is metal-free and its cooling and accretion onto the galactic disk is a
metal dilution process in the cold gas, Equation (\ref{eq:mother}) is reduced
to
\begin{equation}
d\left(M_{\text{cold}}Z_{\text{cold}}\right)=\frac{p}{1-R}d M_*-\frac{1-R+\beta
}{1-R}Z_{\text{cold}}d M_*. \label{eq:bd}
\end{equation}
As mentioned above, here a physical variable as a function of time can be
expressed as a function of $M_*$, and thus the above differential equation has
an analytic solution as follows,
\begin{eqnarray}
Z_{\text{cold}}=\frac{M_{\text{cold},0}Z_{\text{cold},0}}{M_{\text{cold}}}\exp\left(-\int_{M_{*,0}}^{M_*} \frac{1-R+\beta }{1-R} \frac{dM_*'}{M_{\text{cold}}}\right)\nonumber\\
+\frac{p}{1-R}\frac{1}{M_{\text{cold}}}\exp\left(-\int _c^{M_*}\frac{1-R+\beta }{1-R}\frac{dM_*^{'}}{M_{\text{cold}}}\right)\nonumber\\
\cdot\int_{M_{*,0}}^{M_*}\exp\left(\int _c^{M_*^{'}}\frac{1-R+\beta }{1-R}\frac{dM_*^{''}}{M_{\text{cold}}}\right)dM_*^{'},~~~~~~~\label{eq:bi}
\end{eqnarray}
where $M_{\text{cold},0}$ is the initial cold gas mass, $Z_{\text{cold},0}$ is
the initial cold gas metallicity,
$c$ is an arbitrary constant and can be canceled out in the
calculation, $M_*^{'}$ and $M_*^{''}$ are two dummy variables of stellar
masses, and $Z_{\text{cold}}=Z_{\text{cold},0}$ when $M_*=M_{*,0}$ and
$M_{\text{cold}}=M_{\text{cold},0}$.  Below we show that the solutions in
some previous models can be obtained by reducing Equations (\ref{eq:bd}) and
(\ref{eq:bi}) with an assumption that $p$ and $R$ are constant.

\begin{itemize}
\item Leaky-box model. In this case, we set $\dot{M}_{\text{cool}}=0$ and the
feedback efficiency $\beta$ to be constant (which becomes the closed-box model
below if $\beta=0$), and thus Equation (\ref{eq:df4})
becomes
\begin{equation}
d M_*=-\frac{1-R}{1-R+\beta} d M_{\text{cold}}. \label{eq:df5}
\end{equation} 
By applying Equation (\ref{eq:df5}) into Equation (\ref{eq:bi}),
we obtain the solution of the cold gas metallicity as follows,
\begin{equation}
Z_{\text{cold}}=Z_{\text{cold,0}}+\frac{p}{1-R+\beta }\ln  \frac{M_{\text{cold,0}}}{M_{\text{cold}}},\label{eq:leakyc}
\end{equation}
and an application of Equation (\ref{eq:leakyc}) into Equation (\ref{eq:zs})
yields the stellar metallicity as follows,
\begin{eqnarray}
Z_* & = & \frac{M_{*,0}}{M_*}Z_{*,0}+\left(1-\frac{M_{*,0}}{M_*}\right)Z_{\text{cold,0}} \nonumber \\ 
& & +\frac{p(1-R)}{(1-R+\beta )^2} \cdot\frac{M_{\text{cold,0}}}{M_*} \nonumber \\
& & \cdot \left(1-\frac{M_{\text{cold}}}{M_{\text{cold,0}}}+\frac{M_{\text{cold}}}{M_{\text{cold,0}}}\ln  \frac{M_{\text{cold}}}{M_{\text{cold,0}}}\right).~~~~\label{eq:leakys}
\end{eqnarray}
When $M_{\text{cold}}\ll M_{\text{cold,0}}$, we have $M_{\text{cold,0}}\simeq
(M_*-M_{*,0})(1-R+\beta)/(1-R)$, and 
Equation (\ref{eq:leakys}) can be transformed to be:
\begin{eqnarray}
M_*\cdot Z_*=M_{*,0}\cdot Z_{*,0}+\left(M_*-M_{*,0}\right)Z_{\text{cold,0}}\nonumber\\
+\left(M_*-M_{*,0}\right)\frac{p}{1-R+\beta},\label{eq:lk2s0}
\end{eqnarray}
where the term on the left-hand side is the total metal mass in stars, the
first term on the right-hand side is the initial metal mass in stars, the
second term is the metal mass contributed from the initial cold gas to the
newly formed stars (with mass $M_*-M_{*,0}$), and the third term is the
effective metal mass due to SN nucleosynthesis in the newly formed stars
(which is first released into the cold gas reservoir and then returned to the
newly formed stars). Equation (\ref{eq:lk2s0}) can be transformed to be:
\begin{eqnarray}
Z_*-Z_{*,0} = (\frac{p}{1-R+\beta}+Z_{\rm cold,0}-Z_{*,0}) \nonumber \\
 \times (1-M_{*,0}/M_*).
\end{eqnarray}

Then we have $Z_* \rightarrow Z_{\text{cold,0}}+\frac{p}{1-R+\beta}$ if
$M_*\gg M_{*,0}$, and we have $Z_*\rightarrow Z_{*,0}$
if $M_*\gg \left(\frac{1-R}{1-R+\beta}\right) M_{\rm cold,0}$.

In a traditional leaky-box model with $Z_{\text{cold,0}}=Z_{*,0}=0$ and
$M_{*,0}=0$, where the galaxy evolves from a pure gas cloud with a total
baryonic mass $M_0=M_{\text{cold,0}}$, Equations (\ref{eq:leakyc})
and (\ref{eq:leakys}) are reduced to 
\begin{equation}
Z_{\text{cold}}=\frac{p}{1-R+\beta}\ln f_{\text{gas}}^{-1},\label{eq:lk1c}
\end{equation}
\begin{equation}
Z_*=\frac{p}{1-R+\beta }\left(1+\frac{f_{\rm gas}}{1-f_{\rm gas}}\ln f_{\rm gas}\right).\label{eq:lk1s}
\end{equation}
where $f_{\text{gas}}\equiv M_{\text{cold}}/M_{\text{0}}$ is the fraction of the
gas mass relative to the initial total baryonic mass.
When $f_{\rm gas}\ll 1$, the stellar metallicity shown
by Equation (\ref{eq:lk1s}) can be reduced to the following form:
\begin{equation}
Z_*=\frac{p}{1-R+\beta},
\end{equation}
which is usually called the effective yield \citep{Coleetal00}.  If
$\left(\frac{1-R+\beta}{1-R}\right)\frac{M_*}{M_{\rm cold}}\ll 1$, we have
$Z_{\rm cold}=\frac{p}{1-R}\frac{M_*}{M_{\rm cold}}$ and
$Z_*=\frac{1}{2}\frac{p}{1-R}\frac{M_*}{M_{\rm cold}} \propto
\frac{M_*}{M_{\rm cold}}$ (cf., \citealt{DS86}).

\item Closed-box model. If the feedback efficiency $\beta=0$ is set in the above
traditional leaky-box model, Equations (\ref{eq:lk1c}) and (\ref{eq:lk1s}) are
reduced to
\begin{equation}
Z_{\text{cold}}=\frac{p}{1-R}\ln f_{\text{gas}}^{-1}\label{eq:closec}
\end{equation}
and
\begin{equation}
Z_*=\frac{p}{1-R }\left(1+\frac{f_{\rm gas}}{1-f_{\rm gas}}\ln f_{\rm gas}\right).\label{eq:closes}
\end{equation}

\item Accretion model. In \cite{lb75}, gas leaking is not considered (i.e.,
$\beta=0$), but a metal-free accretion is assumed, together with a specified
relation between the cold gas mass and the stellar mass (so that the mass
accretion rate is implicitly given).  The specified relation between the cold
gas mass and the stellar mass in \cite{lb75} is shown below (see eqs.~4.4 and
4.5 therein):
\begin{eqnarray}
M_{\text{cold}}=(\frac{M_*+\Pi}{M_{\infty}+\Pi})(M_{\infty}-M_*),
\label{eq:lbrelation} \\
\Pi=\frac{M_0M_{\infty}}{M_{\infty}-M_0},\quad\quad\quad\quad\quad
\end{eqnarray}
where $M_0$ and $M_\infty$ are the initial and final baryonic masses of the
galaxy, respectively.  In the above relation, when a galaxy just formed with
$M_{*,0}=0$, all the baryonic mass is in the cold gas with $M_{\text{cold,0}}=M_0$ and $Z_{\text{cold,0}}=0$;
and when star formation approaches the end, we have $M_{\text{cold}}=0$ and
$M_*=M_{\infty}$.
In this case, Equation (\ref{eq:bi}) can be reduced to
\begin{eqnarray} Z_{\text{cold}}=\frac{p}{1-R}\frac{1}{M_{\text{cold}}}\exp
\left(-\int _c^{M_*}\frac{dM_*'}{M_{\text{cold}}}\right)\nonumber\\
\cdot\int _0^{M_*}\exp \left(\int _c^{M_*^{\prime }}\frac{dM_*^{\prime\prime
}}{M_{\text{cold}}}\right)dM_*^{\prime }. \label{eq:lbi}
\end{eqnarray}
By applying Equation (\ref{eq:lbrelation}) into Equations
(\ref{eq:lbi}) and (\ref{eq:zs}), the cold gas metallicity (see also eq.~4.8 in
\citealt{lb75}) and the stellar metallicity can be obtained as follows,
\begin{eqnarray}
Z_{\text{cold}}=\frac{p}{1-R}\left(\frac{M_{\infty}+\Pi }{M_*+\Pi }\right)^2~~~~~~~~~~~~~~~~~~~~~~~~\nonumber\label{eq:lbzc}\\
\cdot\left[-\ln (1-M_*/M_{\infty})-\frac{M_*}{M_{\infty}+\Pi }\right],\\
Z_*=\frac{p}{1-R}\cdot \frac{M_{\infty }+\Pi }{M_*+\Pi }~~~~~~~~~~~~~~~~~~~~~~~~~~~~~~~~~\nonumber\label{eq:lbzs}\\
\cdot\left[1+\left(\frac{M_{\infty }}{M_*}-1\right)\ln \left(1-M_*/M_{\infty }\right)\right].
\end{eqnarray}
When $M_*\rightarrow M_\infty$, we have $Z_* \rightarrow \frac{p}{1-R}$.

\item Pre-enriched model. In \cite{pagel97}, the hot halo gas is assumed to have
a constant non-zero metallicity $Z_{\text{hot}}=Z_{\rm hot,0}>0$ (pre-enriched)
before its cooling onto the galaxy disk. Equation (\ref{eq:mother}) becomes
\begin{eqnarray}
d\left[M_{\text{cold}}(Z_{\text{cold}}-Z_{\rm hot,0})\right]=\frac{p}{1-R}d M_*
\nonumber \\
-\frac{1-R+\beta}{1-R}(Z_{\text{cold}}-Z_{\rm hot,0})d M_*, \label{eq:bd2}
\end{eqnarray}
which has a similar form as Equation (\ref{eq:bd}) if $Z_{\text{cold}}-Z_{\rm hot,0}$ is taken as one variable.
The solution of $Z_{\rm cold}$ in
Equation (\ref{eq:bd2}) can be obtained by replacing $Z_{\rm cold}$ and
$Z_{\rm cold,0}$ with $Z_{\rm cold}-Z_{\rm hot,0}$ and $Z_{\rm cold,0}-Z_{\rm
hot,0}$, respectively, in Equation (\ref{eq:bi}).

\item Equilibrium models including gas inflows and outflows.  In the
metallicity evolution model by \citet{FD08}, it is assumed that $dM_{\rm
cold}=0$, $dZ_{\rm cold}=0$, and $Z_{\text{hot}}=\xi Z_{\text{cold}}$ (where
$\xi$ is a parameter representing the ratio of $Z_{\text{hot}}$ to
$Z_{\text{cold}}$), thus Equation (\ref{eq:mother}) is reduced to $Z_{\rm
cold}=\frac{p}{(1-R+\beta)(1-\xi)}$.

In the metallicity evolution model by \citet{Lillyetal13}, it is assumed that
$dM_{\rm cold}\ne 0$, $dZ_{\rm cold}=0$, and $Z_{\rm hot}=Z_{\rm hot,0}$ (a
constant) to obtain an equilibrium metallicity (see eq.~25 therein). Given
those assumptions, Equation (\ref{eq:mother}) can be transformed to be
\begin{equation}
Z_{\rm cold} 
=Z_{\rm hot,0}+\frac{p/(1-R)}{\frac{1-R+\beta}{1-R}+\frac{dM_{\rm cold}/dt}{dM_*/dt}}. 
\label{eq:Lillyetal13}
\end{equation}
Equation (\ref{eq:Lillyetal13}) is consistent with equation (25) or (26) in
\citet{Lillyetal13}. In the model of \citet{Lillyetal13}, the mass ratio of
cold gas to stars is further assumed not to change, i.e., $d(M_{\rm
cold}/M_*)=0$. The assumption of $d(M_{\rm cold}/M_*)=0$ is equivalent to
assuming $dM_{\rm cold}=0$ or $dM_*=0$ if $M_{\rm cold}$ is a function of $M_*$
(as shown in Equation~\ref{eq:gs0} in our model below).

\end{itemize}

\subsection{Our model based on some power-law scaling relations with stellar
masses}
\label{sec:bpg_m}

In this subsection, we introduce a model by assuming that the cold gas mass
$M_{\text{cold}}$ and the stellar feedback efficiency $\beta$ follow power-law
scaling relations with the stellar mass $M_*$ during the growth of a galaxy,
where the effects of gas inflows and outflows are included and there are
no specific assumptions of $dM_{\rm cold}=0$ or $dZ_{\rm cold}=0$ or 
$d(M_{\rm cold}/M_*)=0$.
In reality, the cold gas mass in a galaxy and the feedback efficiency are not
necessarily monotonic functions of $M_*$ during the galaxy evolution. As to be
seen in Section~\ref{sec:scaling} below, both the cold gas mass
$M_{\text{cold}}$ and the feedback efficiency $\beta$ are taken as ``median
values'' in our model, and we will show that this power-law scaling assumption
is plausible and present the quantitative relations obtained from the SAMs. 
In this model, we also relax the assumption of a constant $Z_{\text{hot}}$, as
metals transfer among hot gas, cold gas, and stars through accretion and
ejection.

We assume that $M_{\text{cold}}$ and $\beta$ have the following power-law
scaling relations with $M_*$,
\begin{eqnarray}
M_{\text{cold}}(M_*)=k_{\text{gs}}\cdot M_*^{\alpha_{\text{gs}}},\label{eq:gs0} \\
\beta(M_*)= k_{\text{$\beta$s}}\cdot M_*^{\alpha_{\text{$\beta$s}}}. \label{eq:bs0}
\end{eqnarray}
We assume $Z_{\text{hot}}=\xi Z_{\text{cold}}$ ($0\leq\xi\leq1$), and it is
plausible to have $Z_{\text{cold}}\ge Z_{\text{hot}}$, since the metals
produced by star formation is assumed to be returned to the ISM, not ejected
directly from the stellar disc to the hot gas (i.e., $e=0$ in eqs.~4.6-4.11 in
\citealt{Coleetal00}). Thus, Equation (\ref{eq:mother}) can be transformed to
be the following form:
\begin{eqnarray}
d (M_{\text{cold}}Z_{\text{cold}})=\frac{p}{1-R}d M_* \quad\quad\quad\quad\quad\quad\quad\quad\nonumber\\
-\left[(1-\xi )\frac{1-R+\beta }{1-R}-\xi  \frac{d M_{\text{cold}}}{d M_*}\right]Z_{\text{cold}} d M_*. \label{eq:mother2}
\end{eqnarray}

If $\xi$ is a constant, $Z_{\rm cold}$ in Equation (\ref{eq:mother2}) can be
analytically solved with $Z_{\rm cold,0}=0$ as follows,
\begin{equation}
Z_{\text{cold}}=\frac{p}{1-R}M_{\text{cold}}^{\xi-1}\cdot e^{-f(M_*)}\int _0^{M_*}M_{\text{cold}}^{-\xi}(M_*')\cdot e^{f(M_*')}dM_*^{\prime }, \label{eq:mother3}
\end{equation}
where
\begin{equation}
f\left(M_*\right)\equiv \int _c^{M_*}(1-\xi )\frac{1-R+\beta(M_*')}{1-R}\frac{dM_*^{\prime }}{M_{\text{cold}}(M_*')}.
\end{equation}

Below we discuss the following three simplified cases with different $\xi$
values, in each of which a different term in the right-hand side of Equation
(\ref{eq:mother2}) is simplified to be zero.
\begin{itemize}
\item $\xi=0$: Equations (\ref{eq:mother2}) and (\ref{eq:mother3}) are reduced to Equations (\ref{eq:bd}) and (\ref{eq:bi})
(i.e., metal-free inflows with $Z_{\text{hot}}=0$). By adopting the power-law
scaling relations of
Equations (\ref{eq:gs0}) and (\ref{eq:bs0}) into Equation (\ref{eq:bi}), if
$\beta\gg 1-R$, we have
\begin{align}
 & Z_{\text{cold}} \nonumber \\
= & 
 \begin{cases}
p\beta^{-1} e^{-A}(-A)^{1-\frac{1}{\alpha}}\left[-\gamma(\frac{1}{\alpha},-A)\right],
& {\rm if} \quad \alpha>0, \cr
p\beta^{-1}\left[1+(1-R)\frac{M_{\rm cold}}{\beta M_*}\right]^{-1},
& {\rm if} \quad \alpha=0, \cr
p\beta^{-1}
e^{-A}(-A)^{1-\frac{1}{\alpha}} \Gamma(\frac{1}{\alpha},-A), & {\rm if} \quad \alpha<0,
 \end{cases}
\label{eq:Zcold_betamodel}
\end{align}
where $\alpha=1+\alpha_{\text{$\beta$s}}-\alpha _{\text{gs}}$, 
$A=\frac{1}{1-R}\cdot \frac{1}{\alpha}\cdot\frac{\beta M_*}{M_{\rm cold}}$,
$\gamma(a,z)\equiv\int_0^{z} t^{a-1}e^{-t}dt$ ($\Re a>0$) is the lower incomplete gamma function, and $\Gamma(a,z)\equiv\int_z^\infty t^{a-1}e^{-t}dt$
is the upper incomplete gamma function.
By applying Equations (\ref{eq:gs0}) and (\ref{eq:bs0}) to the $\alpha=0$ case
in Equation (\ref{eq:Zcold_betamodel}), we have
\begin{eqnarray}
Z_{\rm cold}=\frac{p}{\beta}\frac{1}{1+(1-R)k_{\rm gs}/k_{\beta{\rm s}}}, \qquad (\alpha=0).
\label{eq:bmalpha0}
\end{eqnarray}
The stellar metallicity $Z_*$ can be obtained by a numerical integration over
$Z_{\text{cold}}$ when $\alpha \neq0$ (i.e., Eq.~\ref{eq:zs}), and $Z_*=\frac{1}{2-\alpha_{\rm
gs}}Z_{\rm cold}$ when $\alpha=0$ by analytical integration.
The condition of $\beta\gg 1-R$ can be generally satisfied at $M_*\la
10^8\msun$, as to be seen from Figure~\ref{fig:bs_full} below.

If $\beta\ll 1-R$, the solution of $Z_{\rm cold}$ can be obtained by simply
setting $\beta=1-R$ and $\alpha_{\text{$\beta$s}}=0$ in Equation
(\ref{eq:Zcold_betamodel}) and in the expression of $A$ above, as $\beta$ is
involved in Equation (\ref{eq:bi}) or (\ref{eq:mother2}) only through the
term $\frac{1-R+\beta}{1-R}$. 

Although $Z_{\text{hot}}=0$ is assumed in this case as done in many previous
models, both gas inflow and outflow with $\beta\ne 0$ are allowed.  Since in
this case the term of the feedback efficiency $\beta$ is used in Equations
(\ref{eq:mother2})-(\ref{eq:mother3}) and shown explicitly in the solution of
$Z_{\rm cold}$ (Eq.~\ref{eq:Zcold_betamodel}), we call this case the
$\beta$-model.

\item $\xi=1$: in this case, the gas inflow metallicity (hot gas metallicity)
is the same as the gas outflow metallicity (cold gas metallicity). To reach this
scenario, the dynamic timescales of the inflow and
outflow should be both much shorter than the gas accretion timescale onto the
DM halo. With $f\left(M_*\right)=0$, Equation (\ref{eq:mother3})
is reduced to 
\begin{equation}
Z_{\text{cold}}=\frac{p}{1-R}\int_0^{M_*} M_{\text{cold}}^{-1} \, dM_*^{\prime
}\label{eq:gd}
\end{equation}
By adopting the power-law scaling relations, we have
\begin{eqnarray}
Z_{\text{cold}}& = & \frac{p}{1-R}\frac{1}{1-\alpha _{\text{gs}}}\frac{M_*}{M_{\text{cold}}}, \label{eq:gi}\\
Z_* & = & \frac{Z_{\rm cold}}{2-\alpha_{\rm gs}} \nonumber \\
& = & \frac{p}{1-R}\frac{1}{(1-\alpha _{\text{gs}})(2-\alpha _{\text{gs}})}\frac{M_*}{M_{\text{cold}}}. \label{eq:gis}
\end{eqnarray}

Note that in this case, the relation between the gas mass and the stellar
mass is explicitly used in Equations (\ref{eq:mother2})-(\ref{eq:mother3}) (see the term with
$dM_{\rm cold}/dM_*$) and shown in the solution of $Z_{\rm cold}$
(Eq.~\ref{eq:gd}), we call this case the g-model.

\item Specifically, if $\xi$ is set so that the second $dM_*$ term on the
right-hand side of Equation (\ref{eq:mother2}) vanishes, i.e.,
\begin{equation}
(1-\xi)\frac{1-R+\beta}{1-R}-\xi\frac{d M_{\text{cold}}}{d M_*}=0,
\label{eq:xi}
\end{equation}
which means that the inflow metal mass balances the outflow metal mass,
Equation (\ref{eq:mother2}) is reduced to the following simple form:
\begin{equation}
d(M_{\text{cold}}Z_{\text{cold}})=\frac{p}{1-R}d M_*, \label{eq:pd}
\end{equation}
which has the solution as follows,
\begin{align}
&\quad\quad\quad Z_{\text{cold}}=\frac{p}{1-R}\cdot \frac{M_*}{M_{\text{cold}}},\label{eq:pi}\\
&\quad\quad\quad Z_*=\frac{p}{1-R}\cdot\frac{1}{2-\alpha _{\text{gs}}}\cdot\frac{M_*}{M_{\text{cold}}}.\label{eq:pis}
\end{align}
This model is an intermediate case between the $\beta$-model and the g-model,
as we have $0<\xi=(1+\frac{1-R}{1-R+\beta}\frac{dM_{\rm cold}}{dM_*})^{-1}<1$
according to Equation (\ref{eq:xi}).  Note that in this case only the term
with $p$ is used in the right-hand side of Equation (\ref{eq:mother2}), we call
this case the $p$-model.  \end{itemize}

Though the above equations on $Z_{\rm cold}$ and $Z_*$ obtained in this
subsection have some similarity (e.g., proportional to $M_*/M_{\rm cold}$) as
some previous model results as reviewed in Section~\ref{sec:pm}, they are
obtained by relaxing some specific assumptions used in those previous models,
and the ratio of $M_*/M_{\rm cold}$ is a result of the combining effects of
star formation, gas inflows, and outflows.

Note that the expectation from the g-model and the $p$-models should not be taken
as the reference models at $M_*\ga M_{\rm cold}$, as the hot gas metallicity
of the inflow assumed in the models would be unrealistically
too high so that $Z_*$ expected by extrapolating Equations (\ref{eq:gis}) and
(\ref{eq:pis}) to $M_*\ga M_{\rm cold}$ can be even higher than
$\sim p/(1-R)$.

As seen from the results obtained in this subsection, the exponents in the MZRs
are closely related to the exponents in the power-law scaling relations of the
$M_{\rm cold}$--$M_*$ relation and the $\beta$--$M_*$ relation. The
metallicities are solely determined by the $M_{\rm cold}$--$M_*$ relation in
the g-model and the $p$-model. According to Equations (\ref{eq:gis}) and
(\ref{eq:pis}), the slopes of the MZRs obtained with the g- and $p$-models
should be $1-\alpha_{\rm gs}$.  According to the $\alpha=0$ case in Equation
(\ref{eq:bmalpha0}), the slopes of the MZRs obtained with the $\beta$-model
should be $-\alpha_{\beta{\rm s}}=1-\alpha_{\rm gs}$.

In principle, Equations (\ref{eq:df1})--(\ref{eq:df3}) are one part of the SAM,
to describe the chemical evolution of a galaxy in a newly formed halo. Our
analytical model presented above in Section~\ref{sec:bpg_m} has the following
differences from that in the SAM. In our model, we ignore the physical process
of galaxy mergers in the evolution of the stellar metallicity and the stellar
mass.  The effects of the detailed physical processes, such as halo assembly
and gas cooling, are incorporated into the evolution of gas mass with stellar
mass in a statistical way, as shown in Section 4.1.1 below.  The universality
and the simplicity in the expression of the MZR itself suggests that it is
plausible here to simplify the growth of a galaxy as the evolution of several
relevant physical variables (e.g., stellar mass, gas mass, metal mass) and to
use a statistical and analytical way to reveal the origin of the MZR. The
combination of the chemical evolution model constructed in this work with the
results obtained from the SAM provides an efficient way to isolate the effects
of different physical processes and see the dominant reasons leading to the
MZRs.

Note that the chemical enrichment is included in the SAM as described in
Section~\ref{sec:method}, but the metallicities produced from the SAM are not
used in the chemical evolution model described in this section, except that the
SAM metallicity results can provide a support to the assumed range of $\xi$
($0<\xi\la 1$) in the above analysis, as well as a reference to the MZR to be
shown in Figure~\ref{fig:mzr} below.

\section{Results}
\label{sec:result}

In this section, we show the power-law scaling relations between the cold gas
mass and the stellar mass and between the feedback strength and the stellar
mass obtained from the SAMs. Then by applying these scaling relations to the
chemical evolution models presented in Section~\ref{sec:bpg_m}, we reproduce
the MZR and explain its universality. In principle, if these scaling relations
are available from observations, they can also be similarly used to the
chemical evolution models to predict the MZR. The compatibility of the
SAM results with observations (as shown in HYL14 and \citealt{XY19}) suggests
the reasonability to extract some other physical properties (i.e., the
power-law scaling relationships here) from the SAM.

\subsection{Power law scaling relations obtained from the SAM}
\label{sec:scaling}

In this subsection, the relations between the cold gas mass and the stellar
mass and the relation between the feedback strength and the stellar mass during
the evolution of the galaxies and satellites are extracted from the SAM
presented in HYL14 and \citet{XY19}.  In \citet{XY19}, we present some of our
SAM simulation runnings of the SAMs for central galaxies and their satellites
in DM halos ranging from $10^9$ to $10^{15}h^{-1}\msun$ (see table~1
therein).  In this work, we add some runnings for high-mass halos for
better statistics.  The stellar masses of the galaxies and the satellites
range from $10^3M_{\odot}$ to $10^{11}M_{\odot}$ at redshift $z=0$. The SAM
simulations produce the evolution of the physical properties of the galaxies
and their satellites. 

\subsubsection{The cold gas mass--stellar mass scaling relation}
\label{sec:gs}

The gas fraction in a galaxy evolves with inflows, outflows, and star formation
processes. In a closed-box model, the cold gas fraction
$M_{\text{cold}}/(M_{\text{cold}}+M_*)$ decreases monotonically as the stellar
mass increases. In a realistic situation with gas inflow and outflow in the
hierarchical galaxy formation and evolution, a host halo can gain new gas
through accretion or merging with other halos, the newly gained gas can further
cool down onto the galaxy disk, and the cold gas can also be ejected from the
galaxy disk through feedback. Thus, the evolution of the cold gas fraction may
not be monotonic, but fluctuate as the stellar mass increases.

\begin{figure*}[htb]
\centering
  \subfigure{\includegraphics[scale=0.40]{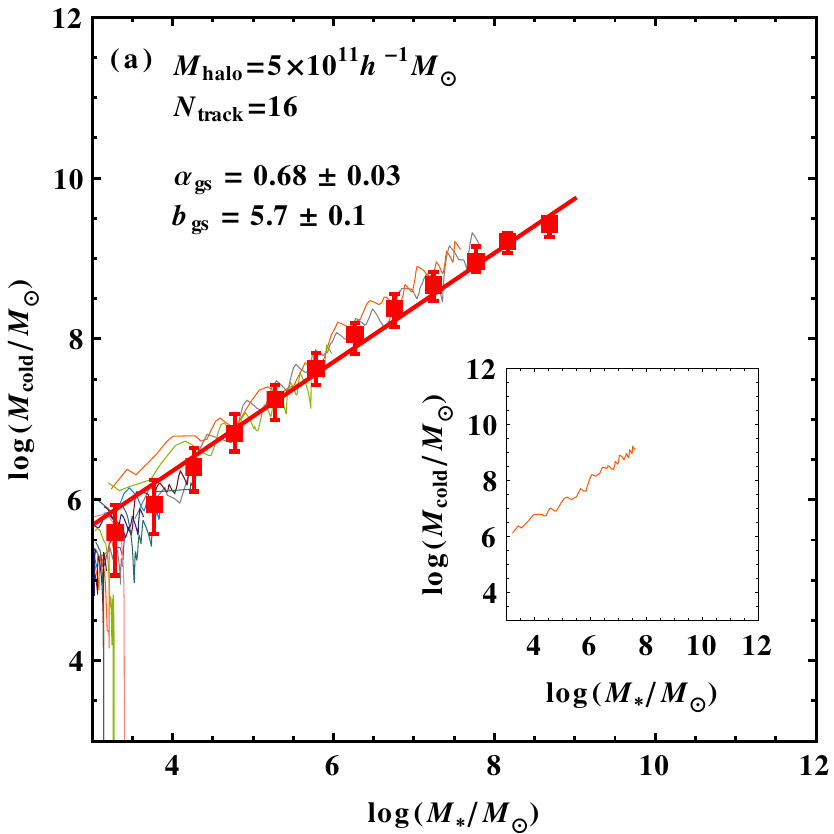}}
  \subfigure{\includegraphics[scale=0.40]{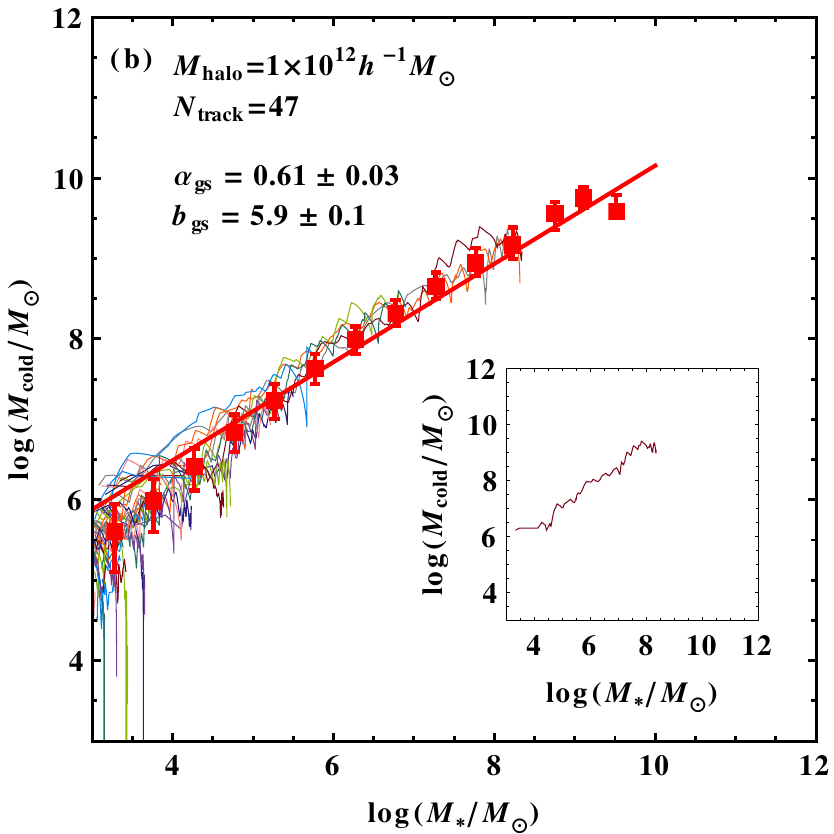}}
  \subfigure{\includegraphics[scale=0.40]{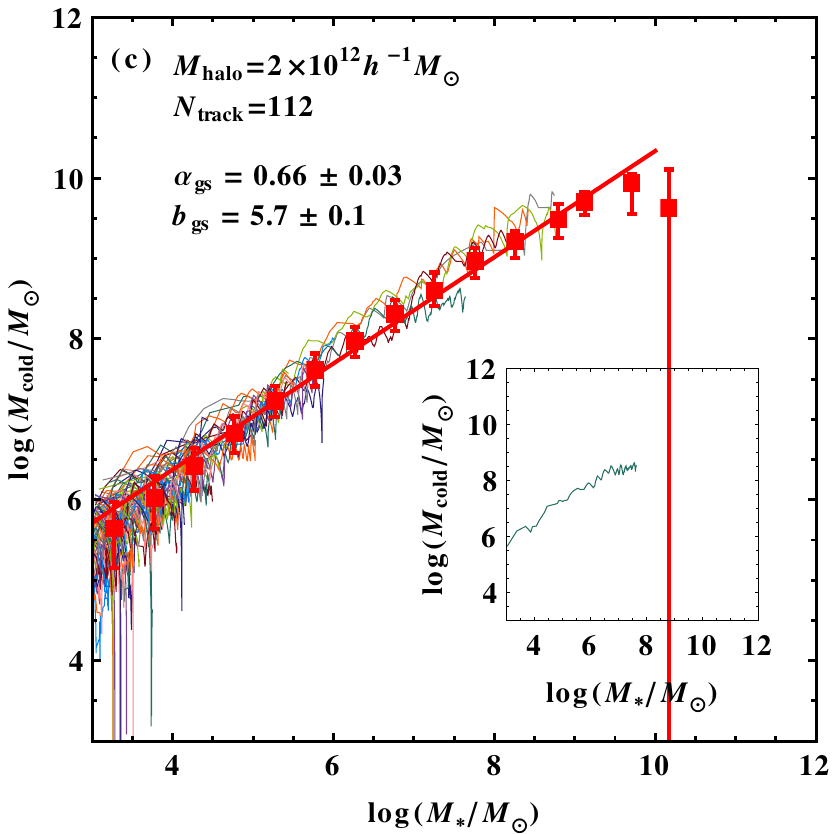}}
  \subfigure{\includegraphics[scale=0.40]{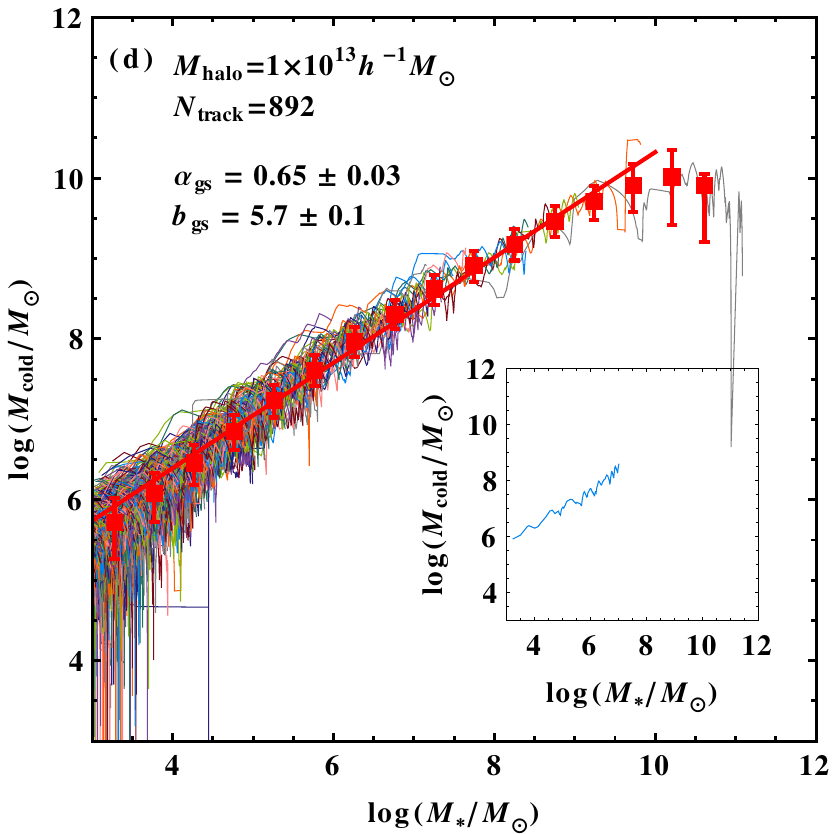}}
  \subfigure{\includegraphics[scale=0.40]{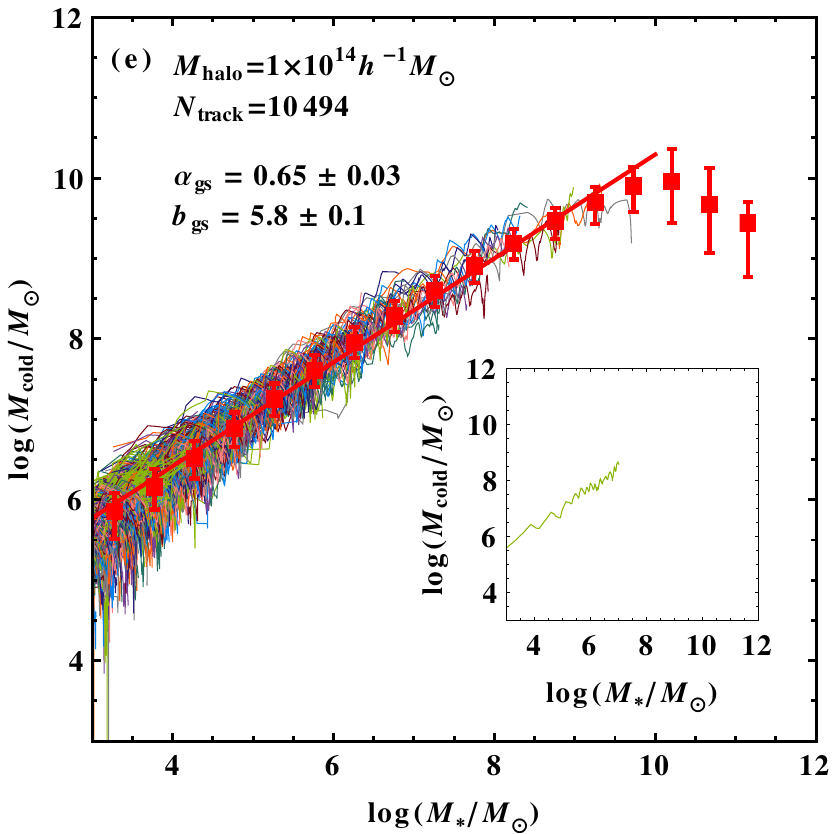}}
  \subfigure{\includegraphics[scale=0.40]{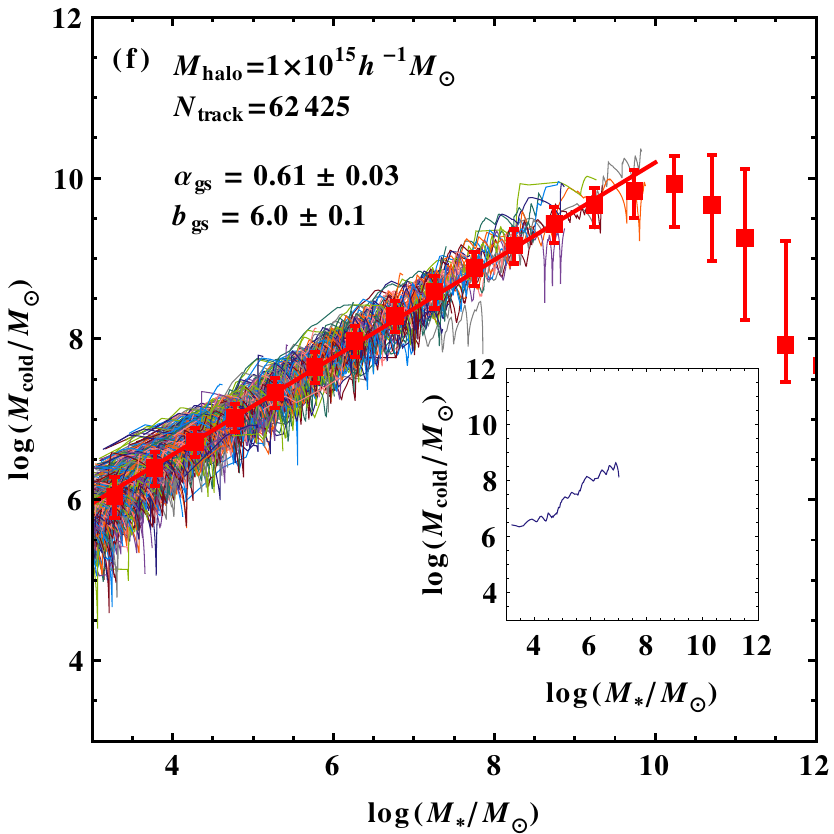}}
  \caption{The median cold gas mass--stellar mass relation of the progenitor
galaxies of satellites in one DM halo. The different panels show the cases for
different halos with 
masses ranging from $5\times10^{11}$ to $1\times10^{15}h^{-1}M_{\odot}$
(no results of lower halo masses are shown due to a poor statistics). In
each panel, each thin solid curve represents the evolution of the cold gas mass
with the stellar mass in the progenitor of one satellite until its infall, and
$N_{\text{track}}$ gives the number of the evolution tracks that ever existed
in the hierarchical merging history of the halo. For view clarity, we only show
1000 randomly selected tracks if $N_{\text{track}}>1000$ in a panel.  The inset
in each panel illustrates one track. The values of ($M_{\text{cold}}, M_*$) are
recorded at the same cosmic time interval in our simulations.  The logarithm of
the stellar mass is divided into some bins with a 0.5-dex interval, starting
from $\log(M_*/\msun)=3$, i.e., [3,3.5],[3.5,4],.... The red squares represent
the median of the recorded ($M_{\text{cold}}, M_*$) of all the tracks in each
bin.  The error bars of the red dots represent the range between the 16th
and 84th percentiles of the distribution of the recorded $M_{\text{cold}}$ in
each bin.  The $M_{\rm cold}$ declines to low
values at the high-$M_*$ end, which is associated with star formation and the
exhaustion of cold gas, and the slow increase of the halo mass in the halo
assembly history.  A linear least-squares fitting to the red dots is
performed in the range of $\log(M_*/M_{\odot})=$3--10, which is shown as the
red solid line.  The values of $\alpha_{\text{gs}}$ and $b_{\text{gs}}$
shown in each panel are the best-fit slope and intercept of the red solid
line, which are in the ranges of $\sim$0.6--0.7 and $\sim$5.6--5.9,
respectively.  See details in Section~\ref{sec:gs}.
}
\label{fig:gs_sat}
\end{figure*}

Figure~\ref{fig:gs_sat} shows the evolution of the cold gas mass as a function
of stellar mass in the progenitor galaxies of the present-day satellites before
their infall into the host halos, which are obtained from our SAM simulations.
As mentioned above, in the assembling history of a halo, the progenitor galaxy
of a present-day satellite is the host central galaxy of a small halo at an
early time before it fell into a big halo. In Figure~\ref{fig:gs_sat}, each
panel presents the results obtained from the progenitor galaxies of the
present-day satellites in one host halo and the different panels have different
host halo masses at $z=0$. Each curve represents the growth history of the
progenitor of one satellite until infall, which we call a track. The number of
the tracks in each panel is denoted by $N_{\rm track}$. The inset in each panel
illustrates only one track.

Each track shown in Figure~\ref{fig:gs_sat} is obtained by tracing the
evolution of the progenitor galaxy back to high redshifts.  Along the
evolution, the progenitor may experience a galaxy merger, and we select the
relatively massive one of the merging galaxies as the progenitor galaxy.

As seen from Figure~\ref{fig:gs_sat}, though fluctuating, the curves show an
overall increasing tendency with increasing stellar masses at $M_*\la
10^{10}\msun$. Below we obtain the ``median'' relation between the cold gas
mass and the stellar mass by a linear fitting to the data. 
The physical properties of the progenitor galaxies are recorded at the same
time interval ($\sim 10^6$~yr) in our simulations.  
Note that the data are not uniformly distributed in the logarithm of the
stellar mass. To avoid some weight bias in the fitting,
we divide the logarithm of the stellar mass uniformly into
some bins starting from $\log(M_*/\msun)=3$ with an interval of 0.5~dex. 
For each evolution track,
we take the medians of their variables ($\log(M_*),\log(M_{\rm cold})$)
in each bin; and then we show the medians of those median values of all the
evolution tracks in each bin by
a red square and show the half of the range between the 16th and 84th
percentiles of those median values of all the evolution
tracks as the error-bar of the red square in Figure~\ref{fig:gs_sat}.
In that way, the different evolution tracks contribute
the same weight to the red square in each bin.
We use the least-squares method to fit the red squares as follows,
\begin{equation}
\log(M_{\text{cold}}/M_{\odot})=\alpha_{\text{gs}}\log(M_*/10^3M_{\odot})+b_{\text{gs}},
\label{eq:gs}
\end{equation}
where $\alpha_{\text{gs}}$ is the slope, and $b_{\text{gs}}$ is the intercept
at $M_*=10^3M_{\odot}$ and related with $k_{\text{gs}}$ in Equation
(\ref{eq:gs0}) by $k_{\text{gs}}=10^{b_{\text{gs}}-3\alpha_{\text{gs}}}$.  
The best-fit results are shown by
the red solid line and the texts in each panel of Figure~\ref{fig:gs_sat}.

\begin{figure*}[htb]
  \subfigure{\includegraphics[scale=0.43]{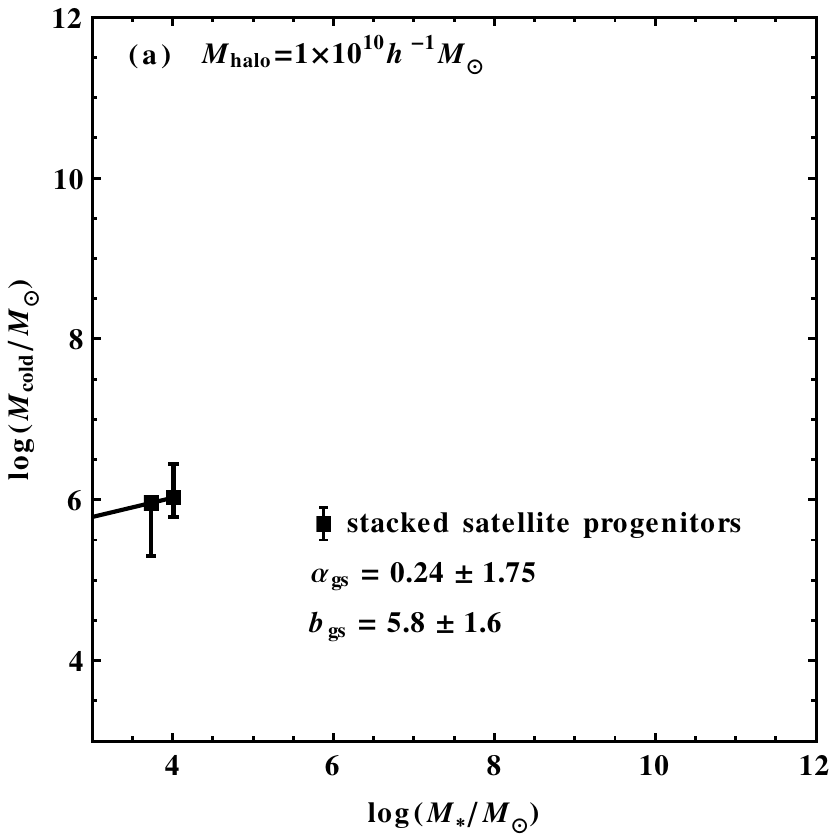}}
  \subfigure{\includegraphics[scale=0.43]{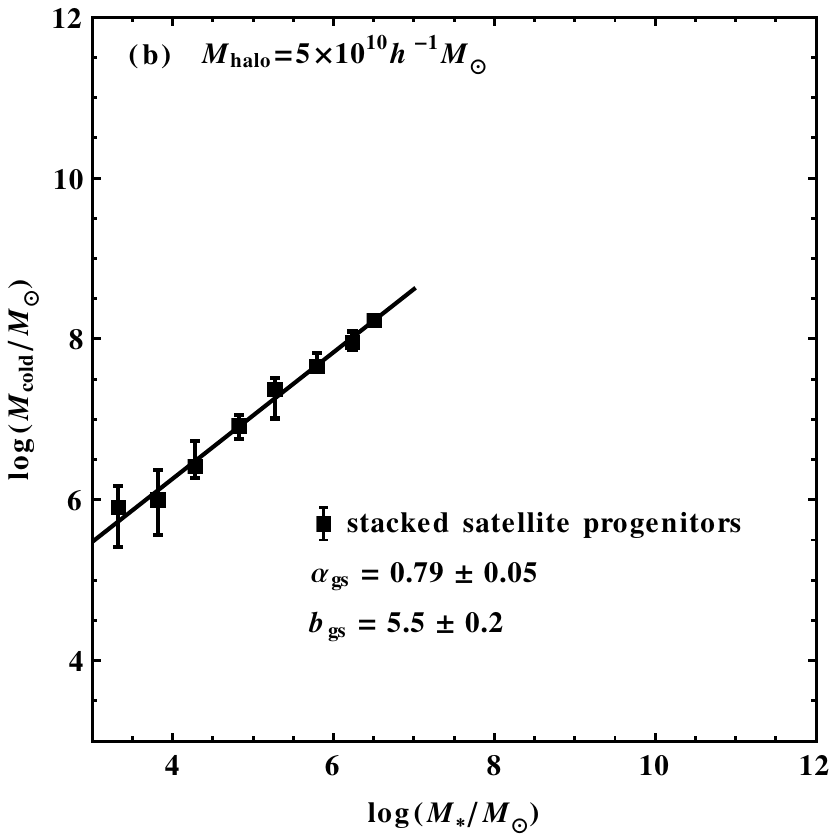}}
  \subfigure{\includegraphics[scale=0.43]{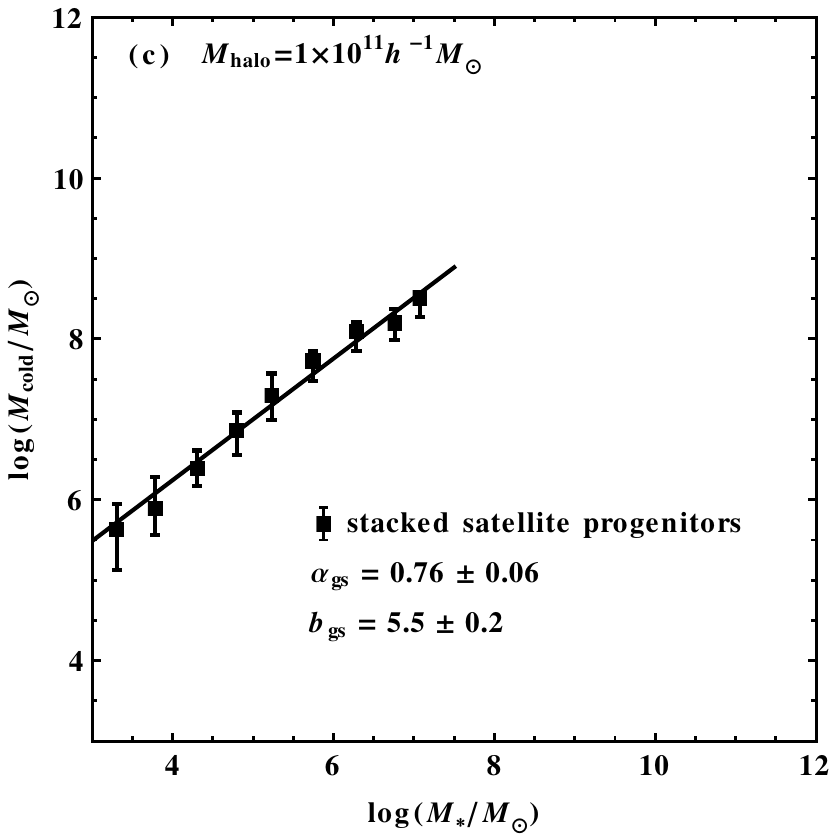}}
  \subfigure{\includegraphics[scale=0.43]{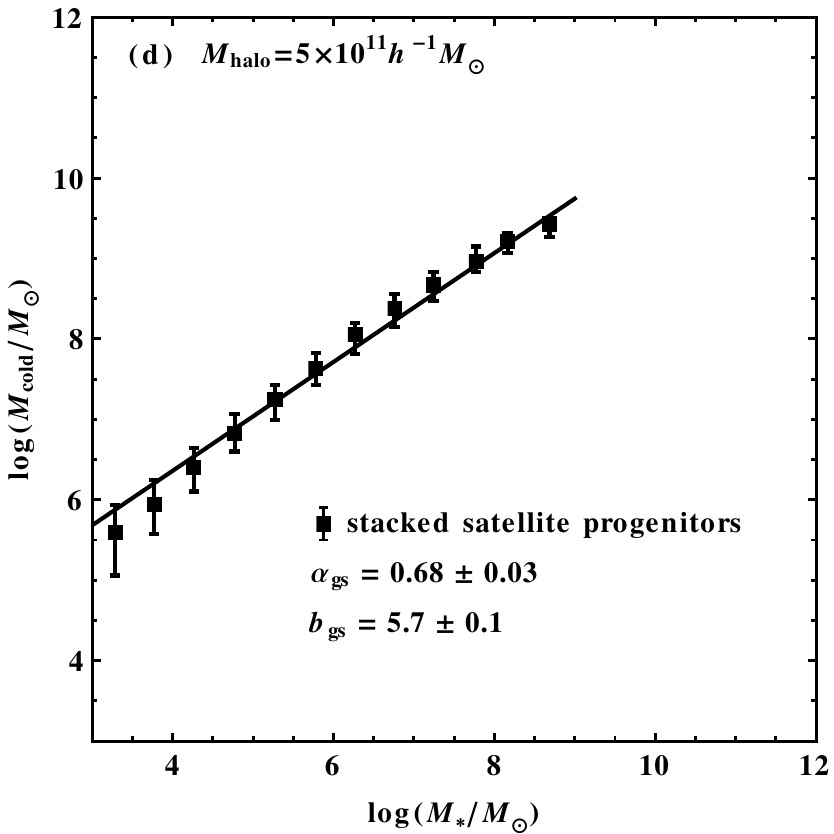}}
  \subfigure{\includegraphics[scale=0.43]{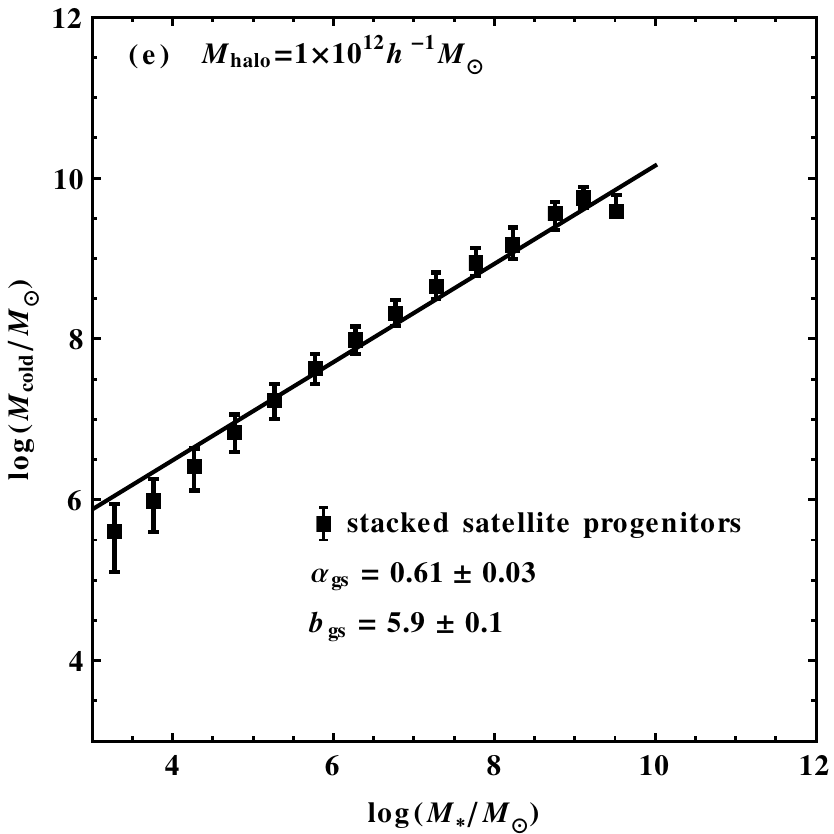}}
  \subfigure{\includegraphics[scale=0.43]{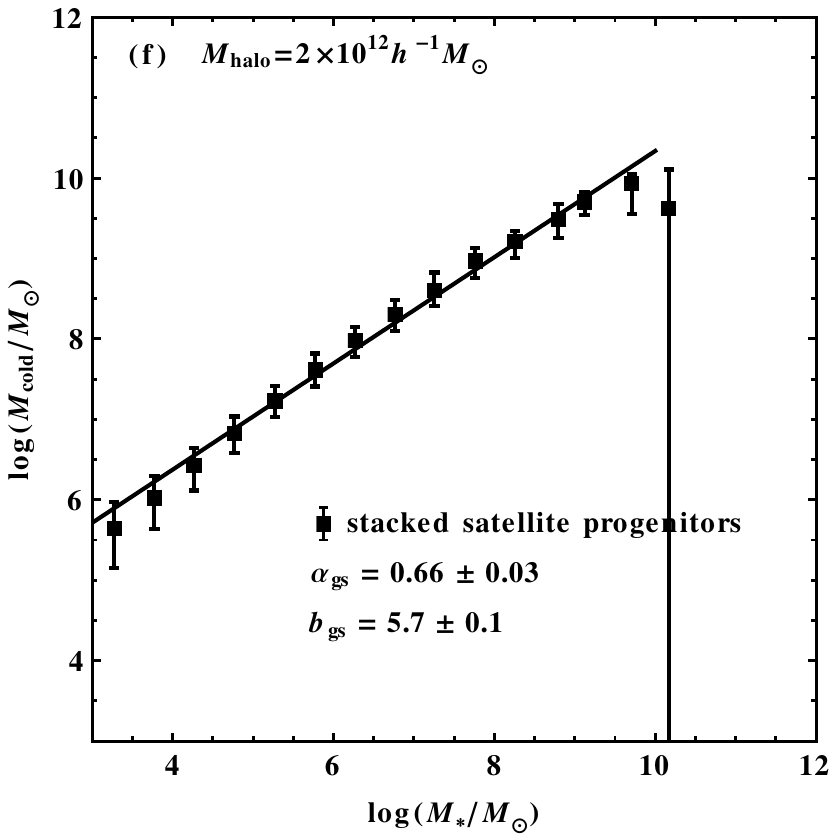}} 
  \subfigure{\includegraphics[scale=0.43]{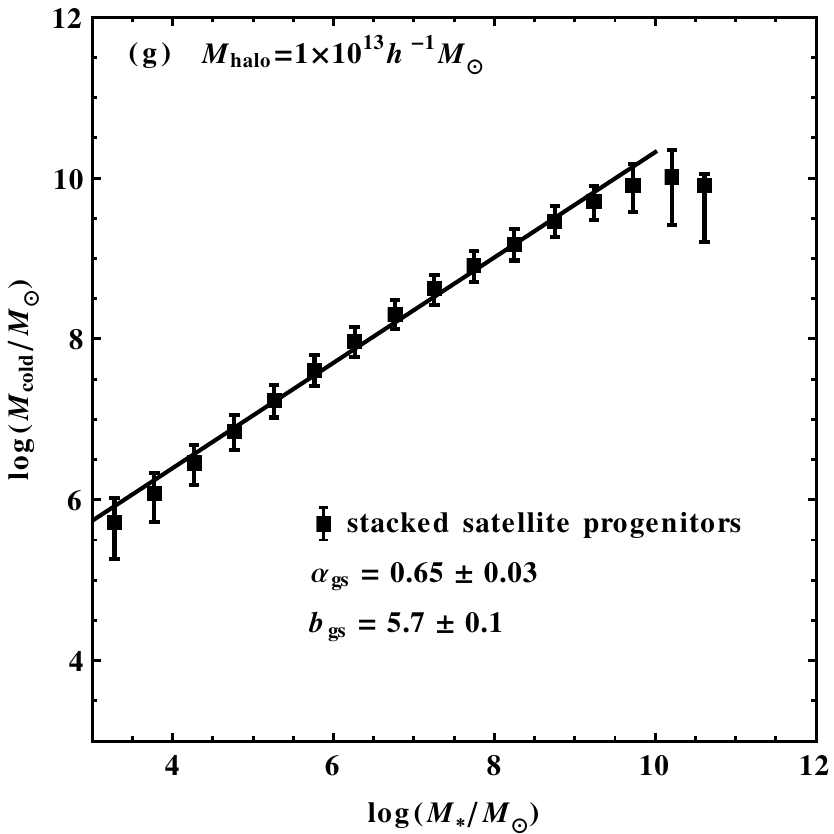}}
  \subfigure{\includegraphics[scale=0.43]{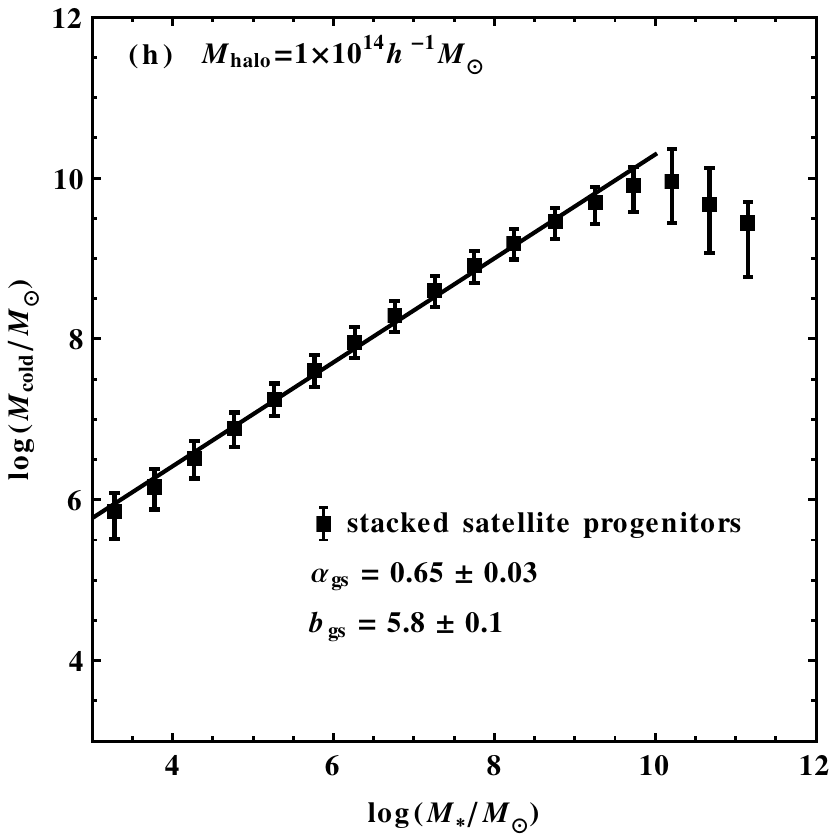}}
  \subfigure{\includegraphics[scale=0.43]{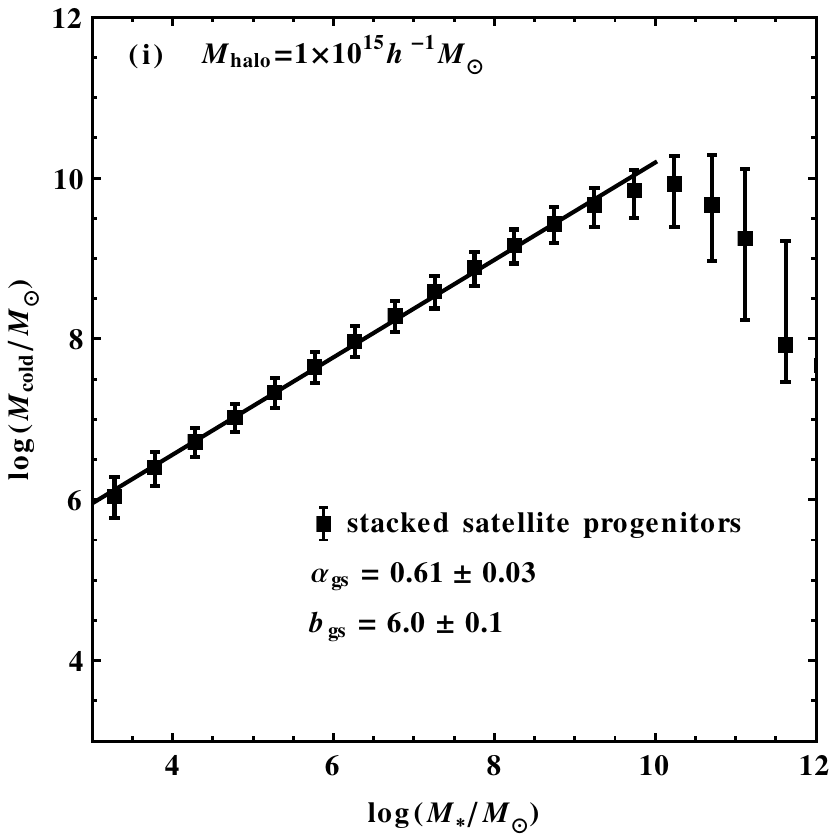}}   
  \caption{The median cold gas mass--stellar mass relation of the progenitor
galaxies of satellites in stacked multiple halos. In each panel, the number of
the stacked halos with the same halo mass is listed in Table~\ref{tab:gs}. The
points and the solid straight line are obtained in the same way as those done
for the red squares and the red straight line shown in
Figure~\ref{fig:gs_sat}, except that the evolution tracks of the satellite
progenitors are obtained from multiple trees in this figure.  The values of
$\alpha_{\text{gs}}$ and $b_{\text{gs}}$ shown in each panel are the best-fit
slope and intercept of the solid line in each panel, which are also listed in
Table~\ref{tab:gs} and in the ranges of $\sim$0.61--0.79 and $\sim$5.5--6.0,
respectively. See Section~\ref{sec:gs}.}\label{fig:gs_sat_stack} \end{figure*}

In order to have a statistically robust result, we stack multiple trees of the
same host halo mass together, and perform a similar fitting to the $M_{\rm
cold}$--$M_*$ relation in the progenitors of their present-day satellites.  Each panel in
Figure~\ref{fig:gs_sat_stack} shows the best-fit results to the progenitors of
the present-day satellites in a stack of ten host halos with the same halo mass. The
fitting method is similar as done for Figure~\ref{fig:gs_sat_stack}.  In each
panel, the evolution curves of the cold gas mass as a function of stellar mass
in all the progenitor galaxies in the ten host halos are used to obtain the
medians of the variables ($\log(M_*),\log(M_{\rm cold})$) (open squares) and
their error-bars. The evolution tendency of the open squares shows that
$M_{\rm cold}$ increases with increasing $M_*$ when $M_*$ is low but
declines at the high-$M_*$ end. The increase of $M_{\rm cold}$ with increasing
$M_*$ is associated with the halo mass buildup of the satellite progenitors.
The decline is associated with star formation, the exhaustion of cold
gas, and the slow increase of the halo mass in the halo assembly history.
We fit the median $M_{\rm cold}$--$M_*$ relation mainly in the increasing
part, and
the stellar mass ranges used in the fitting are listed in Table~\ref{tab:gs}.
The best-fit results are also summarized in Table~\ref{tab:gs}.
As seen from Figure~\ref{fig:gs_sat_stack} and Table~\ref{tab:gs}, where the
host halo masses cover a large range, the slopes $\alpha_{\text{gs}}$ and the
intercepts $b_{\text{gs}}$ of the fit scaling relation are in a narrow space
for $M_{\rm halo}\sim 5\times 10^{10}$--$10^{15}h^{-1}\msun$, with
$\alpha_{\text{gs}}\sim$0.61--0.79 (or $1-\alpha_{\text{gs}}\sim$0.2--0.4) and
$b_{\text{gs}}\sim$5.5--6.0. 

The $\chi^2$ values shown in Table~\ref{tab:gs} are used as a measure of the
goodness of the fit, together with the number of the bins $N_{\rm bin}$ in the
fit. In general the fits are acceptable, and the probability that a random set
of $N_{\rm bin}$ data points drawn from the parent distribution would yield a
value of $\chi^2$ as large as or larger than the tabulated values are mostly in
the range $>$20\%--95\%. Some $\chi^2$ values are somewhat too small, compared
with the degrees of freedom in the fitting ($N_{\rm bin}-2$). Part of that
reason is that a relatively large error has been assigned to the fitting data
(i.e., using the large scatter around the median in each bin).

\begin{figure*}[htb]
  \subfigure{\includegraphics[scale=0.43]{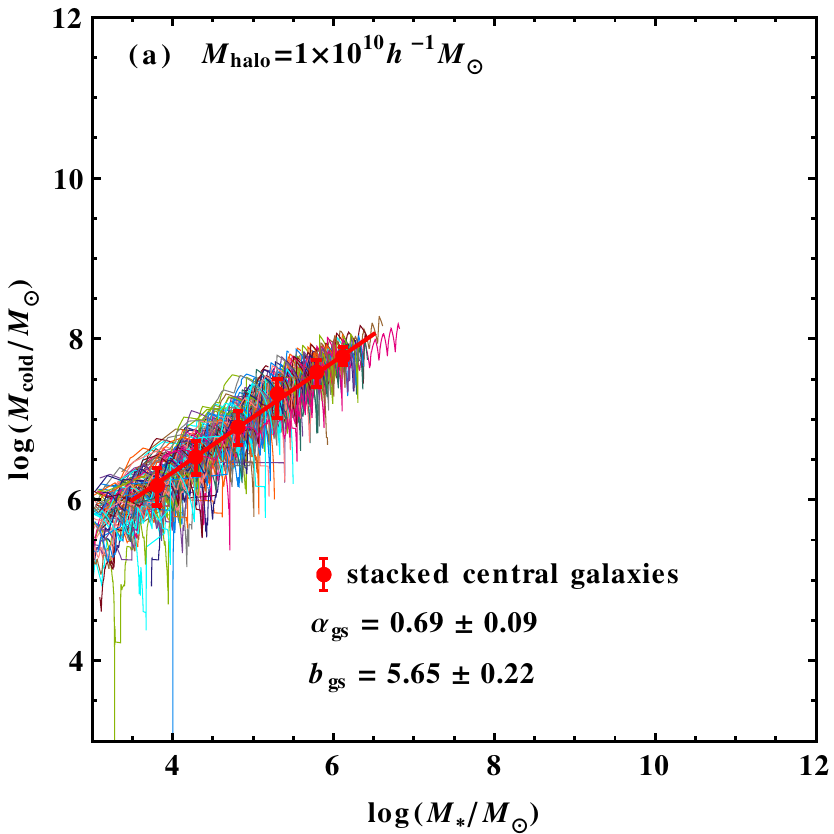}}
  \subfigure{\includegraphics[scale=0.43]{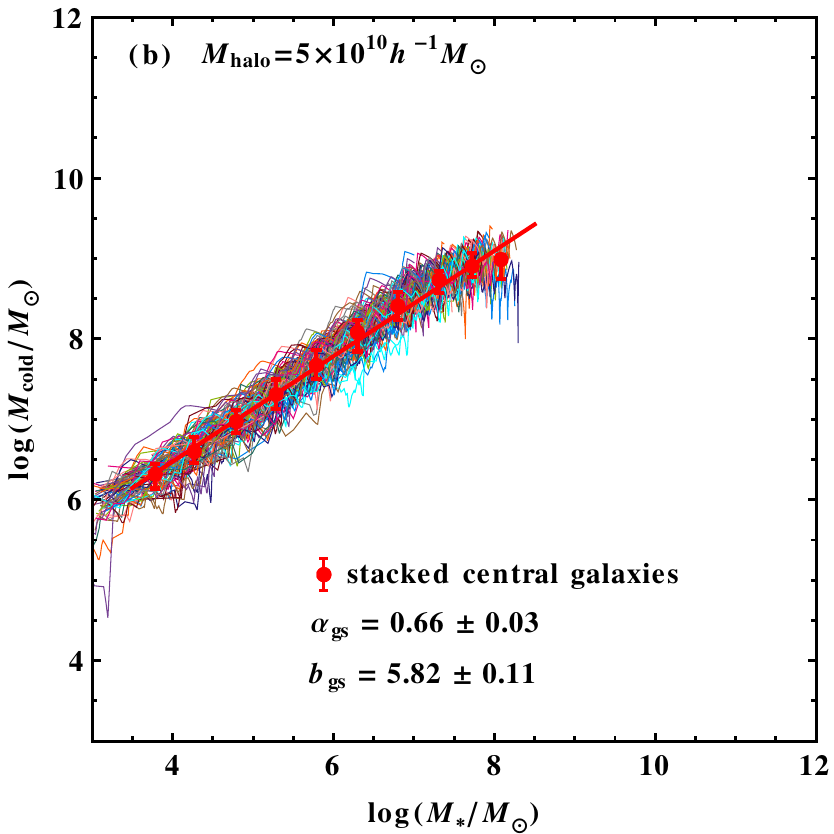}}
  \subfigure{\includegraphics[scale=0.43]{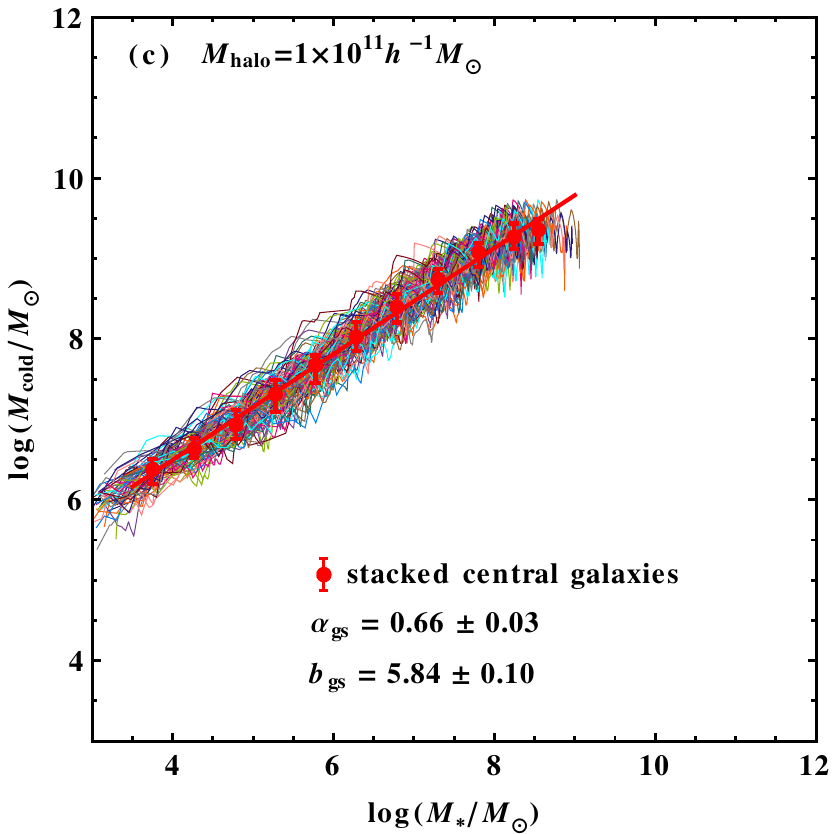}}
  \subfigure{\includegraphics[scale=0.43]{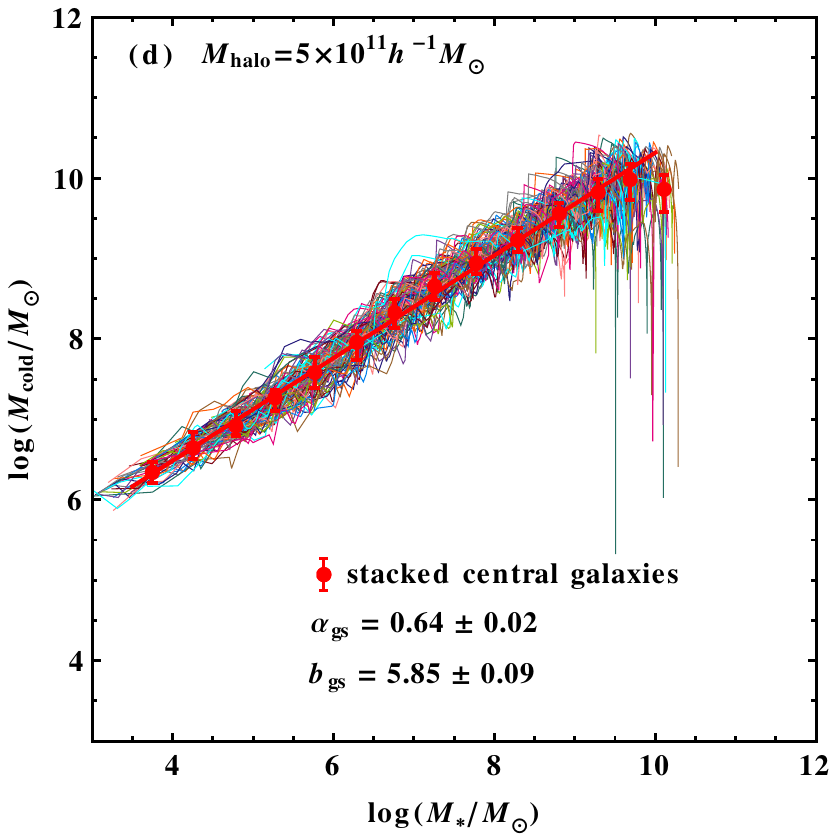}}
  \subfigure{\includegraphics[scale=0.43]{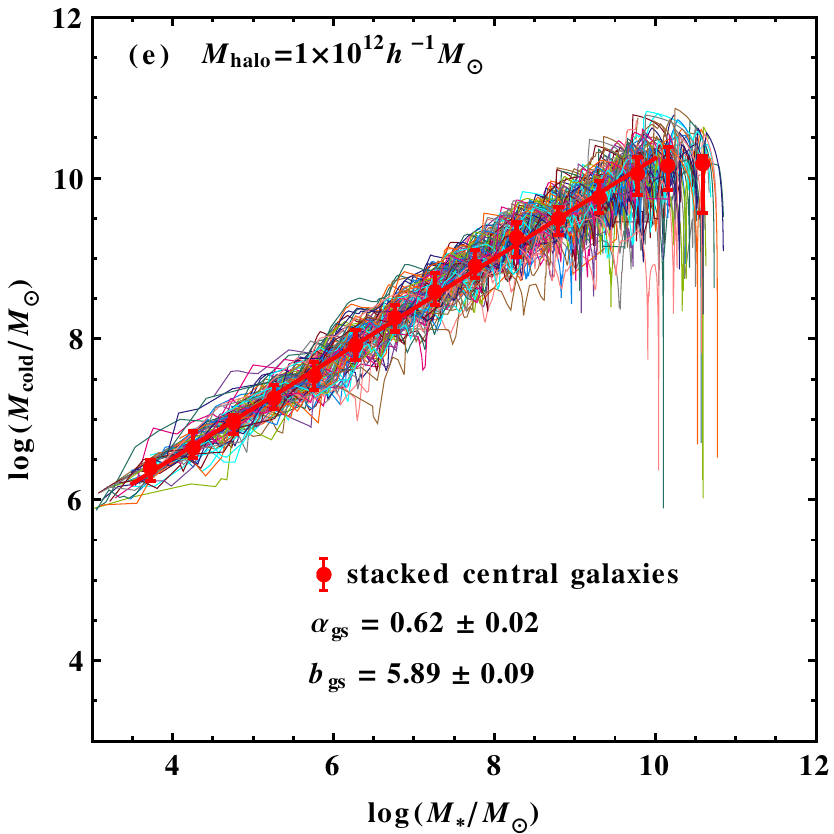}}
  \subfigure{\includegraphics[scale=0.43]{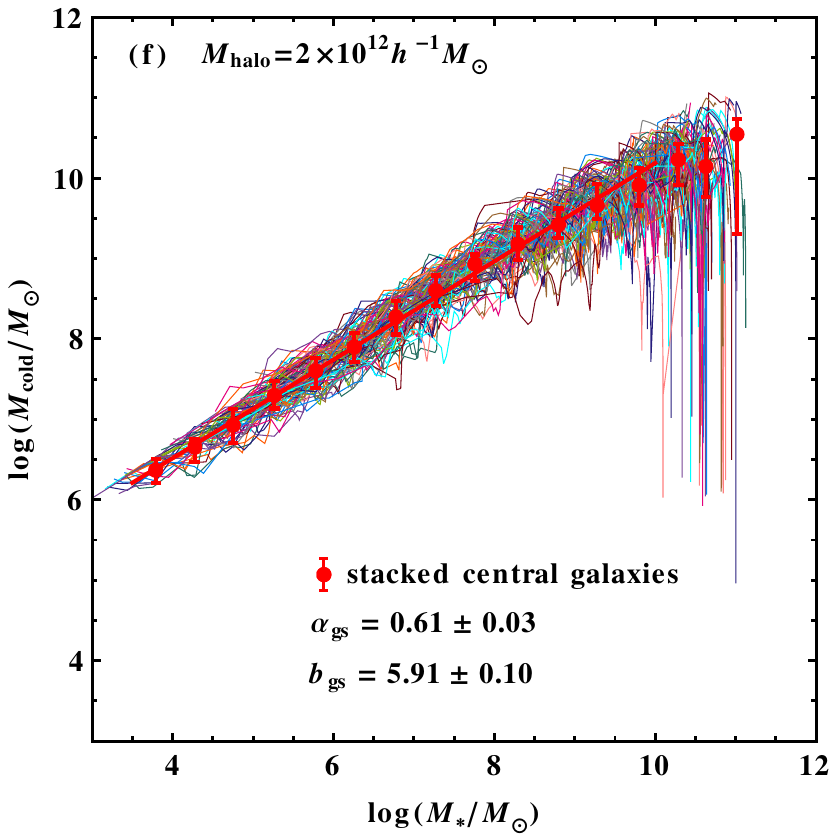}}
  \subfigure{\includegraphics[scale=0.43]{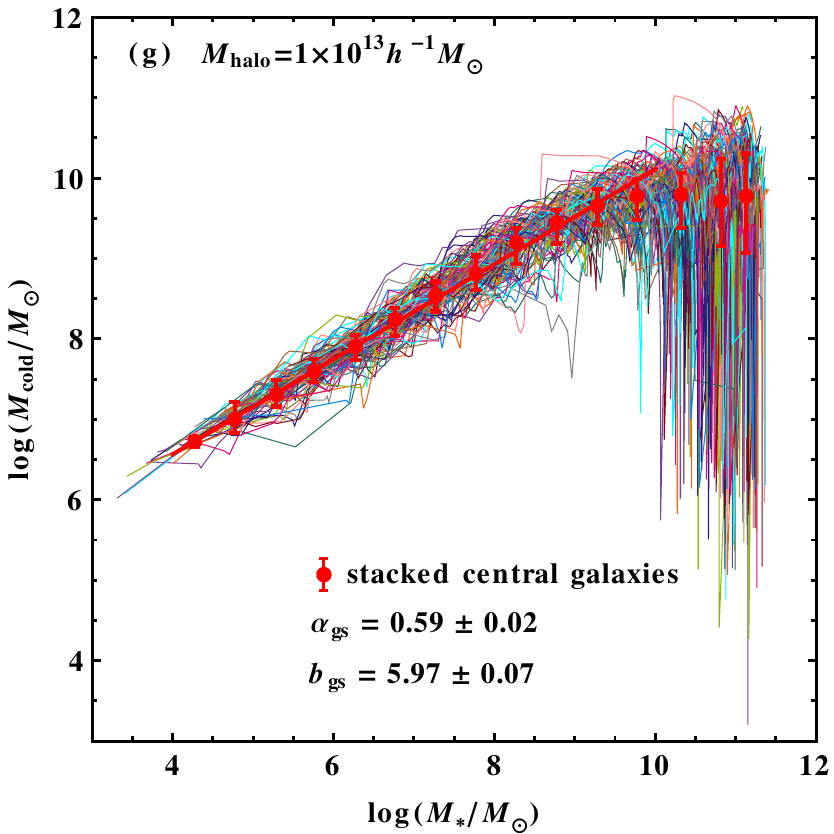}}
  \subfigure{\includegraphics[scale=0.43]{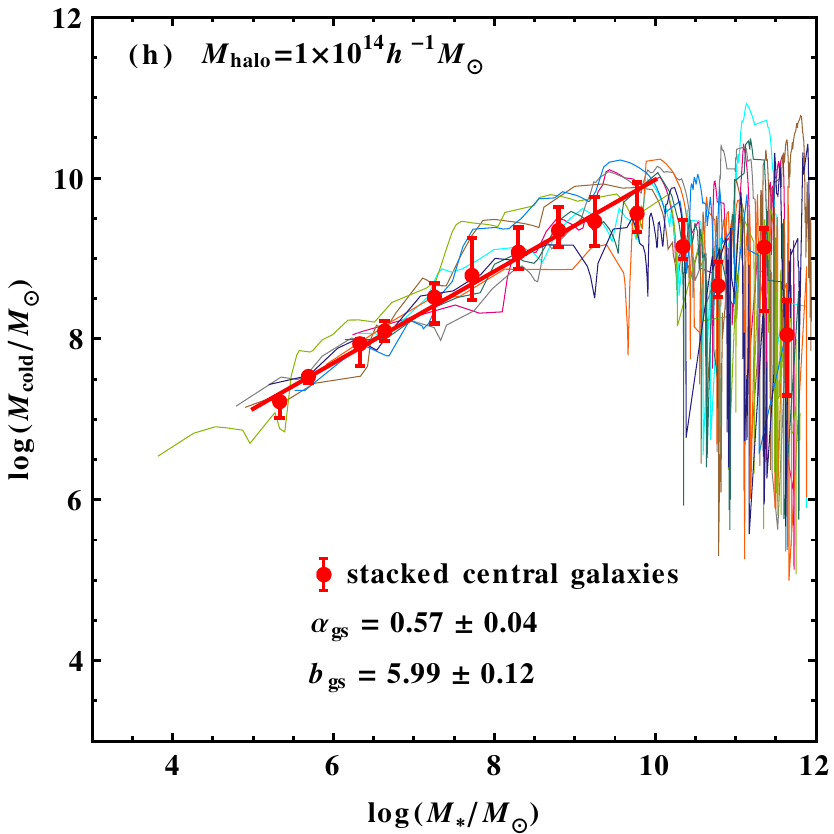}}
  \subfigure{\includegraphics[scale=0.43]{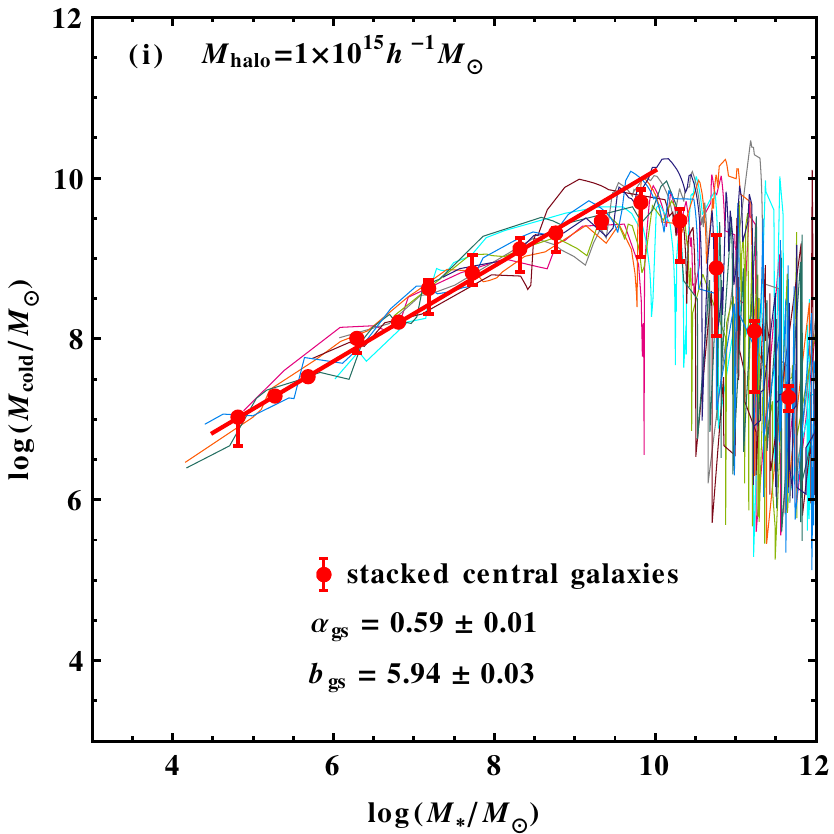}}   
  \caption{The median cold gas mass--stellar mass relation of the central
galaxies in stacked multiple halos. The simulation runnings used are the same
as those done for Figure~\ref{fig:gs_sat_stack}. Each thin solid curve
represents the evolution track of the central galaxy in one halo. The red
points, their error bars, and the red straight line are obtained in a
similar way as those in Figure~\ref{fig:gs_sat}.  The values of
$\alpha_{\text{gs}}$ and $b_{\text{gs}}$ shown in each panel are the best-fit
slope and intercept of the solid line in each panel, which are 
in the ranges of $\sim$0.57--0.69 and $\sim$5.7--6.0,
respectively (as listed in Table~\ref{tab:gs}). See Section~\ref{sec:gs}. }\label{fig:gs_cet} \end{figure*}

To compare the satellites with the central galaxies, we also perform a similar
fitting to the median $M_{\rm cold}$--$M_*$ relation in the central galaxies of
the stacked trees.  Figure~\ref{fig:gs_cet} shows the evolution tracks of the
central galaxies in the stacked host halos. The best-fit results are shown by
the straight solid lines and summarized in Table~\ref{tab:gs}.
As seen from Figure~\ref{fig:gs_cet} and Table~\ref{tab:gs}, though the host
halo masses cover a wide range, the slopes $\alpha_{\text{gs}}$ and the
intercepts $b_{\text{gs}}$ of the best-fit scaling relations are in a universal
range for $M_{\rm halo}\sim 5\times 10^{10}$--$10^{15}h^{-1}\msun$, with
$\alpha_{\text{gs}}\sim$0.57--0.69 (or $1-\alpha_{\text{gs}}\sim$0.3--0.4) and
$b_{\text{gs}}\sim$5.7--6.0.

\begin{figure*}[htb]
\centering
  \subfigure{\includegraphics[scale=0.60]{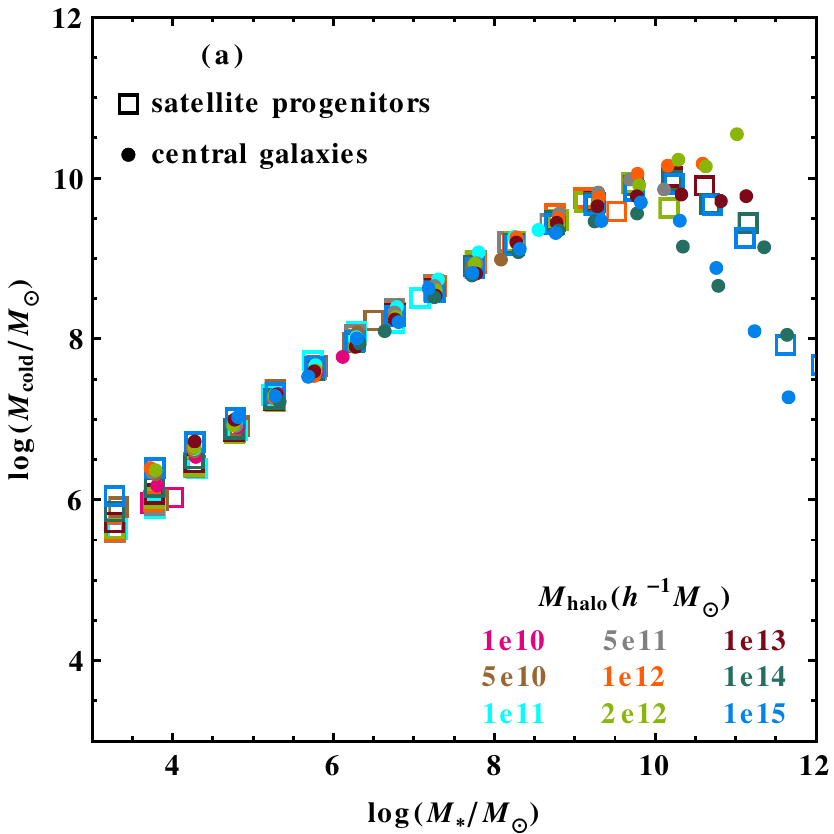}}\\
  \subfigure{\includegraphics[scale=0.60]{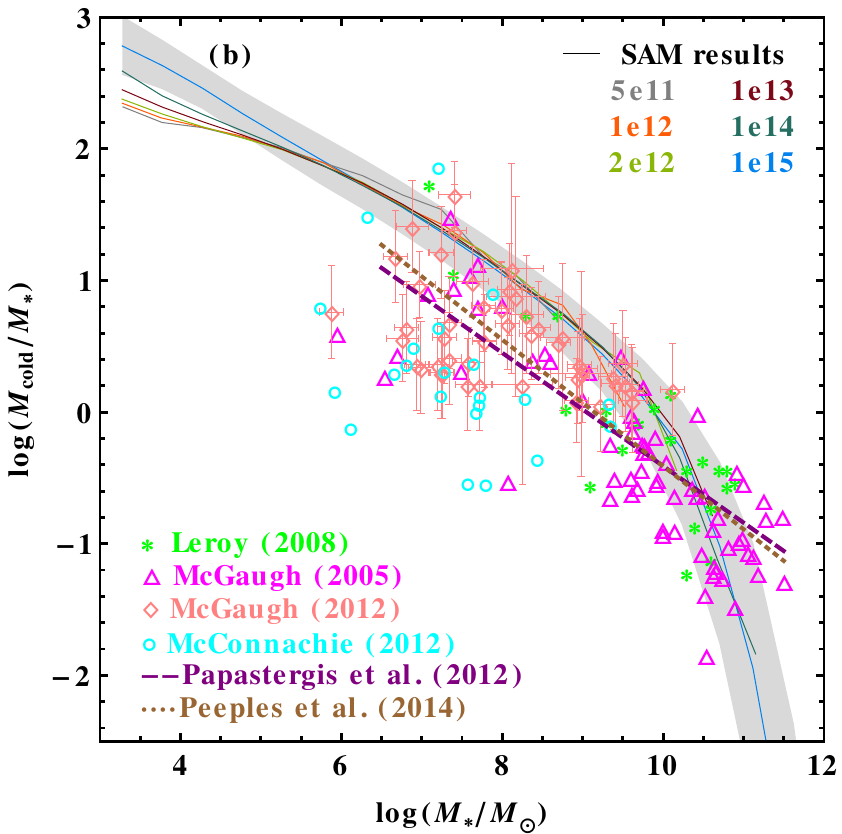}}
  \caption{Panel (a): Comparison of the median cold gas mass--stellar mass
relations obtained for satellite progenitors and central galaxies within
halos with a large range of masses. The values of the points are the same as
those corresponding median points shown in
Figures~\ref{fig:gs_sat_stack}--\ref{fig:gs_cet}. The open squares (i.e., solid
squares in Fig.~\ref{fig:gs_sat_stack}) represent satellite progenitors, and
the filled dots represent central galaxies. The galaxies within halos with
different 
masses $M_{\rm halo}$ are labeled with different colors. The median cold gas
mass--stellar mass relation appears to be universal except for the decline part
at the high-$M_*$ ends.
Panel (b): The ratio of the cold gas mass to the stellar mass. The solid
lines are converted from the SAM results shown in panel (a) by connecting
the corresponding points with the same colors for the growth of satellite progenitors
in $M_{\rm halo}=10^{10}$--$10^{15}h^{-1}\msun$, and the different lines
represent the results obtained with different $M_{\rm halo}$.
The shaded grey region illustrates the $1\sigma$ dispersion around the
blue solid line for the case of $M_{\rm halo}=10^{15}h^{-1}\msun$, converted
from the error-bars shown in Figure~\ref{fig:gs_sat_stack}(i). 
The symbols (stars,
triangles, diamonds, and open circles), the dashed line, and the dotted line
show some observation results for nearby
galaxies ($\la 200\Mpc$) adopted from figure 3 in \citet{peeples14}, which include the cold gas (atomic plus molecular) mass
measurements of disk galaxies by \citet{mcgaugh05,mcgaugh12} and
\citet{leroy08}, a fit to the samples in \citet{mcgaugh05,mcgaugh12} and \citet{leroy08} and the star-forming galaxies in \citet{S11} (dotted line; see eq.~9
in \citealt{peeples14}),
the cold gas mass measurements in Local Group dwarf galaxies
summarized by \citet{mcconnachie12},
a fit to an HI-selected Arecibo Legacy Fast
ALFA (Arecibo L-band Feed Array) survey and an optical-selected SDSS (Sloan Digital
Sky Survey) sample by \citet{papastergis12} (dashed line; see eq.~10
in \citealt{peeples14}).
This panel serves as a reference for the evolution of the mass
ratio in the simulated galaxy evolution history (where
low-$M_*$ progenitors are at relatively high redshifts) to the observational
results of nearby or low-redshift galaxies. See Section~\ref{sec:gs}.  }
\label{fig:gs_full} \end{figure*}

For a comparison of the $M_{\rm cold}$--$M_*$ relations between central galaxies and satellite
progenitors in different halo masses, we combine the points shown in
Figures~\ref{fig:gs_sat_stack} and \ref{fig:gs_cet} together into
Figure~\ref{fig:gs_full}(a). 
This comparison demonstrates that the cold gas mass--stellar mass relations are
universal for both central galaxies and satellite galaxies, and for a wide
range of host halo masses.

In Figure~\ref{fig:gs_full}(b), we show the observational results for galaxies
at redshift $z=0$ as a reference to the $M_{\rm cold}$--$M_*$ relations
obtained during the galaxy growth in this work.  The results are represented by
the ratios of the cold gas mass to the stellar mass.  where the blue solid line
represents the results obtained from our SAMs, and the other points and lines
represent the observational results for galaxies at redshift $z=0$.  Note that
the blue solid line obtained from our SAM simulations is a little higher than
the observational results at relatively low masses ($M_*\la 10^8\msun$), which
is plausible as the low stellar mass range of the blue solid line normally
represents the relatively early evolution stage at high redshifts.

\subsubsection{The stellar feedback efficiency--stellar mass scaling relation}
\label{sec:bs}

In the SAM, the stellar feedback efficiency $\beta$ is obtained through Equation
(\ref{eq:sim_fb}) or (\ref{eq:betaE}).  In this subsection, we obtain the
best-fit results to the scaling relations between the feedback strength $\beta$
and the stellar mass, similarly as done to obtain the scaling relations between
the cold gas mass and the stellar mass in Section~\ref{sec:gs}.

\begin{figure*}[htb]
\centering
  \subfigure{\includegraphics[scale=0.40]{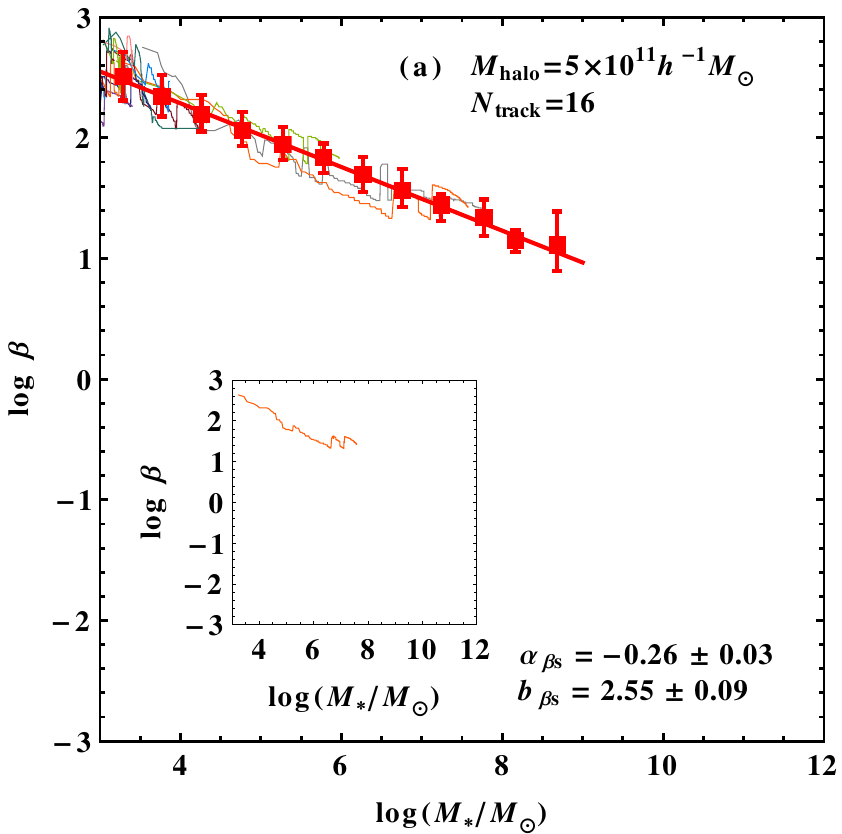}}
  \subfigure{\includegraphics[scale=0.40]{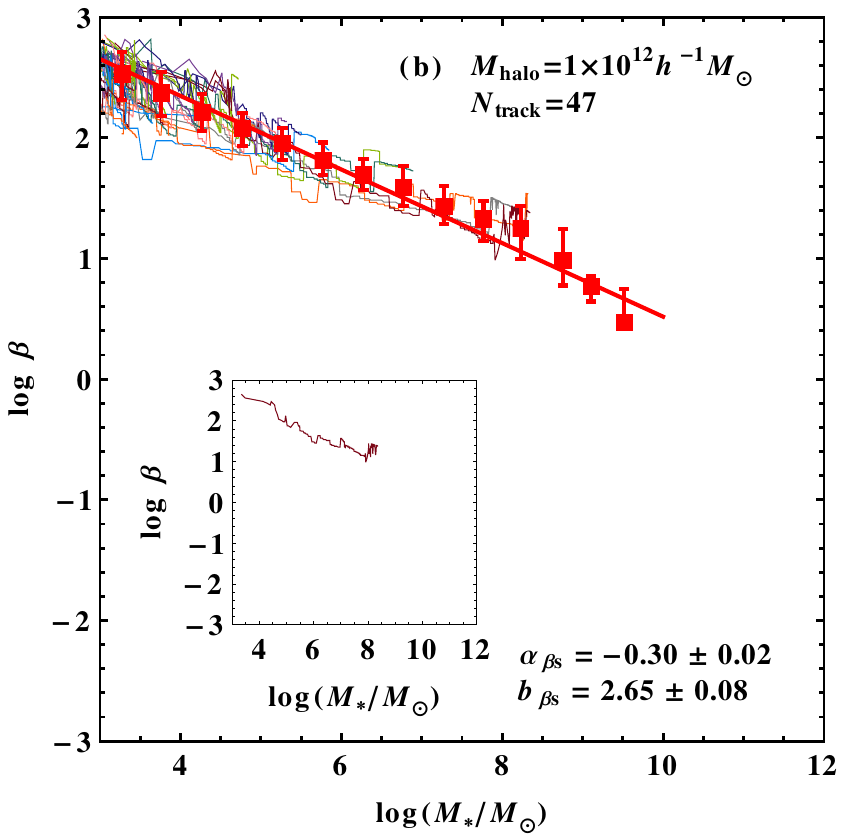}}
  \subfigure{\includegraphics[scale=0.40]{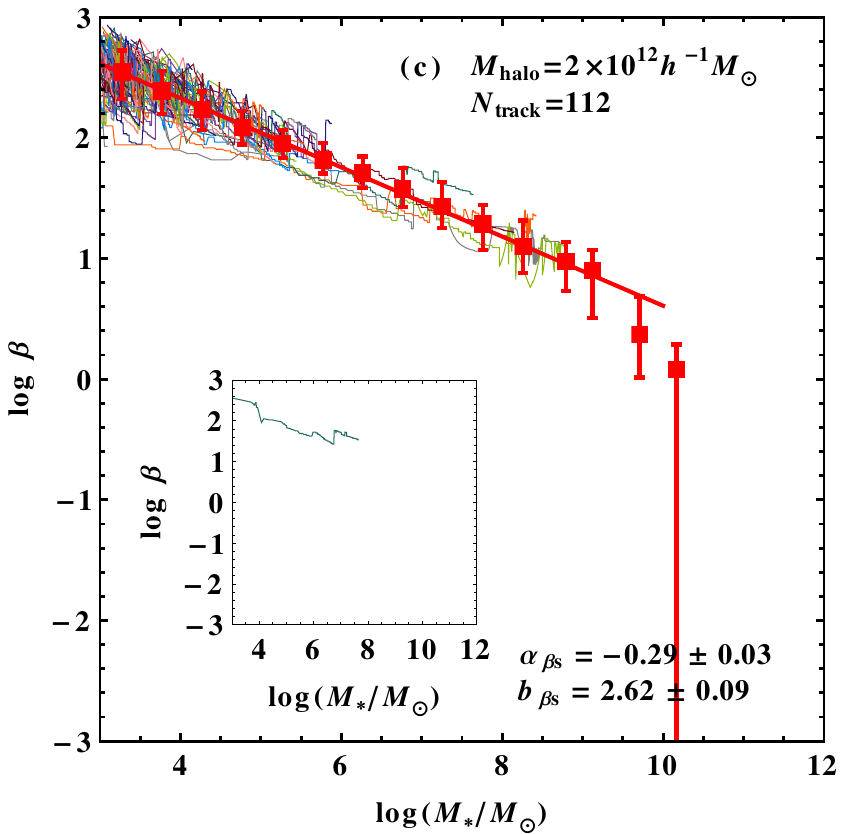}}
  \subfigure{\includegraphics[scale=0.40]{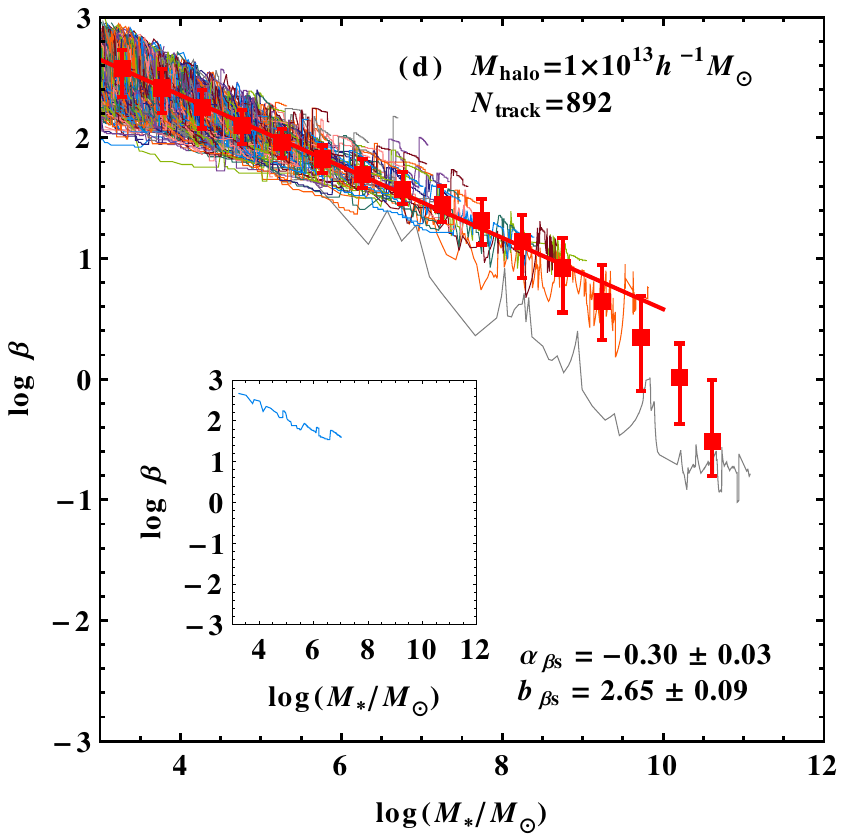}}
  \subfigure{\includegraphics[scale=0.40]{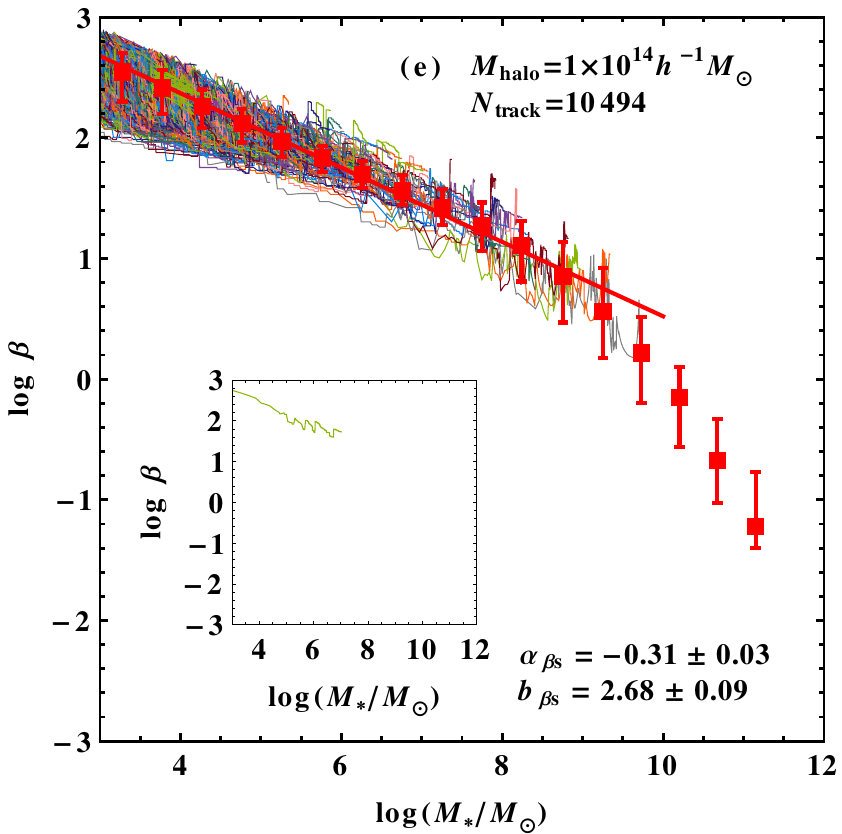}}
  \subfigure{\includegraphics[scale=0.40]{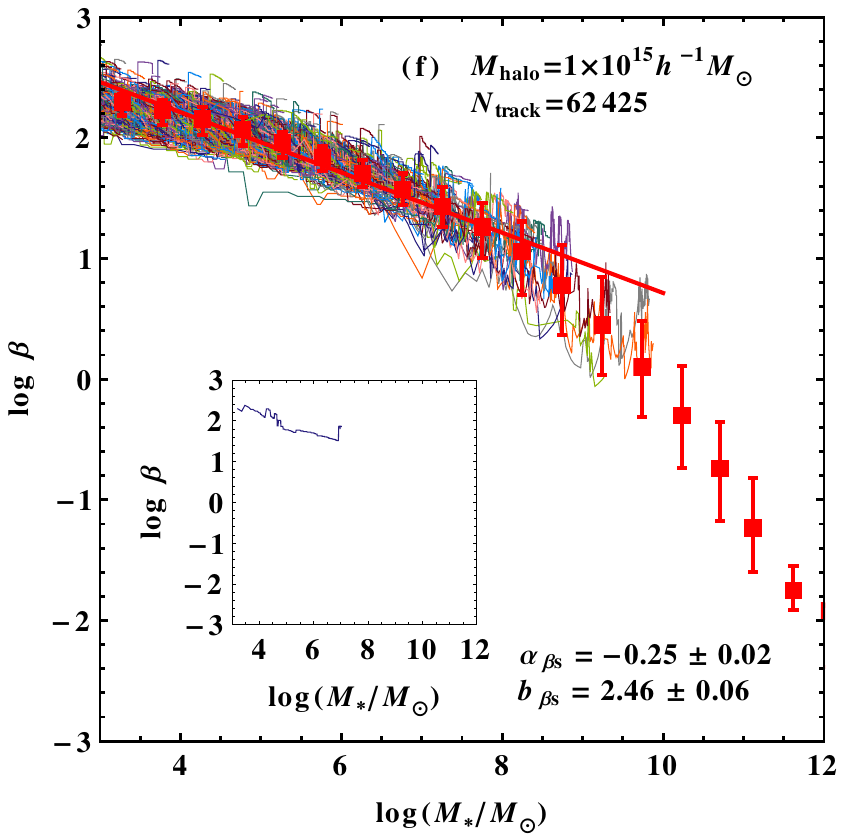}}
  \caption{The stellar feedback efficiency--stellar mass relation of the
progenitor galaxies of satellites in one DM halo. The curves, labels,
and texts have the similar meanings as those in Figure~\ref{fig:gs_sat},
except that the physical variable $M_{\rm cold}$ is replaced by
the stellar feedback efficiency $\beta$. See Section~\ref{sec:bs}.
}\label{fig:bs_sat}
\end{figure*}

\begin{figure*}[htb]
  \subfigure{\includegraphics[scale=0.43]{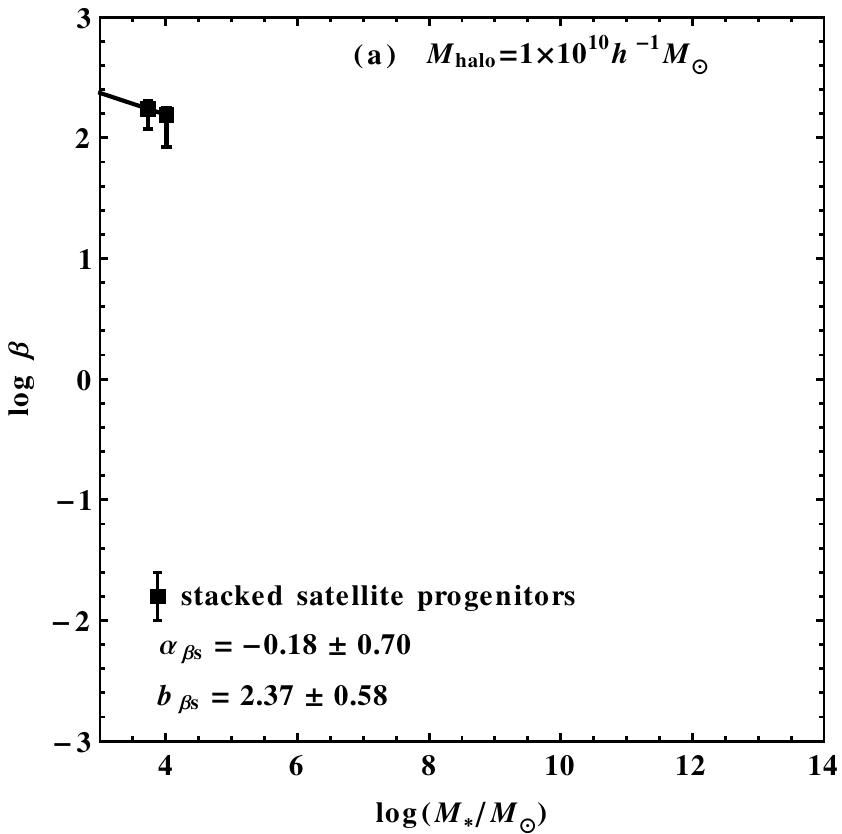}}
  \subfigure{\includegraphics[scale=0.43]{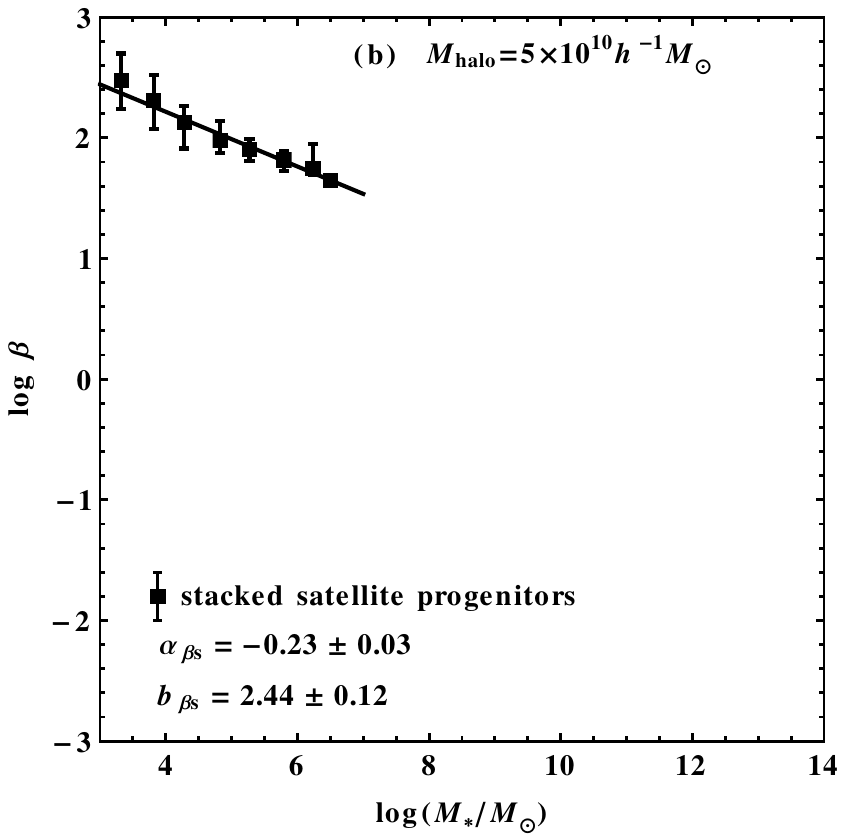}}
  \subfigure{\includegraphics[scale=0.43]{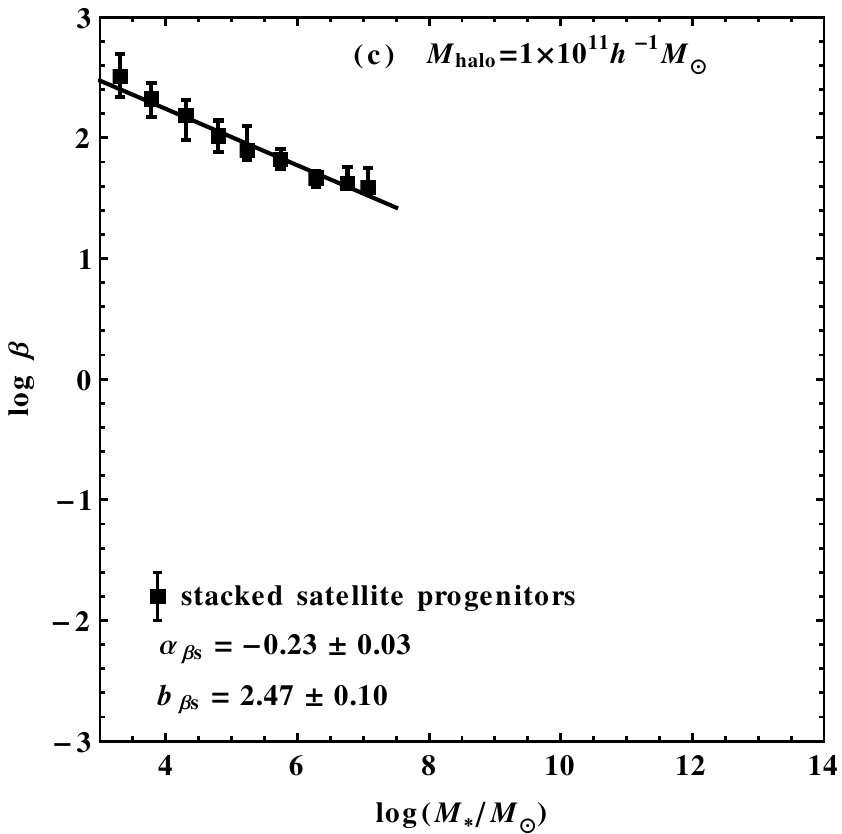}}
  \subfigure{\includegraphics[scale=0.43]{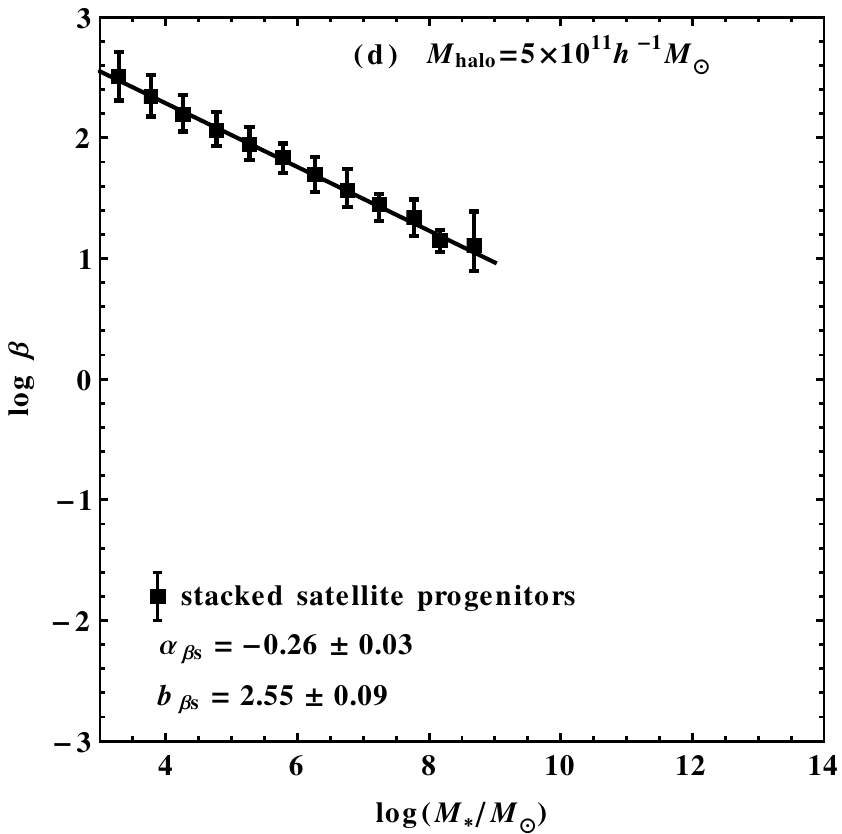}}
  \subfigure{\includegraphics[scale=0.43]{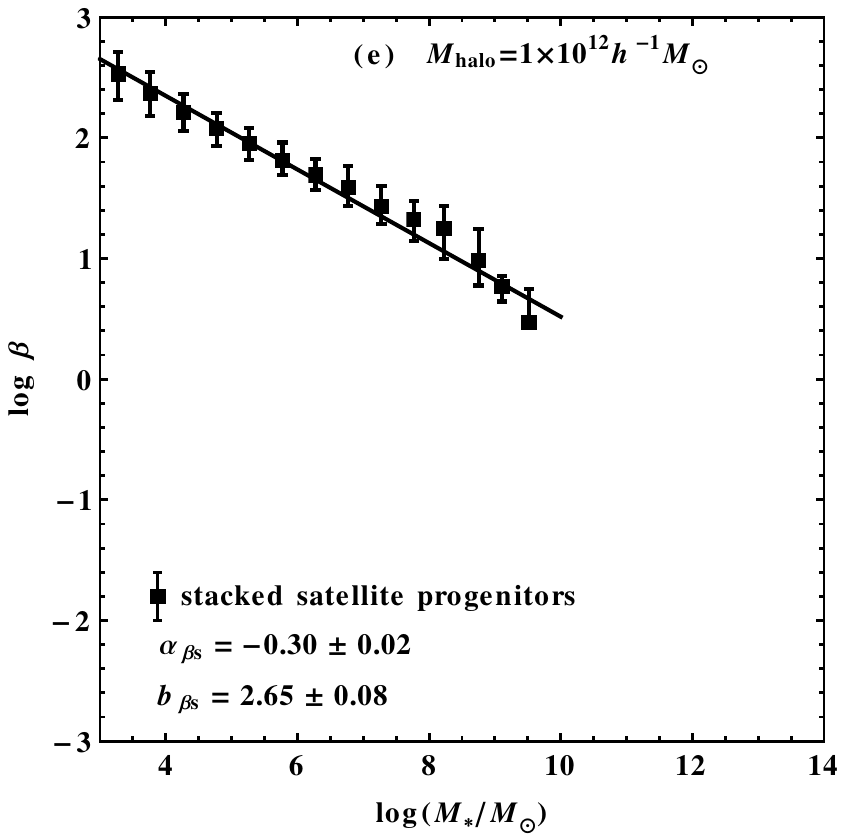}}
  \subfigure{\includegraphics[scale=0.43]{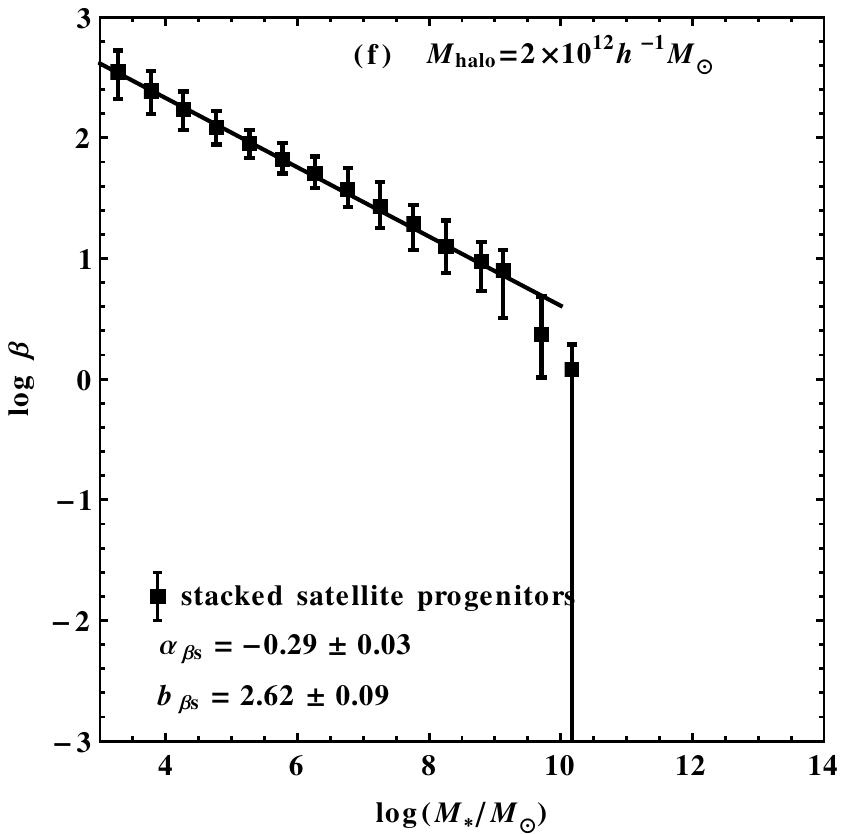}}  
  \subfigure{\includegraphics[scale=0.43]{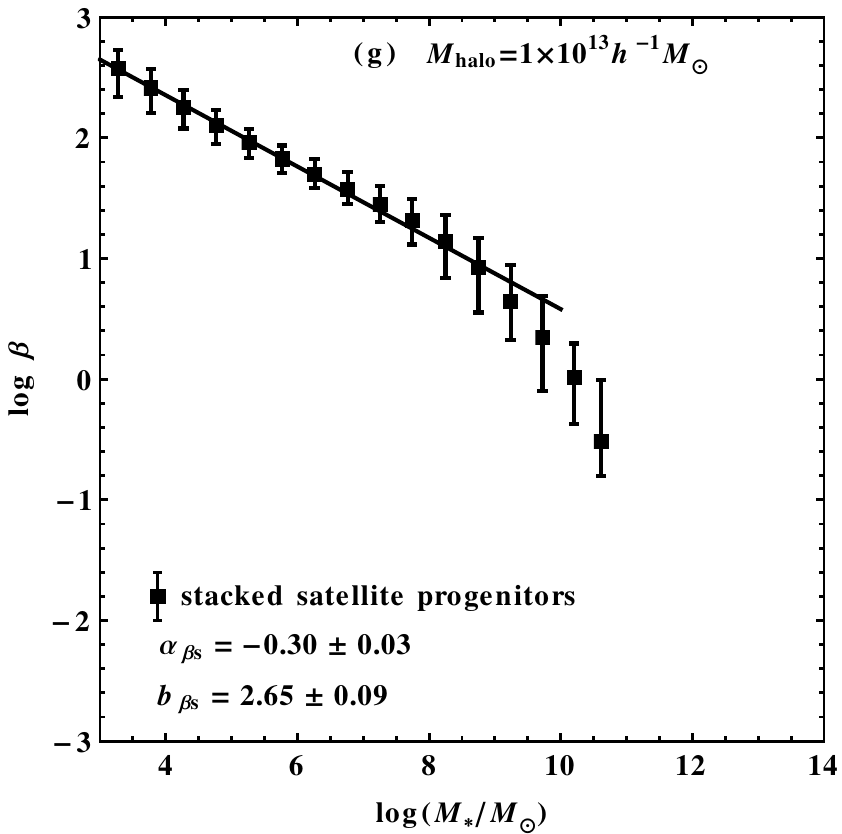}}
  \subfigure{\includegraphics[scale=0.43]{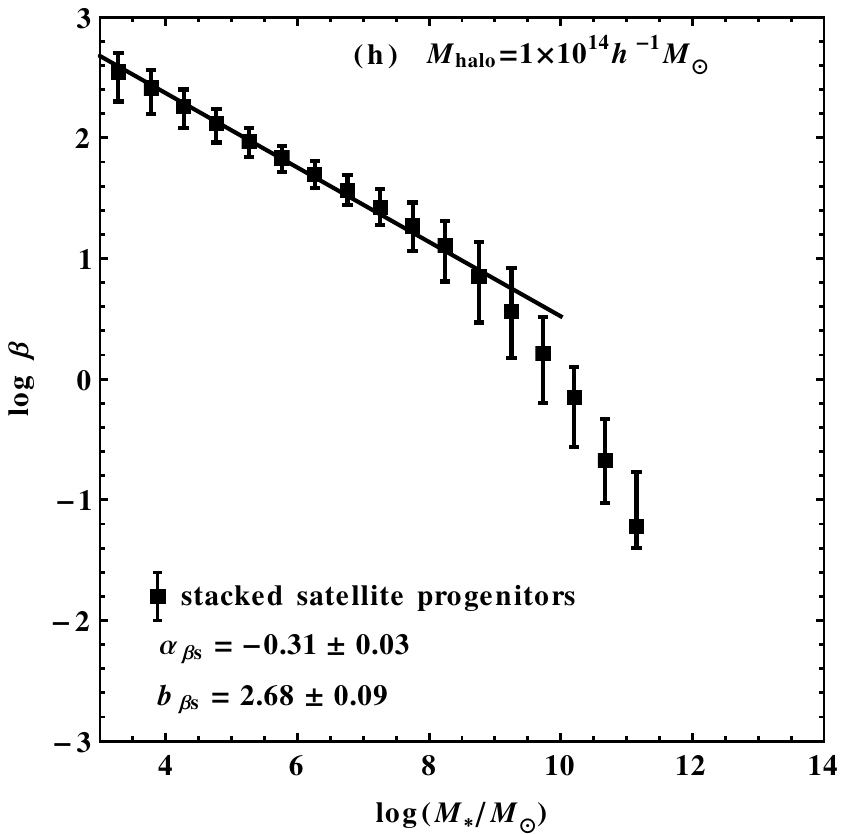}}
  \subfigure{\includegraphics[scale=0.43]{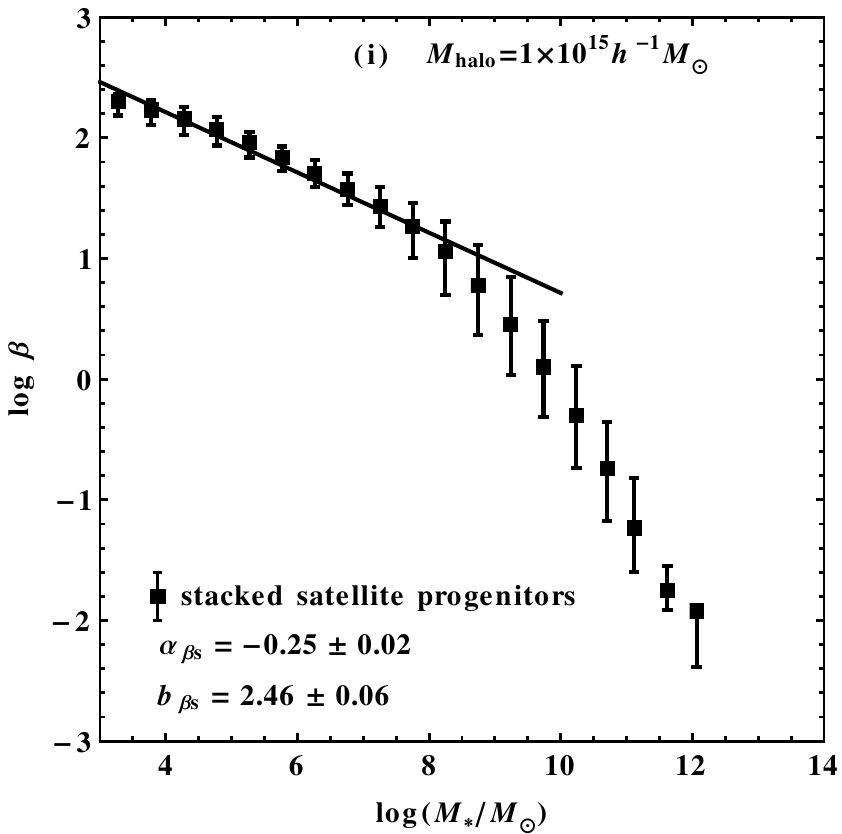}}
  \caption{The stellar feedback efficiency--stellar mass relation of the progenitor
galaxies of satellites in stacked multiple halos. The curves, labels, and
texts have the similar meanings as those in Figure~\ref{fig:gs_sat_stack},
except that the physical variable $M_{\rm cold}$ is replaced by the stellar
feedback efficiency $\beta$. See Section~\ref{sec:bs}.
}\label{fig:bs_sat_stack}
\end{figure*}

\begin{figure*}[htb]
  \subfigure{\includegraphics[scale=0.43]{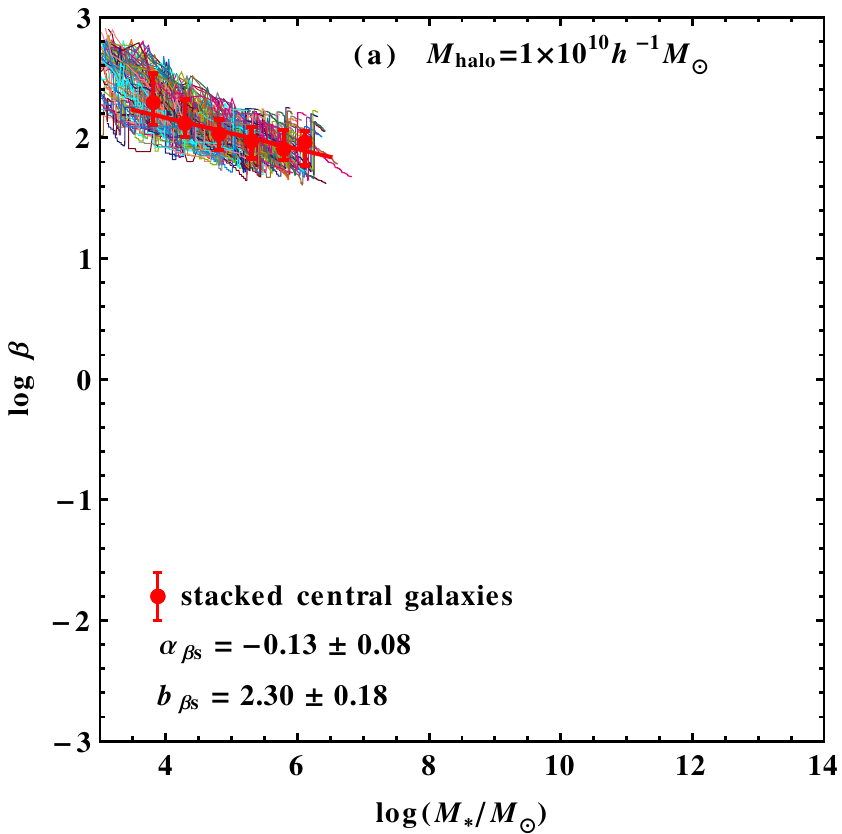}}
  \subfigure{\includegraphics[scale=0.43]{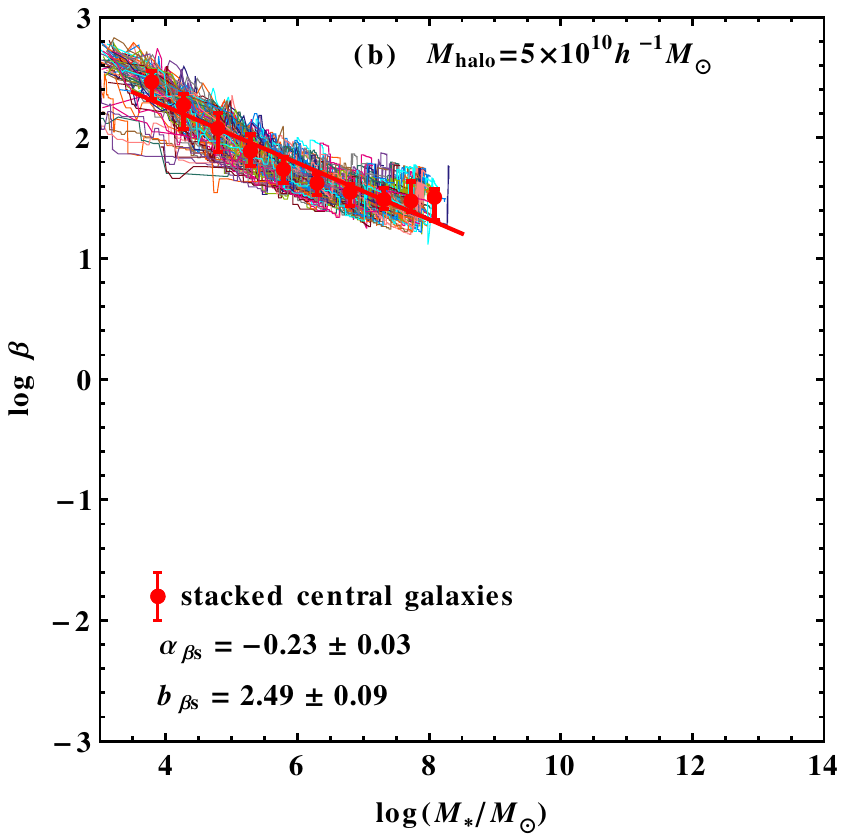}}
  \subfigure{\includegraphics[scale=0.43]{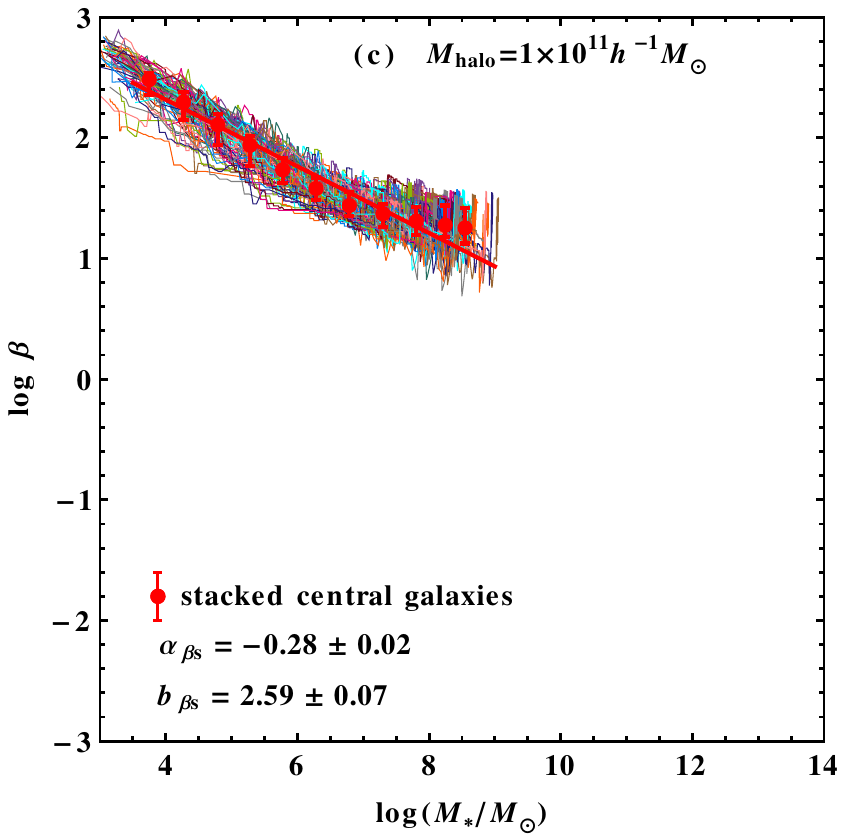}}
  \subfigure{\includegraphics[scale=0.43]{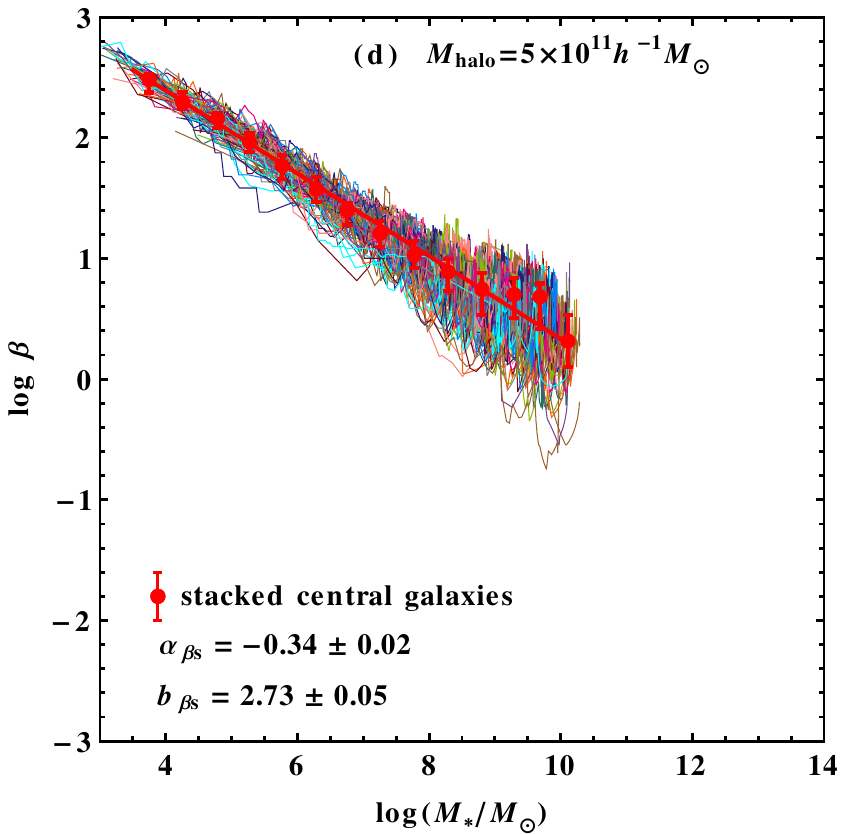}}
  \subfigure{\includegraphics[scale=0.43]{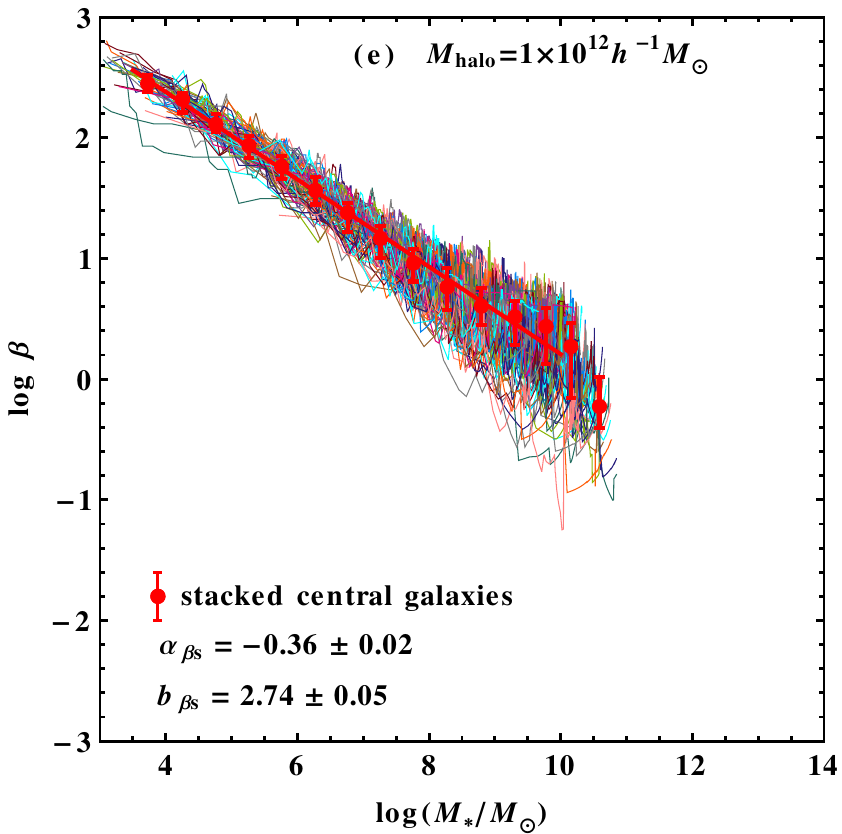}}
  \subfigure{\includegraphics[scale=0.43]{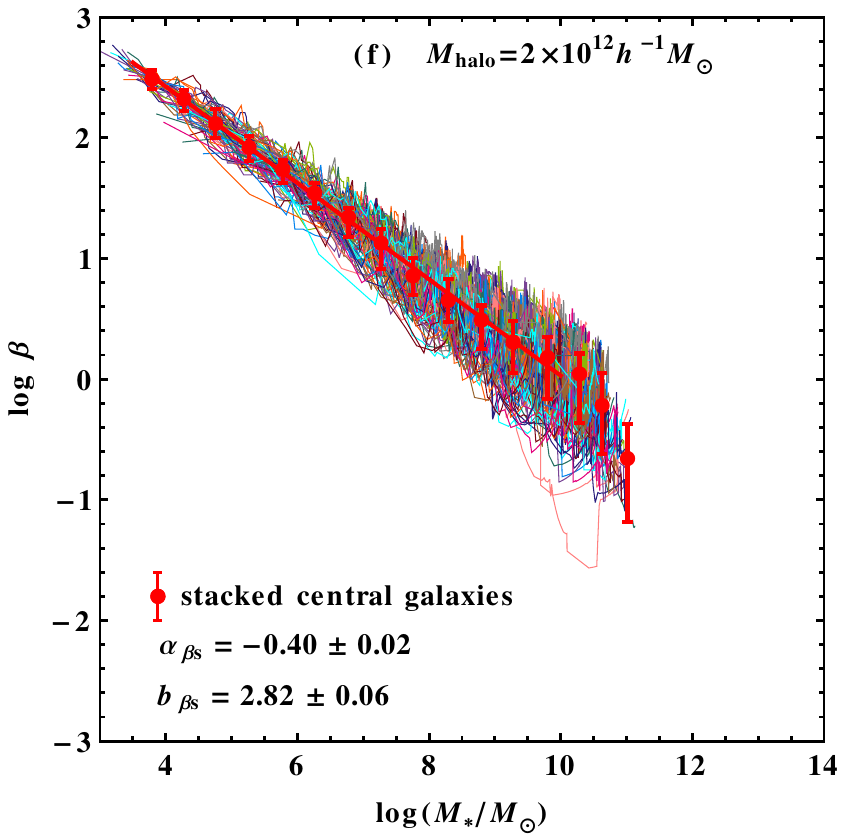}}  
  \subfigure{\includegraphics[scale=0.43]{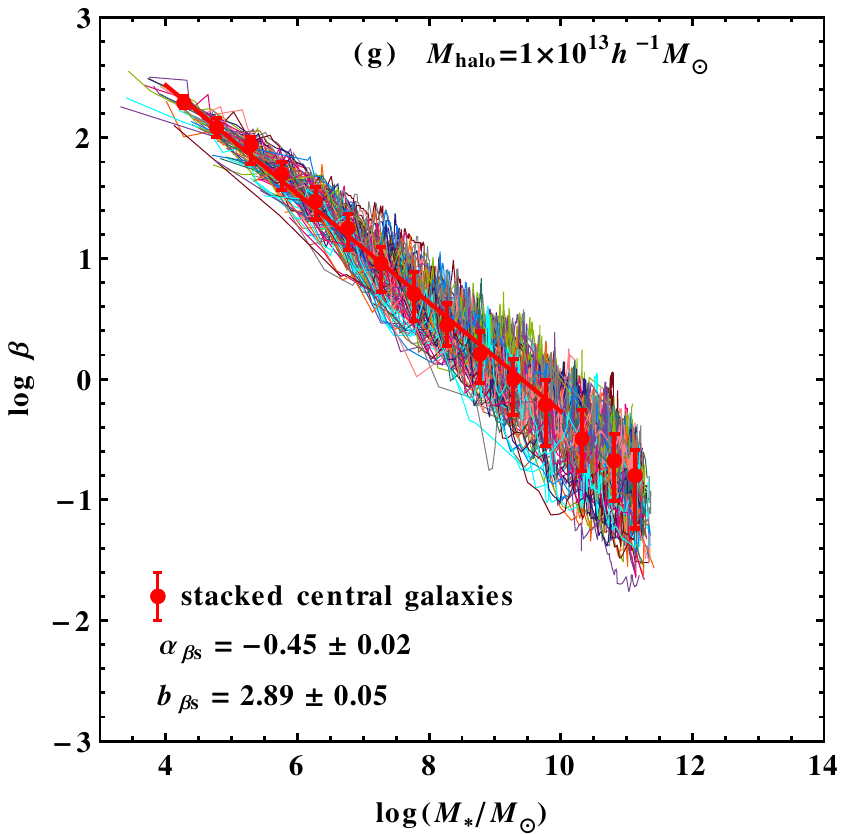}}
  \subfigure{\includegraphics[scale=0.43]{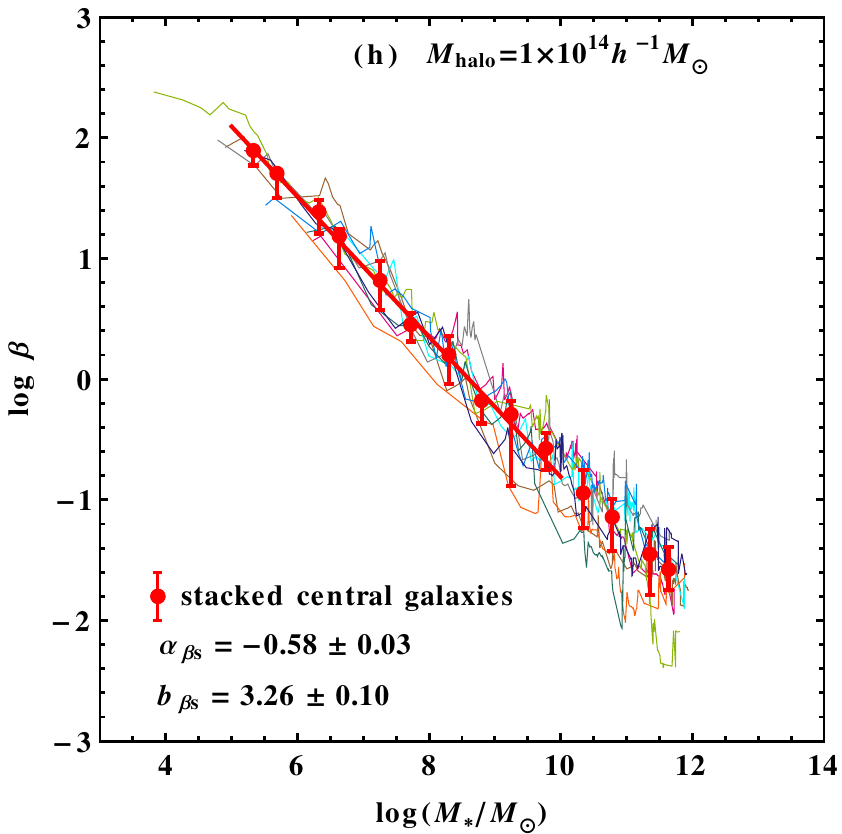}}
  \subfigure{\includegraphics[scale=0.43]{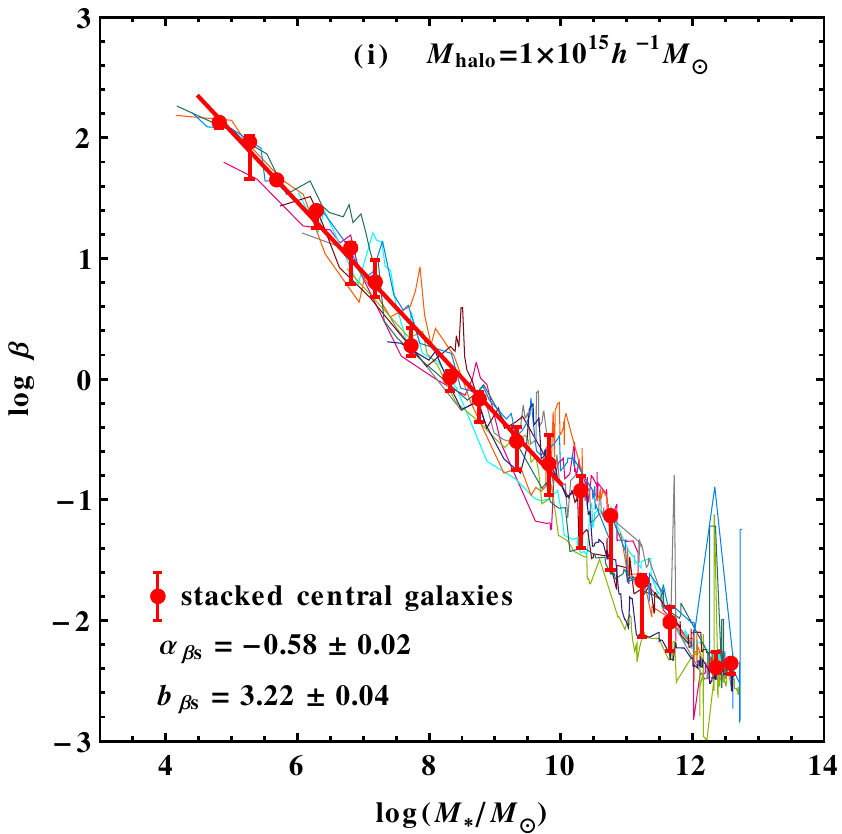}}
  \caption{The stellar feedback efficiency--stellar mass relation of the central
galaxies in stacked multiple halos.  The curves, labels, and texts have the
similar meanings as those in Figure~\ref{fig:gs_cet}, except that the physical
variable $M_{\rm cold}$ is replaced by the stellar feedback efficiency
$\beta$. See Section~\ref{sec:bs}.
}\label{fig:bs_cet}
\end{figure*}

\begin{figure}[htb]
\centering
\includegraphics[scale=0.65]{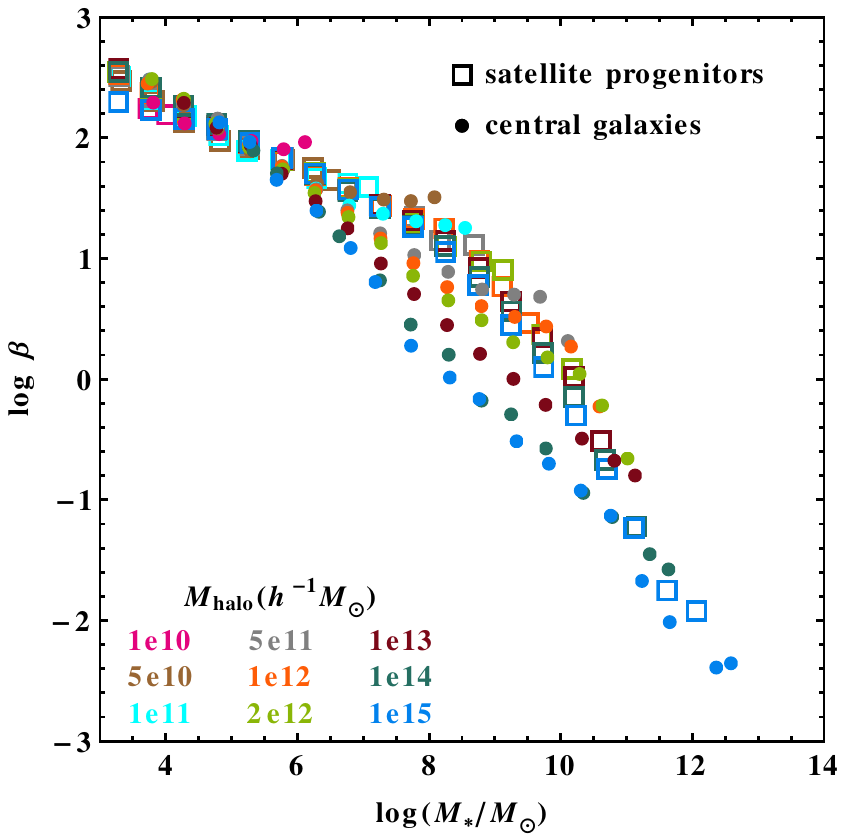}

\caption{Comparison of the median stellar feedback efficiency--stellar mass relations
obtained for satellite progenitors and central galaxies within halos with a
large range of halo masses.  The curves, labels, and texts have the similar
meanings as those in Figure~\ref{fig:gs_full}(a), except that the physical
variable $M_{\rm cold}$ is replaced by the stellar feedback efficiency $\beta$. See
Section~\ref{sec:bs}.  }\label{fig:bs_full} \end{figure}

Figures~\ref{fig:bs_sat}--\ref{fig:bs_full} show the $\beta$--$M_*$ relations
for the satellite progenitors and central galaxies in different halo masses,
corresponding to the same simulation examples shown for the $M_{\rm
cold}$--$M_*$ relations in Figures~\ref{fig:gs_sat}--\ref{fig:gs_full}(a),
respectively.
As seen from the figures, $\beta$ has a decreasing tendency with
increasing stellar masses.
We obtain the power-law scaling relation by fitting the data as follows,
\begin{equation}
\log\beta=\alpha_{\text{$\beta$s}}\log(M_*/10^3M_{\odot})+b_{\text{$\beta$s}},
\label{eq:bs}
\end{equation}
where the stellar mass ranges used in the fitting are the same as
done for the $M_{\rm cold}$--$M_*$ relation, 
$\alpha_{\text{$\beta$s}}$ is the slope, and $b_{\text{$\beta$s}}$ is the intercept
at $M_*=10^3M_{\odot}$ and related with $k_{\text{$\beta$s}}$ in Equation
(\ref{eq:bs0}) by $k_{\text{$\beta$s}}=10^{b_{\text{$\beta$s}}-3\alpha_{\text{$\beta$s}}}$.  
The fitting method for the $\beta$-$M_*$ relation is similar as done for the
$M_{\rm cold}$--$M_*$ relations above.  The best-fit parameters of
$\alpha_{\text{$\beta$s}}$ and $b_{\text{$\beta$s}}$ in
Figures~\ref{fig:bs_sat_stack}--\ref{fig:bs_cet} are listed in
Table~\ref{tab:bs}.
Note that $\beta\ga 1$ at the low-$M_*$ range and $\beta\la 1$ at the high-$M_*$
range. An extrapolation of the fitting formula to higher stellar masses
has little influence on the chemical evolution model results, as
$\beta$ is involved in the chemical evolution model through the term
$1-R+\beta$ (Eqs.~\ref{eq:df2} and \ref{eq:df3}), which is approximately $1-R$
for $\beta\ll 1$.

Figure~\ref{fig:bs_full} shows that the $\beta$--$M_*$ relation is roughly
universal for satellite progenitors. Table~\ref{tab:bs} shows that the slopes
$\alpha_{\text{$\beta$s}}$ and the intercepts $b_{\text{$\beta$s}}$ of the
best-fit scaling relations are in a universal range, with
$\alpha_{\text{$\beta$s}}$ being in the range from  -0.31 to -0.23 and
$b_{\text{$\beta$s}}\sim$2.4--2.7
for $M_{\rm halo}\sim 5\times 10^{10}$--$10^{15}h^{-1}\msun$. 

The median $\beta$--$M_*$ relation for central galaxies in halos with $M_{\rm
halo}\sim 5\times 10^{10}$--$2\times 10^{12}h^{-1}\msun$ is roughly in the same
narrow space, with $\alpha_{\text{$\beta$s}}$ being in the range from -0.40 to
-0.23 and $b_{\text{$\beta$s}}\sim$2.5--2.8. In halos with higher masses
($M_{\rm halo}\sim 10^{13}$--$10^{15}h^{-1}\msun$), Table~\ref{tab:bs} shows
$\alpha_{\text{$\beta$s}}\sim-0.45$ to -0.58 and $b_{\beta{\rm s}}\sim$2.9--3.2.

For a comparison of the $\beta$--$M_*$ relations between central galaxies and
satellite progenitors in different halo masses, we combine the points shown in
Figures~\ref{fig:bs_sat_stack} and \ref{fig:bs_cet} together into
Figure~\ref{fig:bs_full}.
This comparison demonstrates that the $\beta$--$M_*$ relations are roughly universal for central galaxies in $M_{\rm halo}\la
10^{13}h^{-1}\msun$ and satellite galaxies in a wide range of host halo masses
with $10^{10}h^{-1}\msun\la M_{\rm halo}\la 10^{15}h^{-1}\msun$. For high
present-day halo masses ($M_{\rm halo}\ga 10^{13}h^{-1}\msun$), the stellar feedback
efficiencies in central galaxies are relatively lower than those in the
satellite progenitors at the stellar mass range $M_*\ga 10^6\msun$ (which is
because the central galaxies in that stellar mass range are at higher redshifts
and have more compact sizes, and their disk rotation velocities are higher.)
Note that the central galaxies in $M_{\rm halo}\ga 10^{13}h^{-1}\msun$ are
shown to have stellar mass $M_*\ga 10^{11}\msun$ and their stellar metallicities
are close to the yield in \citet{XY19}, while in this paper we focus on the
stellar mass range of $10^3\msun\la M_*\la 10^{11}\msun$ of the MZR.  The stellar
feedback efficiency--stellar mass relation obtained for the central galaxy in
$M_{\rm halo}\ga 10^{13}h^{-1}\msun$ needs not to be used to explain their
present-day stellar metallicity.  Thus the deviation of the stellar feedback
efficiency of the central galaxies in $M_{\rm halo}\ga 10^{13}h^{-1}\msun$
shown in Figure~\ref{fig:bs_full} will not affect our results and conclusions
in this work below.

The $\beta$--$M_*$ power-law scaling relation revealed in this subsection
implies the existence of the $V\vir$--$M_*$ and the $M\halo$--$M_*$ power-law
scaling relations, as $\beta$ is calculated from Equation (\ref{eq:betaE}) with
$\beta\propto V\vir^{-2}$ for a wide range of $M_*$ or $V\vir$; for example,
$\beta$ is taken to be $\beta\E$ at $V\vir\la 200\kms$ in Figure 2 in HYL14.
As to be mentioned in Section~\ref{sec:conclusions}, testing the origin of the
$\beta$--$M_*$ relation (and the $M_{\rm cold}$--$M_*$ relation) is one of the
next steps of the work.

\subsection{The MZRs obtained from our chemical evolution model}
\label{sec:mzr_mod}

By applying the above best-fit $M_{\rm cold}$--$M_*$ and $\beta$--$M_*$ power-law scaling relations (listed in Tables~\ref{tab:gs}--\ref{tab:bs}) to the
analytic chemical evolution model constructed in Section~\ref{sec:bpg_m}, we
obtain the expected MZRs, which is done in details by substituting the scaling
relations of Equations (\ref{eq:gs}) and (\ref{eq:bs}) into Equations
(\ref{eq:mother3}) and (\ref{eq:zs}) for the $\beta$-, g-, and $p$-models in
Section~\ref{sec:bpg_m}. For illustration, we show one example in
Figure~\ref{fig:mzr}, by adopting the best-fit parameters ($\alpha_{\rm gs}$,
$b_{\rm gs}$, $\alpha_{\text{$\beta$s}}$, $b_{\text{$\beta$s}}$) obtained for
the satellite progenitors with $M_{\rm halo}=2\times 10^{12}h^{-1}\msun$. As
the $M_{\rm cold}$--$M_*$ and $\beta$--$M_*$ power-law scaling relations are
shown to be roughly universal, the results for other cases do not differ much.
We show the results of the $\beta$-, g-, and $p$-models with different line
types in Figure~\ref{fig:mzr}.
Note that the MZR results of the $\beta$-model are also obtained numerically
from Equations (\ref{eq:mother3}) and (\ref{eq:zs}), not from the
approximations of $\beta\gg 1-R$ and $\beta \ll 1-R$ analyzed in
Section~\ref{sec:bpg_m}, so that the presented results cover $\beta\sim 1-R$
continuously.
To illustrate the effectiveness
of the chemical evolution model, the MZR results obtained from both the SAM in
\citet{XY19} and observations are also shown in Figure~\ref{fig:mzr}(d) and (e)
almost all of which fall in
the gap between the $g-$ and $\beta$-model results
in the range of $10^3M_{\odot}\leq M_*\leq
10^{10}M_{\odot}$ as shown in Figure~\ref{fig:mzr}(f).
Note that the g-model and the $\beta$-model represent the upper and lower
bounds on the assumption of the infalling hot gas
metallicity, and the differences among the g-, $p$-, and $\beta$-model results
are not larger than 1~dex, so the MZRs are located within a narrow space and
appear universal as shown in \citet{XY19}.

Note that for the MZRs of satellites, the $M_{\rm cold}$--$M_*$ relation and
the $\beta$--$M_*$ relation used to obtain the dotted lines shown in
Figure~\ref{fig:mzr} are obtained from the satellite progenitors (before their
infall into a bigger halo), while the solid lines shown in Figure~\ref{fig:mzr}
are shown for satellites obtained after the satellite progenitors fell into a
bigger halo. Here we argue that the evolution of the MZRs in satellites after
their infall is negligible. As mentioned in Section~\ref{sec:method} and in
HYL14, we apply the energy condition only to a galaxy before it becomes a
satellite. We do not apply it to satellites, but assume that the reheated gas
from satellites (with mass expected by Equations~\ref{eq:dMreheat} and
\ref{eq:sim_fb}) is expelled into the big host halo, as the original halos of
the satellites are largely tidally disrupted along their motion in the big host
halo, and the tidal field induced by the big host halo also helps to keep those
expelled materials out of the satellites.  According to the stellar feedback
model described in Section~\ref{sec:method}, after the infall, the effective
stellar feedback efficiency changes from $\beta\E$ shown by Equation
(\ref{eq:betaE}) to $\beta$ shown by Equation (\ref{eq:sim_fb}), which
increases significantly (mostly in the low-$M_*$ range; see also figure 1 in
HYL14). The chemical evolution model of a satellite after its infall can be
approximated by the leaky-box model described in Section~\ref{sec:model}.  Thus
the change of the stellar mass after the infall is not significant due to the
significant increase of $\beta$ (see Eq.~\ref{eq:df5}), and neither is the
change of the stellar metallicity (see Eq.~\ref{eq:lk2s0}).

As mentioned above, the $\beta$-, g-, and $p$-models are based on some simple
assumptions on the ratio of the hot gas metallicity $Z_{\rm hot}$ to the cold
gas metallicity $Z_{\rm cold}$.  In Figure~\ref{fig:mzr}, we also demonstrate
how the MZRs expected from the chemical evolution model with other different
ratios $\xi$ are distributed between the $g$- and the $\beta$-model results.
As seen from the distribution, the expected MZR when $\xi\la 0.1$ is quite
close to the expectation from the $\beta$-model.  The realistic ratios of $\xi$
obtained for the satellite progenitors in our SAMs are mostly distributed in
the range of 0.1--0.9 at their infall into bigger host halos.

As mentioned above in Section~\ref{sec:bpg_m}, the exponents in the MZRs are
expected to be closely related to the exponents in the power-law scaling
relations of the $M_{\rm cold}$--$M_*$ relation and the $\beta$-$M_*$ relation,
$1-\alpha_{\rm gs}$ or $-\alpha_{\beta{\rm s}}$, which is supported by the SAM
results and the exponents shown in Tables~\ref{tab:gs} and \ref{tab:bs}, as
summarized below.
\begin{itemize}
\item The slopes of the MZRs obtained from the SAM in \citet{XY19} are in the
range of $\sim 0.2$--0.4, as mentioned at the end of Section~\ref{sec:method}. 
\item For the $\beta$-model, the slope of the MZR expected by Equation
(\ref{eq:bmalpha0}) with $\alpha=0$ equals to $-\alpha_{\beta{\rm s}}$, which
is shown to be in the range of 0.23 to 0.31 for $M_{\rm halo}\sim 5\times
10^{10}$--$10^{15}h^{-1}\msun$ in Table~\ref{tab:bs}, consistent with
the SAM results.  Note that the exponents in Tables~\ref{tab:gs} and
\ref{tab:bs} are generally satisfied with $\alpha=1-\alpha_{\rm
gs}+\alpha_{\beta{\rm s}} \sim -0.14$ to 0.18 and
$\frac{k_{\rm
gs}}{k_{\beta{\rm s}}}\cdot 10^{-3(1-\alpha_{\rm gs}+\alpha_{\beta{\rm
s}})}=10^{b_{{\rm gs}}-b_{\beta{\rm s}}-3}\sim$0.5--3 for $M_{\rm halo}\sim
5\times 10^{10}$--$10^{15}h^{-1}\msun$, and correspondingly
$\frac{1-R+\beta}{1-R}\cdot\frac{M_*}{M_{\rm cold}}\sim$0.5--5.
\item For the $p$-model and the g-model, the slope of the MZR expected by
Equations (\ref{eq:gis}) and (\ref{eq:pis}) equals to $1-\alpha_{\rm gs}$,
which is shown to be in the range of 0.21--0.39 for $M_{\rm halo}\sim 5\times
10^{10}$--$10^{15}h^{-1}\msun$ in Table~\ref{tab:gs}, consistent with
the SAM result.
\end{itemize}

\section{Summary and Discussion}
\label{sec:conclusions}

In this work, we investigated the origin of the stellar mass--stellar
metallicity relations.  We constructed a chemical evolution model, based on the
continuity equations in baryonic mass and metal mass conservations in the
galaxy formation and evolution model.  We applied some scaling relations
between the cold gas mass and the stellar mass ($M_{\rm cold}$--$M_*$) and
between the feedback efficiency and the stellar mass ($\beta$--$M_*$) to the
model and obtained some analytical solutions in the MZRs.  The solutions agree
well with the simulation results obtained from the SAMs and the MZR
observations. The range between the upper and the lower bounds in MZRs
predicted in our models (through the upper and lower bounds in the ratio of the
hot gas metallicity to the cold gas metallicity) is narrow, which provides an
explanation to the universality in the MZRs revealed in the study by
\citet{XY19}. The exponents in the MZRs are closely connected with the
exponents in the $M_{\rm cold}$--$M_*$ and $\beta$--$M_*$ power-law scaling
relations.
 
The $M_{\rm cold}$--$M_*$ and $\beta$--$M_*$ relations in this work represent
the ``average'' evolutional tracks with the increase of the stellar masses in a
galaxy, which appears to be universal for diverse star formation histories of
different galaxies.  Our work shows that the formation of the MZR is a result
caused jointly by that the cold gas mass fraction decreases with increasing
$M_*$ and that the stellar feedback efficiency decreases with increasing $M_*$ in
the galaxy growth. The MZR represents an ``average'' evolutional track for the
stellar metallicity of a galaxy.

In the chemical evolution model constructed in this work, both the effects of
gas inflows and outflows are considered, and some specific assumptions used in
the previous models are removed (e.g., those in the ``closed-box'' model, the
``leaky-box'' model, cold gas equilibrium, metal mass equilibrium as discussed
in Section~\ref{sec:model}). The power-law scaling relations applied to the
model are obtained from the SAM in HYL14 and \citet{XY19}, which give
how the cold gas mass and the feedback efficiency change with the increase of
the stellar mass during the growth of a galaxy. The effects of halo
assemblies and the physical processes of gas cooling are implicitly included in
the scaling relations extracted from the SAM, and they result in star formation
and the growth of the stellar mass and the gas mass in a galaxy.

Our SAM simulation results show that the median relations between the cold gas
mass and the stellar mass ($M_{\rm cold}$--$M_*$) and between the stellar feedback
efficiency and the stellar mass ($\beta$--$M_*$) are distributed within a
narrow space during the growth of $M_*$.  During the growth of a galaxy, the
median relations between the cold gas mass and the stellar mass appear to
increase when the stellar mass is small and decline when the stellar mass is
sufficiently large.  The increasing part of the relation appears to be
universal in a power-law scaling relationship, with $\log(M_{\rm
cold}/M_{\odot})=\alpha_{\rm gs}\log(M_*/10^3\msun)+b_{\rm gs}$ with
$\alpha_{\rm gs}\sim$0.6--0.8 and $b_{\rm gs}\sim$5.5--6.0 over a large range
of halo masses, during the growth of both central galaxies and satellite
progenitors. The $\beta$--$M_*$ relation is also roughly the same for the
progenitor satellites within $10^{10}h^{-1}\msun\la M_{\rm halo}\la
10^{15}h^{-1}\msun$ and central galaxies in $M_{\rm halo}\la 2\times
10^{12}h^{-1}\msun$, following a power-law scaling relationship with
$\beta=\alpha_{\beta{\rm s}}\log(M_*/10^3\msun)+b_{\beta{\rm s}}$,
$\alpha_{\beta{\rm s}}\sim -0.4$ to $-0.2$ and $b_{\beta{\rm s}}\sim$2.5--2.8.
The exponents of the MZRs are close to the exponents of $1-\alpha_{\rm gs}$ or
$-\alpha_{\beta{\rm s}}$.  

The following reasons have also been proposed in the literature to explain the
increase of the stellar metallicity with increasing stellar mass revealed in
nearby galaxies by using (1) outflows, where $\beta$ decreases with increasing
$M_*$; or (2) lower specific star formation rates in high-mass systems; or (3)
changing of the yield $p$ due to different IMFs in different-mass systems.  In
this work, the evolution of the ratio $M_*/M_{\rm cold}$ and the formation of
the MZR have the contribution from both outflows and inflows, where inflows are
associated with halo growth.  Point (2) can also be a consequence of a
relatively high $M_*/M_{\rm cold}$ ratio in high-$M_*$ systems. The change of
$p$ is not necessary for this work to match the observational MZR.

Some assumptions in some previous equilibrium models involving both inflows and
outflows have been relaxed in this work. In some special cases, e.g., $Z_{\rm
hot}=0$, there exists some similarity (but not exactly the same) between the
form of the solution obtained in this work and the form obtained in those
previous equilibrium models. If the expectations by some equilibrium models are
compatible with observations, it may imply some kind of reasonability in the
assumption that a local equilibrium at a given stellar mass is close to being
reached in practice. The general solution obtained in our model can be reduced
to a local equilibrium model at a given stellar mass, which requires that the
average stellar mass is sufficiently massive compared to the average cold gas
mass, in consideration of the effect of stellar feedback.  However, in general, the
model in this work involves a dynamical and secular evolution of the average
stellar mass and the average gas mass during the growth of a galaxy, as well
as the associated evolution of inflow gas metallicity, where the mass growth
of a galaxy can span a vast mass range.

\citet{XY19} show that a double power law exists in the MZR relations for both
central galaxies and stacked satellites.  In this paper, we do not investigate
the details on non-linear effects of the MZRs revealed in \citet{XY19}, which
can be related to the non-linear relations in the $M_{\rm cold}$--$M_*$
relations and in the $\beta$--$M_*$ relations.

Some next important steps of this work are to investigate the evolution of the
MZR with redshift (e.g., \citealt{Maetal16}, where the evolution of the MZR
with redshifts is attributed to the redshift evolution of the gas mass fraction
within a halo, illustrated through cosmological zoom-in simulations and a
``closed-box'' model), connect it to cold gas phase metallicities revealed in
observations (e.g., \citealt{tremonti04,mannucci10,Leeetal06}) and the hot gas
metallicities in DM halos, test the $M_{\rm cold}$--$M_*$ relation and the
$\beta$--$M_*$ relation and their origin, investigate the non-linear effects in
the correlations, and explore whether there exists a possible dependence on
three or more parameters (e.g., star formation rate, as shown in
\citealt{Ellisonetal08,mannucci10,BB17}) etc.

We thank Sandra Faber, Youjun Lu, Filippo Mannucci, and Ying-jie Peng for helpful discussions.
This work was supported in part by the National Natural Science Foundation of
China under Nos. 11673001, 11273004, 10973001, 11721303, the National Key R \& D
Program of China (Grant No. 2016YFA0400703), and the Strategic Priority Program
of the Chinese Academy of Sciences (grant No. 23040100).

\addcontentsline{toc}{section}{REFERENCES}
\bibliography{reference}

\begin{thebibliography}{}

\bibitem[Barrera-Ballesteros et al.(2017)]{BB17} Barrera-Ballesteros, J.~K.,
S{\'a}nchez, S.~F., Heckman, T., Blanc, G.~A., \& The MaNGA Team 2017, \apj,
844, 80

\bibitem[Benson(2010)]{Benson10} Benson, A.~J.\ 2010, \physrep, 495, 33 

\bibitem[Bond et al.(1991)]{Bondetal91} Bond, J.~R., Cole, S., Efstathiou, G., \& Kaiser, N.\ 1991, \apj, 379, 440

\bibitem[Brooks et al.(2007)]{Brooksetal07} Brooks, A.~M., Governato, F., Booth, C.~M., et al.\ 2007, \apjl, 655, L17

\bibitem[Cole et al.(2000)]{Coleetal00} Cole, S., Lacey, C.~G., Baugh, C.~M., \& Frenk, C.~S.\ 2000, \mnras, 319, 168 

\bibitem[Dalcanton(2007)]{dalcanton07} Dalcanton, J.~J.\ 2007, \apj, 658, 941 

\bibitem[Dekel \& Silk(1986)]{DS86} Dekel, A., \& Silk, J.\ 1986, \apj, 303, 39
 
\bibitem[Ellison et al.(2008)]{Ellisonetal08} Ellison, S.\ L., Patton, D.\ R., Simard, L., \& McConnachie, A.\ W.\ 2008, ApJL, 672, L107

\bibitem[Finlator(2016)]{F16} Finlator, K.\ 2016, arXiv:1612.00802

\bibitem[Finlator \& Dav{\'e}(2008)]{FD08} Finlator, K., \& Dav{\'e}, R.\ 2008, \mnras, 385, 2181

\bibitem[Font et al.(2011)]{font2011}
  Font, A.~S., Benson, A.~J., Bower, R.~G., et al.\ 2011, \mnras, 417, 1260

\bibitem[Gallazzi et al.(2005)]{gallazzi05} Gallazzi, A., Charlot, S., Brinchmann, J., White, S.~D.~M., \& Tremonti, C.~A.\ 2005, \mnras, 362, 41 

\bibitem[Gallazzi et al.(2006)]{gallazzi06} Gallazzi, A., Charlot, S., Brinchmann, J., \& White, S.~D.~M.\ 2006, \mnras, 370, 1106 

\bibitem[Galli \& Palla(1998)]{galli1998}
  Galli, D., \& Palla, F.\ 1998, \aap, 335, 403

\bibitem[Gnedin(2000)]{gnedin00} Gnedin, N.~Y.\ 2000, \apj, 542, 535 

\bibitem[Gonz{\'a}lez Delgado et al.(2014)]{GDetal14} Gonz{\'a}lez Delgado,
R.~M., Cid Fernandes, R., Garc{\'{\i}}a-Benito, R., et al.\ 2014, \apjl, 791,
L16

\bibitem[Guo et al.(2011)]{Guoetal11} Guo, Q., White, S., Boylan-Kolchin, M., et al.\ 2011, \mnras, 413, 101

\bibitem[Hou et al.(2014)]{HYL14} Hou, J., Yu, Q., \& Lu, Y.\
2014, ApJ, 791, 8 (HYL14)

\bibitem[Iwamoto et al.(1999)]{SN_Ia_pattern} Iwamoto, K., Brachwitz, F.,
Nomoto, K., et al.\ 1999, \apjs, 125, 439

\bibitem[Kauffmann et al.(1993)]{Kauffmann93} Kauffmann, G., White,
S.~D.~M., \& Guiderdoni, B.\ 1993, \mnras, 264, 201

\bibitem[Kirby et al.(2013)]{Kirbyetal13} Kirby, E.~N., Cohen, J.~G., Guhathakurta, P., et al.\ 2013, \apj, 779, 102 

\bibitem[Kobayashi et al.(2007)]{kobayashi07} Kobayashi, C., Springel, V., \& White, S.~D.~M.\ 2007, \mnras, 376, 1465 

\bibitem[K{\"o}ppen et al.(2007)]{koppen07} K{\"o}ppen, J., Weidner, C., \& Kroupa, P.\ 2007, \mnras, 375, 673 

\bibitem[Kravtsov et al.(2004)]{kravtsov04} Kravtsov, A.~V., Gnedin, O.~Y., \& Klypin, A.~A.\ 2004, \apj, 609, 482 

\bibitem[Lacey \& Cole(1993)]{LC93} Lacey, C., \& Cole, S.\ 1993, \mnras, 262, 627

\bibitem[Lee et al.(2006)]{Leeetal06} Lee, H., Skillman, E.~D., Cannon, J.~M., et al.\ 2006, \apj, 647, 970

\bibitem[Leroy et al.(2008)]{leroy08} Leroy, A.~K., Walter, F., Brinks, E., et al.\ 2008, \aj, 136, 2782 

\bibitem[Li et al.(2010)]{Lietal10} Li, Y.-S., De Lucia, G., \& Helmi, A.\ 2010, \mnras, 401, 2036

\bibitem[Lilly et al.(2013)]{Lillyetal13} Lilly, S.~J., Carollo, C.~M., Pipino, A., Renzini, A., \& Peng, Y.\ 2013, \apj, 772, 119
 
\bibitem[Lu et al.(2014)]{Luetal14} Lu, Y., Wechsler, R.~H., Somerville, R.~S., et al.\ 2014, \apj, 795, 123

\bibitem[Lu et al.(2017)]{Luetal17} Lu, Y., Benson, A., Wetzel, A., et al.\ 2017, \apj, 846, 66

\bibitem[Lynden-Bell(1975)]{lb75} Lynden-Bell, D.\ 1975, Vistas in Astronomy, 19, 299 

\bibitem[Ma et al.(2016)]{Maetal16} Ma, X., Hopkins, P.~F., Faucher-Gigu{\`e}re, C.-A., et al.\ 2016, \mnras, 456, 2140 

\bibitem[Maiolino \& Mannucci(2019)]{MM19} Maiolino, R., \& Mannucci, F.\ 2019, \aapr, 27, 3.

\bibitem[Mannucci et al.(2010)]{mannucci10} Mannucci, F., Cresci, G., Maiolino, R., Marconi, A., \& Gnerucci, A.\ 2010, \mnras, 408, 2115 

\bibitem[Maoz et al.(2010)]{MSG10} Maoz, D., Sharon, K., \& Gal-Yam, A.\ 2010, \apj, 722, 1879

\bibitem[McConnachie(2012)]{mcconnachie12} McConnachie, A.~W.\ 2012, \aj, 144, 4 
\bibitem[McGaugh(2005)]{mcgaugh05} McGaugh, S.~S.\ 2005, \apj, 632, 859 

\bibitem[McGaugh(2012)]{mcgaugh12} McGaugh, S.~S.\ 2012, \aj, 143, 40 

\bibitem[Nomoto et al.(2006)]{SN_II_pattern} Nomoto, K., Tominaga, N., Umeda,
H., Kobayashi, C., \& Maeda, K.\ 2006, Nuclear Physics A, 777, 424

\bibitem[Okamoto et al.(2008)]{Okamoto08} Okamoto, T., Gao, L., \& Theuns, T.\ 2008, \mnras, 390, 920 

\bibitem[Pagel(1997)]{pagel97} Pagel, B.~E.~J.\ 1997, Nucleosynthesis and Chemical Evolution of Galaxies, by Bernard E.~J.~Pagel, pp.~392.~ISBN 0521550610.~Cambridge, UK: Cambridge University Press, October 1997., 392 

\bibitem[Panter et al.(2008)]{Panteretal08} Panter, B., Jimenez, R., Heavens,
A.~F., \& Charlot, S.\ 2008, \mnras, 391, 1117

\bibitem[Papastergis et al.(2012)]{papastergis12} Papastergis, E., Cattaneo, A., Huang, S., Giovanelli, R., \& Haynes, M.~P.\ 2012, \apj, 759, 138 

\bibitem[Parkinson et al.(2008)]{parkinson08} Parkinson, H., Cole, S., \& Helly, J.\ 2008, \mnras, 383, 557 

\bibitem[Peeples et al.(2014)]{peeples14} Peeples, M.~S., Werk, J.~K., Tumlinson, J., et al.\ 2014, \apj, 786, 54 

\bibitem[Peng \& Maiolino(2014)]{PM14} Peng, Y.-j., \& Maiolino, R.\ 2014, \mnras, 443, 3643 

\bibitem[Press \& Schechter(1974)]{PS74} Press, W.~H., \& Schechter, P.\ 1974, \apj, 187, 425

\bibitem[Saintonge et al.(2011)]{S11} Saintonge, A., Kauffmann, G., Kramer, C.,
et al.\ 2011, MNRAS, 415, 32

\bibitem[Schmidt(1963)]{schmidt63} Schmidt, M.\ 1963, \apj, 137, 758 

\bibitem[Searle \& Sargent(1972)]{SS72} Searle, L., \& Sargent, W.~L.~W.\ 1972, \apj, 173, 25 

\bibitem[Somerville et al.(2008)]{Somervilleetal08} Somerville, R.~S.,
Hopkins, P.~F., Cox, T.~J., Robertson, B.~E., \& Hernquist, L.\ 2008, \mnras,
391, 481

\bibitem[Somerville et al.(2015)]{Somervilleetal15} Somerville, R.~S., Popping, G., \& Trager, S.~C.\ 2015, \mnras, 453, 4337

\bibitem[Somerville \& Kolatt(1999)]{SK99} Somerville, R.~S., \& Kolatt, T.~S.\ 1999, \mnras, 305, 1

\bibitem[Somerville \& Primack(1999)]{Somerville99} Somerville, R.~S., \& Primack, J.~R.\ 1999, \mnras, 310, 1087

\bibitem[Starkenburg et al.(2013)]{S13} Starkenburg, E., Helmi, A., De Lucia, G., et al.\ 2013, \mnras, 429, 725

\bibitem[Sutherland \& Dopita(1993)]{SD93} Sutherland, R.~S., \& Dopita, M.~A.\ 1993, \apjs, 88, 253

\bibitem[Talbot \& Arnett(1971)]{TA71} Talbot, R.~J., Jr., \& Arnett, W.~D.\ 1971, \apj, 170, 409 

\bibitem[Tremonti et al.(2004)]{tremonti04} Tremonti, C.~A., Heckman, T.~M., Kauffmann, G., et al.\ 2004, \apj, 613, 898 

\bibitem[Vale Asari et al.(2009)]{VAetal09} Vale Asari, N., Stasi{\'n}ska, G., Cid Fernandes, R., et al.\ 2009, \mnras, 396, L71

\bibitem[White \& Frenk(1991)]{WhiteFrenk91} White, S.~D.~M., \& Frenk, C.~S.\ 1991, \apj, 379, 52

\bibitem[Wolcott-Green et al.(2017)]{W17} Wolcott-Green, J., Haiman, Z., \& Bryan, G.~L.\ 2017, \mnras, 469, 3329 

\bibitem[Xia \& Yu(2019)]{XY19} Xia, M., \& Yu, Q.\ 2019, ApJ, 874, 105

\bibitem[Yates et al.(2013)]{Y13} Yates, R.~M., Henriques,
B., Thomas, P.~A., et al.\ 2013, \mnras, 435, 3500

\end{thebibliography}

\begin{table*}[b]
\centering
\caption{Best-fit results for the median $M_{\rm cold}$--$M_*$ scaling relations}
\begin{tabular}{c|c|c|c|c|c|c|c|c|c}
\hline
\hline
Objects & $M_{\text{halo}}/(h^{-1}M_{\odot})$ & $\log(M_*/M_{\odot})$ & $\alpha_{\text{gs}}$ & $b_{\text{gs}}$ & $\chi^2$ & $N_{\text{bin}}$  & $N_{\text{tree}}$ & $N_{\text{galaxy}}$ & $M_{\rm res}/(h^{-1}M_{\odot})$ \\
\hline
& $1\times10^{10}$ &	$3-4$	 & $0.24\pm1.75$ & $5.8\pm1.6$ & - & 2 & 100 & 9 & $10^4$ \\
& $5\times10^{10}$ &	$3-7$	 & $0.79\pm0.05$ & $5.5\pm0.2$ & 0.9 & 8 & 100 & 62 & $10^4$ \\
satellite & $1\times10^{11}$ &	$3-7.5$	 & $0.76\pm0.06$ & $5.5\pm0.2$ & 2.6 & 9 & 100 & 165 & $10^4$ \\
progenitors & $5\times10^{11}$ &  $3-9$	 & $0.68\pm0.03$ & $5.7\pm0.1$ & 4.3 & 14 & 100 & 2163 & $10^4$ \\
& $1\times10^{12}$	 &	$3-10$	 & $0.61\pm0.03$ & $5.9\pm0.1$ & 15.7 & 14 & 100 & 5970 & $10^5$ \\
& $2\times10^{12}$	 &	$3-10$	 & $0.66\pm0.03$ & $5.7\pm0.1$ & 3.3 & 14 & 100 & 15265 & $10^5$ \\
& $1\times10^{13}$ &	$3-10$	 & $	0.65\pm0.03$ & $5.7\pm0.1$ & 2.9 & 14 & 100 & 102164 & $10^6$ \\
& $1\times10^{14}$ &	$3-10$	 & $0.65\pm0.03$ & $5.8\pm0.1$ & 2.0 & 14 & 10 & 85563 & $10^7$ \\
& $1\times10^{15}$	 &	$3-10$	 & $0.61\pm0.03$ & $6.0\pm0.1$ & 0.9 & 14 & 10 & 543679 & $10^8$ \\
\hline
& $1\times10^{10}$ &	$3.5-6.5$	 & $0.69\pm0.09$ & $5.65\pm0.22$ & 0.2 & 6 & 100 & 100 & $10^4$ \\
& $5\times10^{10}$ &	$3.5-8.5$	 & $0.66\pm0.03$ & $5.82\pm0.11$ & 2.4 & 10 & 100 & 100 & $10^4$ \\
central & $1\times10^{11}$ &	$3.5-9$	 & $0.66\pm0.03$ & $5.84\pm0.10$ & 1.9 & 11 & 100 & 100 & $10^4$\\
galaxies & $5\times10^{11}$ &	$3.5-10$	 & $0.64\pm0.02$ & $5.85\pm0.09$ & 1.3 & 13 & 100 & 100 & $10^4$\\
& $1\times10^{12}$	 &	$3.5-10$	 & $0.62\pm0.03$ & $5.89\pm0.09$ & 0.9 & 13 & 100 & 100 & $10^5$ \\
& $2\times10^{12}$	 &	$3.5-10$	 & $0.61\pm0.03$ & $5.91\pm0.10$ & 1.6 & 13 & 100 & 100 & $10^5$ \\
& $1\times10^{13}$ &	$4-10$	 & $	0.59\pm0.02$ & $5.97\pm0.07$ & 1.1 & 12 & 100 & 100 & $10^6$ \\
& $1\times10^{14}$ &	$5-10$	 & $0.57\pm0.04$ & $5.99\pm0.12$ & 2.5 & 10 & 10 & 10 & $10^7$ \\
& $1\times10^{15}$	 &	$4.5-10$	 & $0.59\pm0.01$ & $5.94\pm0.03$ & 9.2 & 11 & 10 & 10 & $10^8$ \\
\hline
\end{tabular}
\tablecomments{A summary of the linear least-squares fitting results for the
median cold gas mass--stellar mass relations.  The $M_{\text{halo}}$ is the host
DM halo mass at redshift zero, the column of $\log(M_*/\msun)$ gives
the stellar mass range in the linear fitting, $\alpha_{\text{gs}}$ \&
$b_{\text{gs}}$ are the best-fit slopes and intercepts obtained from the
fitting, respectively (see Eq.~\ref{eq:gs}), $\chi^2$ is the chi-square value
of the best-fit, $N_{\text{bin}}$ is the number of the bins used in each
fitting, $N_{\text{tree}}$ is the number of the trees stacked together for the
fitting, $N_{\text{galaxy}}$ is the number of evolution tracks of the
corresponding galaxy objects in stacked halos, and
$M_{\rm res}$ is the minimum progenitor halo mass set in the halo merger trees.
The best-fit results are also
shown in Figures~\ref{fig:gs_sat_stack}--\ref{fig:gs_cet}.
See also Section~\ref{sec:gs}.
}
\label{tab:gs}
\end{table*}

\begin{table*}[b]
\centering
\caption{Best-fit results for the median $\beta$-$M_*$ scaling relations}
\begin{tabular}{c|c|c|c|c|c|c|c|c|c}
\hline
\hline
Objects & $M_{\text{halo}}/h^{-1}M_{\odot}$ & $\log(M_*/M_{\odot})$ & $\alpha_{\beta{\rm s}}$ & $b_{\beta{\rm s}}$ & $\chi^2$ & $N_{\text{bin}}$  & $N_{\text{tree}}$ & $N_{\text{galaxy}}$ & $M_{\rm res}/(h^{-1}M_{\odot})$ \\
\hline
& $1\times10^{10}$ &    $3-4$    & $-0.18\pm0.70$ & $2.37\pm0.58$ & - & 2 & 100 & 9 & $10^4$ \\
& $5\times10^{10}$ &    $3-7$    & $-0.23\pm0.03$ & $2.44\pm0.12$ & 0.04 & 8 & 100 & 62 & $10^4$ \\
satellite & $1\times10^{11}$ &  $3-7.5$  & $-0.23\pm0.03$ & $2.47\pm0.10$ & 0.10 & 9 & 100 & 165 & $10^4$ \\
progenitors & $5\times10^{11}$ &  $3-9$  & $-0.26\pm0.03$ & $2.55\pm0.09$ & 0.03 & 14 & 100 & 2163 & $10^4$ \\
& $1\times10^{12}$       &      $3-10$   & $-0.30\pm0.02$ & $2.65\pm0.08$ & 0.28 & 14 & 100 & 5970 & $10^5$ \\
& $2\times10^{12}$       &      $3-10$   & $-0.29\pm0.03$ & $2.62\pm0.09$ & 0.07 & 14 & 100 & 15265 & $10^5$ \\
& $1\times10^{13}$ &    $3-10$   & $-0.30\pm0.03$ & $2.65\pm0.09$ & 0.09 & 14 & 100 & 102164 & $10^6$ \\
& $1\times10^{14}$ &    $3-10$   & $-0.31\pm0.03$ & $2.68\pm0.09$ & 0.12 & 14 & 10 & 85563 & $10^7$ \\
& $1\times10^{15}$       &      $3-10$   & $-0.25\pm0.02$ & $2.46\pm0.06$ & 0.46 & 14 & 10 & 543679 & $10^8$ \\
\hline
& $1\times10^{10}$ &    $3.5-6.5$        & $-0.13\pm0.08$ & $2.30\pm0.18$ & 0.04 & 6 & 100 & 100 & $10^4$ \\
& $5\times10^{10}$ &    $3.5-8.5$        & $-0.23\pm0.03$ & $2.49\pm0.09$ & 0.48 & 10 & 100 & 100 & $10^4$ \\
central & $1\times10^{11}$ &    $3.5-9$  & $-0.28\pm0.02$ & $2.59\pm0.07$ & 0.47 & 11 & 100 & 100 & $10^4$ \\
galaxies & $5\times10^{11}$ &   $3.5-10$         & $-0.34\pm0.02$ & $2.73\pm0.05$ & 0.24 & 13 & 100 & 100 & $10^4$ \\
& $1\times10^{12}$       &      $3.5-10$         & $-0.36\pm0.02$ & $2.74\pm0.05$ & 0.09 & 13 & 100 & 100 & $10^5$ \\
& $2\times10^{12}$       &      $3.5-10$         & $-0.40\pm0.02$ & $2.82\pm0.06$ & 0.04 & 13 & 100 & 100 & $10^5$ \\
& $1\times10^{13}$ &    $4-10$   & $-0.45\pm0.02$ & $2.89\pm0.06$ & 0.12 & 12 & 100 & 100 & $10^6$ \\
& $1\times10^{14}$ &    $5-10$   & $-0.58\pm0.03$ & $3.26\pm0.10$ & 0.10 & 10 & 10 & 10  & $10^7$ \\
& $1\times10^{15}$       &      $4.5-10$   & $-0.58\pm0.02$ & $3.21\pm0.04$ & 0.53 & 11 & 10 & 10 & $10^8$ \\
\hline
\end{tabular}

\tablecomments{A summary of the linear least-squares fitting results for the
median stellar feedback efficiency--stellar mass relations. This table is similar
as Table~\ref{tab:gs}, except that the physical variable $M_{\rm cold}$ is
replaced by the stellar feedback efficiency $\beta$.  The fitting results are
also shown in Figures~\ref{fig:bs_sat_stack}--\ref{fig:bs_cet}.  
See also Section~\ref{sec:bs}.
}
\label{tab:bs} \end{table*}

\begin{figure}[htb]
\includegraphics[width=0.45\linewidth]{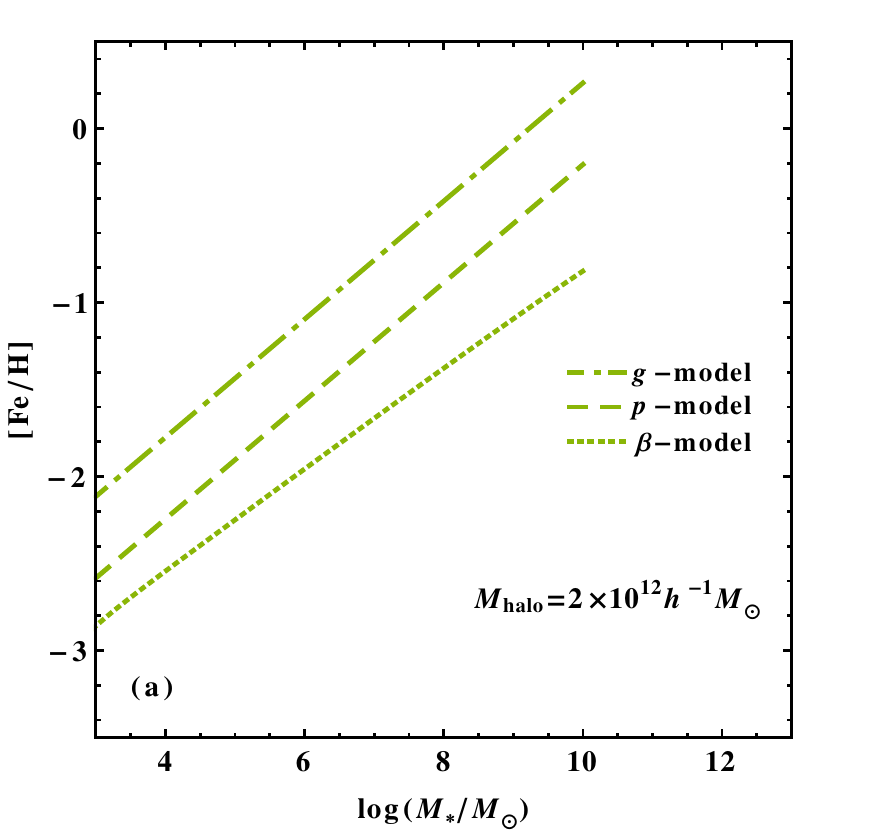}
\includegraphics[width=0.45\linewidth]{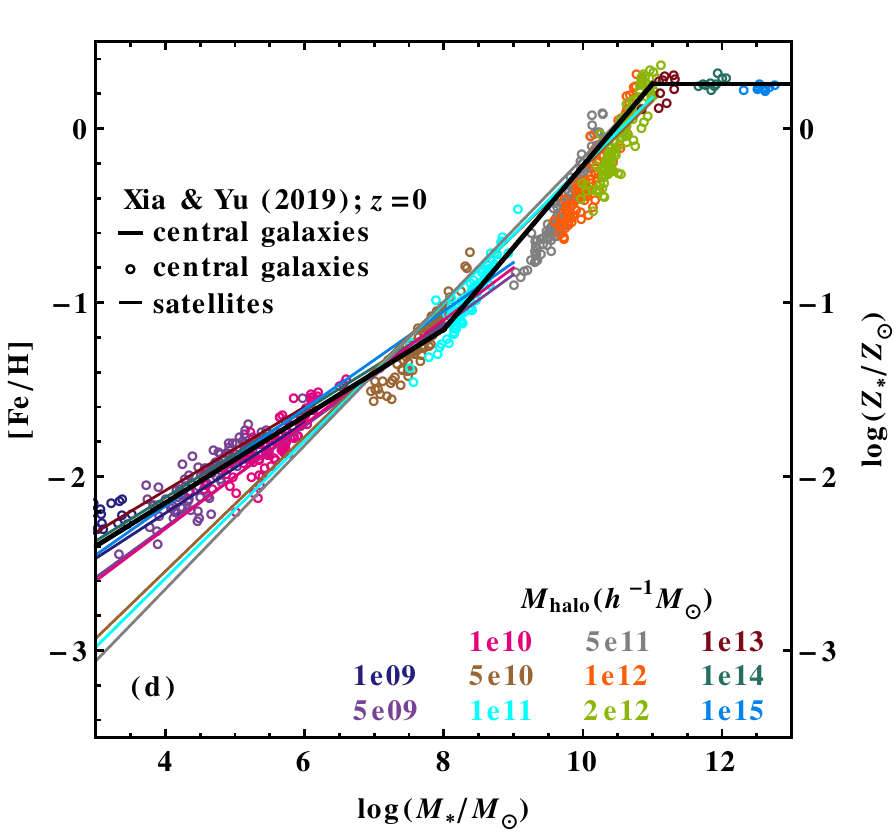} \\
\includegraphics[width=0.45\linewidth]{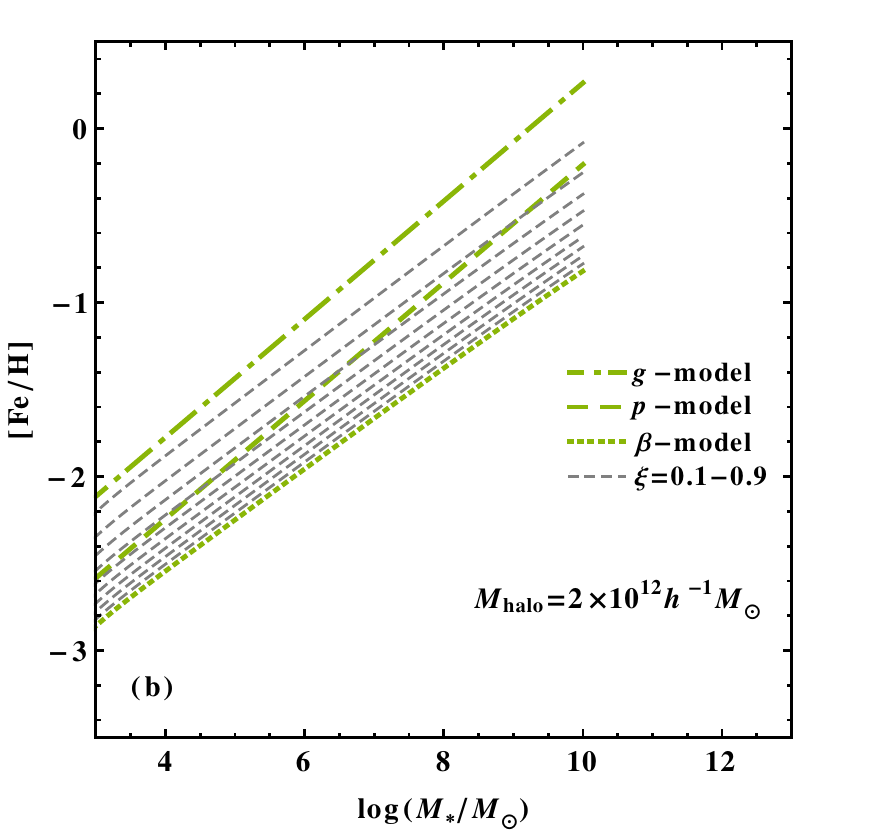}
\includegraphics[width=0.45\linewidth]{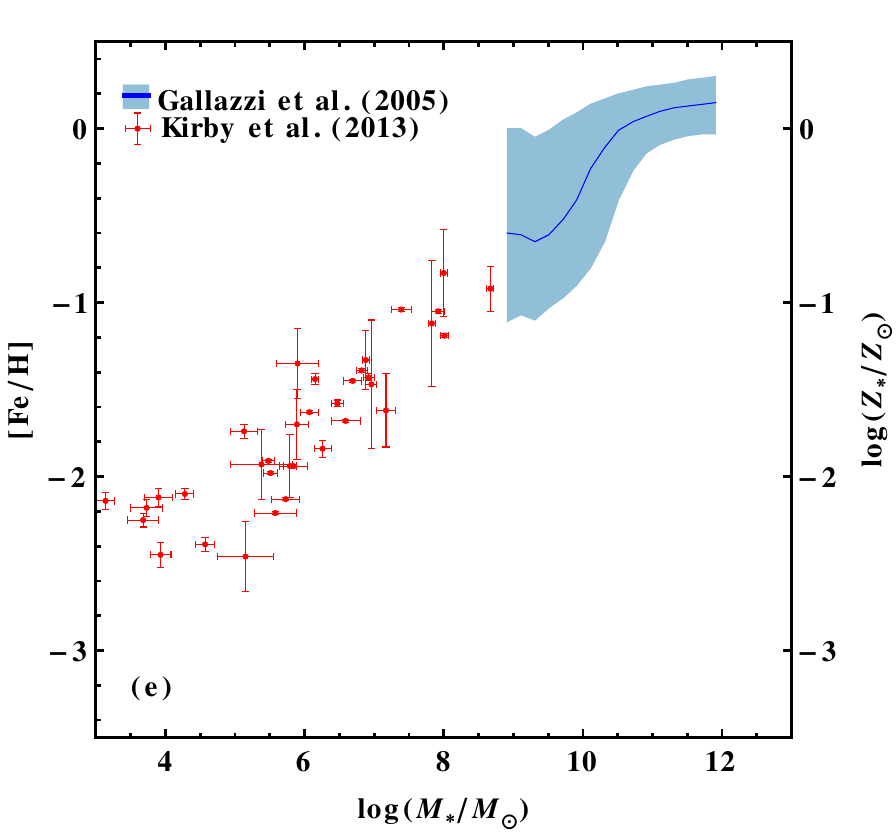} \\
\includegraphics[width=0.45\linewidth]{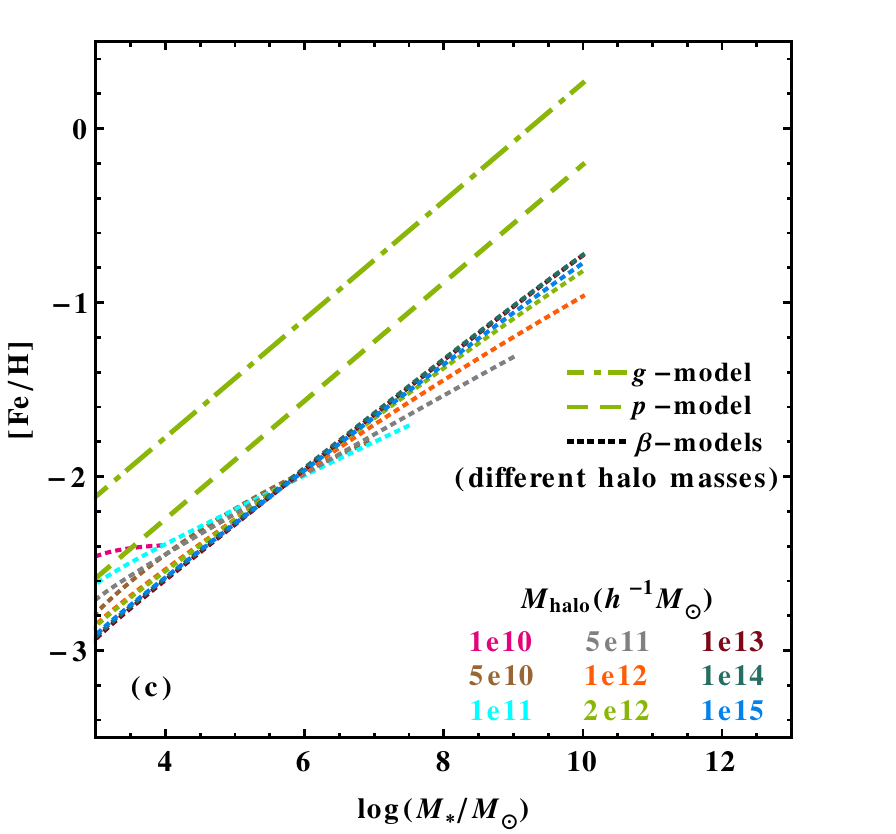}
\includegraphics[width=0.45\linewidth]{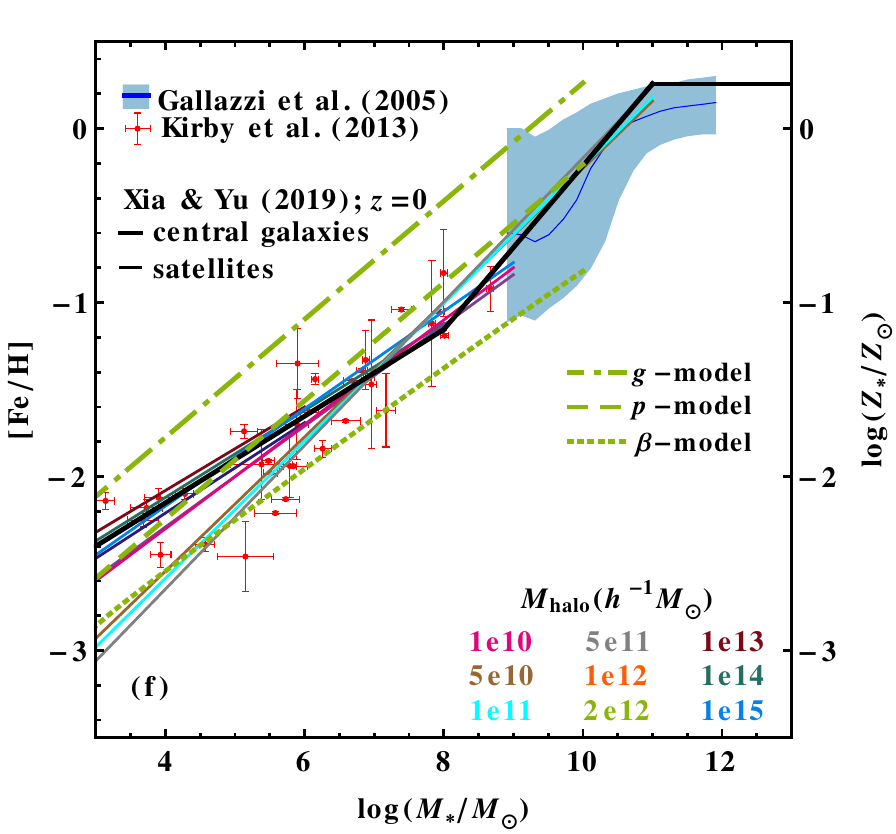}
\caption{}
\end{figure}

\begin{figure}[htb]
\contcaption{Panel (a): The MZRs predicted by our analytical chemical evolution
models constructed in Section~\ref{sec:model}. The MZRs are obtained by
applying the best-fit $M_{\rm cold}$--$M_*$ and $\beta$--$M_*$ scaling
relations to the analytical solution of our chemical evolution model. The
dotted, the dashed, and the dot-dashed lines represent the results obtained by
the $\beta$-, $p$-, and g-models, respectively, which have different
assumptions on the ratio of $\xi=Z_{\rm hot}/Z_{\rm cold}$ in
Section~\ref{sec:model}. The best-fit $M_{\rm cold}$--$M_*$ and $\beta$--$M_*$
scaling relations are adopted from those obtained for satellite progenitors in
halos with $M_{\rm halo}=2\times 10^{12} h^{-1}\msun$, with $\alpha_{\rm gs}=0.66\pm0.03$ and $\alpha_{\beta{\rm s}}=-0.29\pm0.03$ as shown in
Tables~\ref{tab:gs} and \ref{tab:bs}, respectively. The slopes of the three
lines are $\sim 0.3$ ($\sim -\alpha_{\beta{\rm s}}$ or $1-\alpha_{\rm gs}$).
Panel (b): The MZRs predicted by our analytical chemical evolution models with
other different $\xi$ are shown by the grey dashed lines,
where the same example as shown in panel (a) is used (for satellite progenitors in
halos with $M_{\rm halo}=2\times 10^{12} h^{-1}\msun$).
Panel (c): The $\beta$-model results obtained for different halo masses are
shown by the dotted lines with different colors, where the best-fit $M_{\rm
cold}$--$M_*$ and $\beta$--$M_*$ scaling relations obtained for the satellite
progenitors with the corresponding halo masses are applied. We do not show the
g- and $p$-model results for other different halo masses, as the model results
depend on the ratio of $M_*/M_{\rm cold}$ (see Eqs.~\ref{eq:gis} and \ref{eq:pis})
the $M_{\rm cold}$--$M_*$ relations appear quite universal for the different
halo masses as shown in Figure~\ref{fig:gs_full}(a).
Panel (d): The MZRs obtained from our SAM simulations in \citet{XY19}.
The open circles represent central galaxies within DM halos with different
masses, which are adopted
from figure 4(a) in \citet{XY19}. The thick solid lines represent the fitting
results to the open circles, which are the same as the dotted lines shown
in figure 4(a) in \citet{XY19}.
The thin solid lines represent our simulation results
for the satellites within different DM halos, which are adopted from the solid
lines in figure 4(c) in \citet{XY19}. 
Panel (e): Observational MZRs, which are the same as figure 4(b) in
\citet{XY19}.  The red filled squares are the observational results for dwarf
galaxies in the Local Group (see Figure 9 in \citealt{Kirbyetal13}). The blue
solid line gives the median of the MZR for 44,254 late-type galaxies drawn from
SDSS DR2 \citep{gallazzi05}, and the light blue region is between 16th and 84th
percentiles of the distribution.
Panel (f): combination of panels (a), (d), and (e), where the open circles
shown in panel (d) are removed for view clarity.  See
Section~\ref{sec:mzr_mod}.
This figure shows that both the MZR results obtained from the SAM in \citet{XY19}
and observations fall in the gap between the g- and $\beta$-model results in 
the range of $10^3M_{\odot}\leq M_*\leq 10^{10}M_{\odot}$ as shown in panel (f).
Note that the g-model and the $\beta$-model represent the upper and lower bounds
on the assumption of the infalling hot gas metallicity (with $\xi=1$ and
0, respectively), and the differences among the g-, $p$-, and $\beta$-model results
are not larger than 1 dex, so the MZRs are located within a narrow space and appear
universal as shown in \citet{XY19}.
\label{fig:mzr}} \end{figure}
\end{document}